\documentclass[
 reprint,
 showpacs,preprintnumbers,
 amsmath,amssymb,
 aps,
 pra,
 floatfix,
]{revtex4-1}

\usepackage{graphicx}
\usepackage{tikz}
\usetikzlibrary{arrows}

\makeatletter
\newcommand{\colorcaption}[2][]{%
  \begingroup%
  \renewcommand{\@caption@fignum@sep}{.}%
  \caption[#1]{#2}%
  \endgroup%
}
\makeatother

\usepackage{dcolumn}
\usepackage{bm}
\usepackage{bbm}
\usepackage{upgreek}
\usepackage{hyperref}
\usepackage{blindtext}

\newcommand{\im}{\mathrm{i}}

\newcommand{\slfrac}[2]{\left.#1\middle/#2\right.}

\begin{document}

	\preprint{}

	\title{Magnetoanisotropic Josephson effect due to interfacial spin-orbit fields in superconductor/ferromagnet/superconductor junctions}
	
	\author{Andreas Costa}%
	\email[Corresponding author: ]{andreas.costa@physik.uni-regensburg.de}
 	\affiliation{Institute for Theoretical Physics, University of Regensburg, 93040 Regensburg, Germany}
 	
	\author{Petra H{\"o}gl}%
 	\affiliation{Institute for Theoretical Physics, University of Regensburg, 93040 Regensburg, Germany}
 	
	\author{Jaroslav Fabian}%
 	\affiliation{Institute for Theoretical Physics, University of Regensburg, 93040 Regensburg, Germany}
 
	\date{\today}

\begin{abstract}
We study theoretically the effects of interfacial Rashba and Dresselhaus spin-orbit~coupling in superconductor/ferromagnet/superconductor~(S/F/S)~Josephson~junctions\textemdash with allowing for tunneling~barriers between the ferromagnetic and superconducting~layers\textemdash by solving the Bogoljubov--de~Gennes~equation for realistic heterostructures and applying the Furusaki-Tsukada~technique to calculate the electric current at a finite temperature. 
The presence of spin-orbit~couplings leads to out- and in-plane~magnetoanisotropies of the Josephson~current, which are giant in comparison to current magnetoanisotropies in similar normal-state ferromagnet/normal~metal~(F/N)~junctions. Especially huge anisotropies appear in the vicinity of $ 0 $-$ \pi $~transitions, caused by the exchange-split~bands in the ferromagnetic metal layer. We also show that the direction of the Josephson~critical~current can be controlled (inducing $ 0 $-$ \pi $~transitions) by the strength of the spin-orbit~coupling and, more crucial, by the orientation of the magnetization. Such a control can bring new functionalities into 
Josephson~junction~devices.  
\end{abstract}


	\maketitle

\section{Introduction\label{SectionI}}

The interplay of superconductivity and ferromagnetism can bring spectacular effects~\cite{Eschrig2011, Linder2015, Gingrich2016}. Perhaps most striking is the emergence of $ \pi $~states~\cite{Golubov2004, Buzdin2005, Bergeret2005, Bulaevskii1977, *Bulaevskii1977alt, Buzdin1982, *Buzdin1982alt} in S/F/S~junctions. The exchange~coupling in the ferromagnetic layer can add an extra $ \pi $~shift to the superconducting phase~difference and lead to a reversal of the Josephson~current, compared to the usual state~($ 0 $~state) of the junction. The initial demonstration of the $ \pi $~state in Nb/CuNi/Nb~trilayers~\cite{Ryazanov2001}, and subsequent experimental studies~\cite{Kontos2002, Robinson2006, Feofanov2010, Gingrich2016}, have boosted hopes for finding ways to control 0-$ \pi $~transitions, thereby controlling the direction of the supercurrent. Such a control could not only be important for manipulating proposed superconducting $\pi $~qubits~\cite{Yamashita2005}, but also for bringing spintronics functionalities~\cite{Fabian2004, Fabian2007} into superconducting quantum computing circuits~\cite{Ioffe1999, Mooij1999, Devoret2013} and Josephson~junction~technology~\cite{Likharev1979, Likharev2012, Feofanov2010}.

Contact interfaces invariably introduce spin-orbit~fields into the constituent regions. There is always the Rashba~(or Bychkov-Rashba) field~\cite{Bychkov1984}, which is present due to the space~inversion~asymmetry of the heterostructure. If also bulk~inversion~symmetry is broken, as it is the case with tunneling~barriers of III-V~zinc-blende~semiconductors such as GaAs~\cite{Gmitra2013}, there will additionally be a spin-orbit~field of the Dresselhaus~type~\cite{Dresselhaus1955}. The interference of both spin-orbit~fields results in a $ C_{2v} $~symmetric field~\cite{Fabian2007,MatosAbiague2009}, which reflects the symmetry of the corresponding interface. In normal-state F/N~junctions, these spin-orbit~fields are responsible for transport magnetoanisotropies, exemplified by the tunneling anisotropic magnetoresistance~(TAMR)~\cite{Brey2004}, which has already been observed in epitaxial-quality Fe/GaAs/Au~tunnel~junctions~\cite{Moser2007}. Much larger anisotropies were recently predicted for the differential conductance of superconducting F/S~junctions~\cite{Hoegl2015}, mainly caused by unconventional Andreev~reflection of incoming electrons at the F/S~interface. Therefore, the related effect was termed magnetoanisotropic Andreev~reflection~(MAAR).

Many unique phenomena are bound to occur when spin-orbit~fields couple with magnetism and superconductivity. This topic is driven mainly by the research of Majorana~states, which are believed to appear in the presence of spin-orbit~fields in superconducting proximity regions~\cite{Oreg2010, Mourik2012, Rokhinson2012}, even in the presence of a magnetic order~\cite{Duckheim2011, Nadj-Perge2014}. In F/S~junctions, a supercurrent can be spin~polarized due to the formation of Cooper~pair~triplets~\cite{Bergeret2001, Volkov2003, Halterman2007, Eschrig2008,Sun2015}. It has been proposed that spin-orbit~coupling can facilitate the triplets~formation, leading to a long-range proximity~effect in ferromagnets~\cite{Bergeret2013, Bergeret2014, Jacobsen2015}. Spin-orbit~fields can even induce superconducting proximity~effects in half~metals~\cite{Duckheim2011}. Moreover, magnetic anisotropies of the critical~current with respect to the orientation of the present spin-orbit~fields have been predicted to occur in lateral S/nanowire/S~Josephson~junctions with a Zeeman~splitting~\cite{Yokoyama2014, Yokoyama2014a, Arjoranta2016}, as well as in diffusive vertical S/F/S~Josephson~junctions~\cite{Jacobsen2016}.

In this paper we investigate the (dc)~Josephson~effect~\cite{Josephson1962, Josephson1964} in ballistic vertical S/F/S~junctions in the presence of interfacial Rashba~\cite{Bychkov1984} and Dresselhaus~\cite{Dresselhaus1955}~spin-orbit~fields. We are particularly interested in the unique signatures of the interplay of the Josephson~effect, spin-orbit~fields, and ferromagnetism. The paper is structured in the following way. The theoretical model used for our studies is introduced step by step in~Sec.~\ref{SectionII}. We construct the Bogoljubov--de~Gennes scattering~states in the different regions of the Josephson~junction for the injection of electronlike and holelike~quasiparticles from the left superconducting electrode, and apply the Furusaki-Tsukada~method~\cite{Furusaki1991} to express the total Josephson~current in terms of scattering~coefficients in the Bogoljubov--de~Gennes scattering~states. In Sec.~\ref{SectionIII} we first concentrate on S/F/S~Josephson~junctions in which the interfacial spin-orbit~fields are absent. We numerically evaluate the Josephson~current for realistic model~junctions and recover previously obtained results~\cite{Radovic2003}, suggesting that transitions between $ 0 $ and $ \pi $~states can be controlled by altering the thickness of the metallic interlayer. In the following section we study the impact of interfacial SOC on the Josephson~current~flow. We exemplarily focus on the effects caused by the presence of Rashba~spin-orbit~fields. On the one hand, our calculations confirm that interfacial spin-orbit~fields can indeed convert spin-singlet into spin-triplet~Cooper~pairs via interfacial spin~flips and remarkably enhance the Josephson~current, as already observed in diffusive $ \mathrm{NbTiN} $/$ \mathrm{CrO}_2 $/$ \mathrm{NbTiN} $~Josephson~junctions~\cite{Keizer2006}. On the other hand, we predict that modulating the strengths of the Rashba~fields may also induce a switching between $ 0 $ and $ \pi $~states, without changing the thickness of the metallic interlayer. Section~\ref{SectionV} is devoted to the impact of differing Fermi~wave~vectors or effective~masses in the superconducting and ferromagnetic components on the Josephson~current~flow. Furthermore, we show in Sec.~\ref{SectionVI} that interfacial Rashba~spin-orbit~fields give rise to marked out-of-plane~magnetoanisotropies of the Josephson~current, while the interference of Rashba and Dresselhaus~fields leads to in-plane~magnetoanisotropies. The amplitudes of this \textit{magnetoanisotropic~Josephson~current~(MAJC)} are not only giant when compared to the normal-state~TAMR~\cite{Moser2007, MatosAbiague2009}, but even larger than the recently predicted giant MAAR~ratios in single F/S~tunnel~junctions~\cite{Hoegl2015}. Finally, we demonstrate that also the magnetization~orientation can be used in experiments to manipulate $ 0 $ to $ \pi $~transitions in Sec.~\ref{SectionVII}. Additional calculations to clarify the influence of the interlayer~thickness or the strength of the exchange~splitting in the ferromagnet on the outcomes are attached in the Appendices.

\section{Theoretical model\label{SectionII}}

\begin{figure}
	\includegraphics[width=0.48\textwidth]{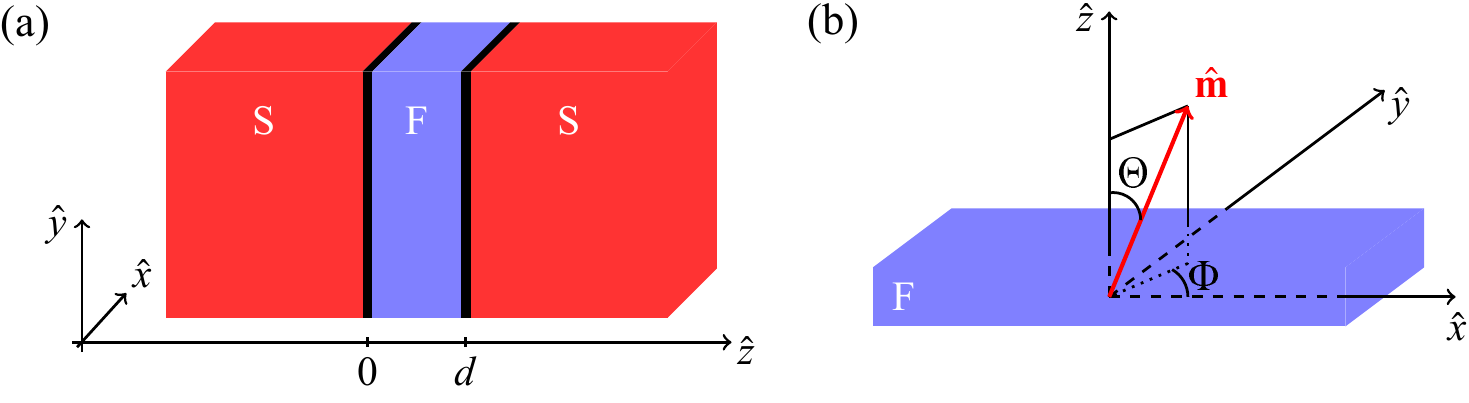}
	\colorcaption{\label{FigSystem}  (a)~Sketch of the considered S/F/S~Josephson~junction, using in general $ C_{2v} $ principal crystallographic orientations~$ \hat{x}=[110] $, $ \hat{y}=[\overline{1}10] $, and $ \hat{z}=[001] $. (b)~The direction of the magnetization~vector~$ \hat{\mathbf{m}} $ in the ferromagnetic layer of the Josephson~junction is determined by the polar~angle~$ \Theta $ and azimuthal~angle~$ \Phi $.}
\end{figure}

The considered S/F/S~Josephson~junction consists of two semi-infinite superconducting regions~($ z<0 $ and $ z>d $), which are weakly coupled by a ferromagnetic link with thickness~$ d $. The system is schematically shown in Fig.~\ref{FigSystem}(a). 
At the flat interfaces between the different parts of the system, ultrathin tunneling~barriers introduce potential and spin-orbit~coupling~(SOC)~scattering. To compute the (dc)~Josephson~current flowing across the Josephson~junction, we generalize the Furusaki-Tsukada~technique~\cite{Furusaki1991}, which allows us to relate the Josephson~current to the Andreev~reflection~coefficients in the Bogoljubov--de~Gennes scattering~states for incoming quasiparticles from the left superconducting electrode. The addressed scattering~states~$ \Psi(\mathbf{r}) $ for quasiparticles with excitation~energy~$ E $ are obtained by solving the stationary Bogoljubov--de~Gennes~equation~\cite{DeGennes1989, Zutic1999, Zutic2000}
\begin{equation}
	\label{EqBogoljubovdeGennes}
	\left[ \begin{matrix} \hat{H}_\mathrm{e} & \hat{\Delta}_\mathrm{S} \\ \hat{\Delta}_\mathrm{S}^\dagger & \hat{H}_\mathrm{h} \end{matrix} \right] \, \Psi(\mathbf{r}) = E \, \Psi(\mathbf{r}) ,
\end{equation}
where the single-particle~Hamiltonian for electrons reads $ \hat{H}_\mathrm{e} = \left\{  - \left( \hbar^2/2 \right) \boldsymbol{\nabla} \left[ 1/m(z) \right] \boldsymbol{\nabla} - \mu(z) \right\} \mathbbm{1}_{2 \times 2} - (\Delta_\mathrm{XC}/2) \Theta(z) \Theta(z- d) (\hat{\mathbf{m}} \cdot \hat{\boldsymbol{\sigma}}) + \hat{H}^\mathrm{int}_\mathrm{L} + \hat{H}^\mathrm{int}_\mathrm{R} $, whereas the one for holes is $ \hat{H}_\mathrm{h} = -\hat{\sigma}_y \hat{H}_\mathrm{e}^* \hat{\sigma}_y $. The effective~masses of quasiparticles $ m(z) $ are $ m_\mathrm{S} $ in the two superconducting regions~($ z < 0 $ and $ z > d $) and $ m_\mathrm{F} $ in the ferromagnet~($ 0 < z < d $). Accordingly, the chemical~potentials~$ \mu(z) $ are given by $ \mu_\mathrm{S} $ and $ \mu_\mathrm{F} $. The ferromagnetic material is described by the Stoner~band~model with the exchange~energy~gap~$ \Delta_\mathrm{XC} $. The magnetization~direction is determined by the unit~vector~$ \hat{\mathbf{m}} = ( \sin \Theta \cos \Phi, \, \sin \Theta \sin \Phi, \, \cos \Theta ) $~[see Fig.~\ref{FigSystem}(b)] and $ \hat{\boldsymbol{\sigma}} $ comprises the Pauli~spin~matrices. The potential and SOC~scattering at the left~(L) and right~(R) interfaces are modeled by $ \hat{H}^\mathrm{int}_\mathrm{L}= \left( V_\mathrm{L} d_\mathrm{L} \mathbbm{1}_{2 \times 2} + \boldsymbol{\Omega}_\mathbf{L} \cdot \hat{\boldsymbol{\sigma}} \right) \, \delta(z) $ and $ \hat{H}^\mathrm{int}_\mathrm{R}= \left( V_\mathrm{R} d_\mathrm{R} \mathbbm{1}_{2 \times 2} + \boldsymbol{\Omega}_\mathbf{R} \cdot \hat{\boldsymbol{\sigma}} \right) \, \delta(z-d) $, where $ V_\mathrm{L \; (R)} $ and $ d_\mathrm{L \; (R)} $ are the heights and widths of the deltalike~barriers, respectively, while the effective interfacial spin-orbit~fields~$ \boldsymbol{\Omega}_\mathbf{L}=\left[ (\alpha_\mathrm{L} -\beta_\mathrm{L}) \, k_y , \, -(\alpha_\mathrm{L} +\beta_\mathrm{L}) \, k_x , \, 0 \right] $ and $ \boldsymbol{\Omega}_\mathbf{R}=-\left[ (\alpha_\mathrm{R} -\beta_\mathrm{R}) \, k_y , \, -(\alpha_\mathrm{R} +\beta_\mathrm{R}) \, k_x , \, 0 \right] $ include both Rashba and linear Dresselhaus terms~\cite{Fabian2004, Fabian2007}, parametrized by $ \alpha_\mathrm{L \; (R)} $ and $ \beta_\mathrm{L \; (R)} $. Neglecting the proximity~effect in the ferromagnetic layer, the superconducting~pairing~potential can be approximated by the steplike behavior $ \hat{\Delta}_\mathrm{S} = \left[ \left| \Delta_\mathrm{S} \right| \Theta(-z) + \left| \Delta_\mathrm{S} \right| \, \mathrm{e}^{\mathrm{i} \phi_\mathrm{S}} \Theta(z-d) \right] \mathbbm{1}_{2 \times 2} $~(the accuracy of this approximation was discussed, for~instance, by  Likharev~\cite{Likharev1979} and Beenakker~\cite{Beenakker1997} earlier), where $ \left| \Delta_\mathrm{S} \right| $ is the isotropic energy~gap in the two $ s $-wave~superconductors and $ \phi_\mathrm{S} $ is the macroscopic phase~difference across the junction.

We first solve the Bogoljubov--de~Gennes~equation in the three regions of the Josephson~junction separately to obtain the corresponding scattering~states~$ \Psi(\mathbf{r}) $ for the injection of electronlike and holelike~quasiparticles from the left superconducting electrode. Since the wave~vector $ \mathbf{k}_\parallel = \left[ k_x, \, k_y, \, 0 \right]^\mathrm{T} $ parallel to the interfaces is conserved, we can substitute $ \Psi(\mathbf{r}) = \Psi(z) \mathrm{e}^{\mathrm{i} \mathbf{k}_\parallel \cdot \mathbf{r}_\parallel} $~($ \mathbf{r}_\parallel = \left[ x, \, y, \, 0 \right]^\mathrm{T} $) in~Eq.~\eqref{EqBogoljubovdeGennes} to reduce the scattering~problem to an effective one-dimensional description for the unknown states~$ \Psi(z) $. For an incident electronlike~quasiparticle with spin~up from the left superconducting lead, the solutions of the reduced Bogoljubov--de~Gennes~equation in the three components of the junction are
\begin{eqnarray}
	\Psi^{(1)} (z < 0) &&= \mathrm{e}^{\mathrm{i} q_\mathrm{ez} z} \left[ \begin{matrix} u \\ 0 \\ v \\ 0 \end{matrix} \right] \nonumber \\
	&&+ a^{(1)} \mathrm{e}^{-\mathrm{i} q_\mathrm{ez} z} \left[ \begin{matrix} u \\ 0 \\ v \\ 0 \end{matrix} \right] + b^{(1)} \mathrm{e}^{-\mathrm{i} q_\mathrm{ez} z} \left[ \begin{matrix} 0 \\ u \\ 0 \\ v \end{matrix} \right] \nonumber \\
	&&+ c^{(1)} \mathrm{e}^{\mathrm{i} q_\mathrm{hz} z} \left[ \begin{matrix} 0 \\ v \\ 0 \\ u \end{matrix} \right] + d^{(1)} \mathrm{e}^{\mathrm{i} q_\mathrm{hz} z} \left[ \begin{matrix} v \\ 0 \\ u \\ 0 \end{matrix} \right] ,
\end{eqnarray}
\begin{eqnarray}
	\label{EqStatesFM}
	\Psi^{(1)} (0 < z < d) &&= e^{(1)} \mathrm{e}^{\mathrm{i} k_\mathrm{ez}^\uparrow z} \chi_\mathrm{e}^\uparrow + f^{(1)} \mathrm{e}^{\im k_\mathrm{ez}^\downarrow z} \chi_\mathrm{e}^\downarrow \nonumber \\
	&&+ g^{(1)} \mathrm{e}^{-\mathrm{i} k_\mathrm{hz}^\uparrow z} \chi_\mathrm{h}^\uparrow + h^{(1)} \mathrm{e}^ {-\mathrm{i} k_\mathrm{hz}^\downarrow z} \chi_\mathrm{h}^\downarrow \nonumber \\
	&&+ i^{(1)} \mathrm{e}^{-\mathrm{i} k_\mathrm{ez}^\uparrow z} \chi_\mathrm{e}^\uparrow + j^{(1)} \mathrm{e}^{-\mathrm{i} k_\mathrm{ez}^\downarrow z} \chi_\mathrm{e}^\downarrow \nonumber \\
	&&+ k^{(1)} \mathrm{e}^{\mathrm{i} k_\mathrm{hz}^\uparrow z} \chi_\mathrm{h}^\uparrow + l^{(1)} \mathrm{e}^{\mathrm{i} k_\mathrm{hz}^\downarrow z} \chi_\mathrm{h}^\downarrow ,
\end{eqnarray}
and
\begin{eqnarray}
	\label{EqStatesSC}
	\Psi^{(1)} (z > d) &&= m^{(1)} \mathrm{e}^{\mathrm{i} q_\mathrm{ez} z} \left[ \begin{matrix} u \mathrm{e}^{\mathrm{i} \phi_\mathrm{S}} \\ 0 \\ v \\ 0 \end{matrix} \right] + n^{(1)} \mathrm{e}^{\mathrm{i} q_\mathrm{ez} z} \left[ \begin{matrix} 0 \\ u \mathrm{e}^{\mathrm{i} \phi_\mathrm{S}} \\ 0 \\ v \end{matrix} \right] \nonumber \\
	&&+ o^{(1)} \mathrm{e}^{-\mathrm{i} q_\mathrm{hz} z} \left[ \begin{matrix} v \mathrm{e}^{\mathrm{i} \phi_\mathrm{S}} \\ 0 \\ u \\ 0 \end{matrix} \right] + p^{(1)} \mathrm{e}^{-\mathrm{i} q_\mathrm{hz} z} \left[ \begin{matrix} 0 \\ v \mathrm{e}^{\mathrm{i} \phi_\mathrm{S}} \\ 0 \\ u \end{matrix} \right] , \nonumber \\
\end{eqnarray}
with the BCS~coherence~factors
\begin{equation}
	u \, (v) = \sqrt{ \frac{1}{2} \left( 1 + (-) \frac{\sqrt{E^2-\left| \Delta_\mathrm{S} \right|^2}}{E} \right)} .
\end{equation}
The $ \hat{z} $-components of the wave~vectors for electronlike~(holelike)~quasiparticles in the superconducting regions can be written as $ q_\mathrm{ez \, (hz)} = \sqrt{q_\mathrm{F}^2 + (-) 2 m_\mathrm{S}/\hbar^2 \sqrt{E^2-\left| \Delta_\mathrm{S} \right|^ 2} - \left| \mathbf{k}_\parallel \right|^2} $, whereas the spin-resolved wave~vectors for electrons and holes in the Stoner~ferromagnet with a spin parallel~($ \uparrow $) or antiparallel~($ \downarrow $) to the magnetization~direction~$ \hat{\mathbf{m}} $ are $ k_\mathrm{ez}^{\uparrow \, (\downarrow)} = \sqrt{k_\mathrm{F}^2 + 2 m_\mathrm{F} / \hbar^2 \left( E+(-) \Delta_\mathrm{XC}/2 \right) - \left| \mathbf{k}_\parallel \right|^2} $ as well as $ k_\mathrm{hz}^{\uparrow \, (\downarrow)} = \sqrt{k_\mathrm{F}^2 + 2 m_\mathrm{F} / \hbar^2 (-E+(-) \Delta_\mathrm{XC}/2) - \left| \mathbf{k}_\parallel \right|^2} $, respectively. Thereby, $ q_\mathrm{F} $ and $ k_\mathrm{F} $ denote the Fermi~wave~vectors in the superconducting and ferromagnetic constituents of the Josephson~junction. The spinors for electrons and holes in the ferromagnetic region have the form $ \chi_\mathrm{e}^{\uparrow \, (\downarrow)} = \left[ \chi^{\uparrow \, (\downarrow)}, \, 0 \right]^ \mathrm{T} $ and $ \chi_\mathrm{h}^{\uparrow \, (\downarrow)} = \left[ 0, \, \chi^{\downarrow \, (\uparrow)} \right]^\mathrm{T} $, both containing
\begin{equation}
	\chi^{\uparrow \, (\downarrow)} = \frac{1} {\sqrt{2}} \left[ \begin{matrix} (-) \sqrt{1+ (-) \cos \Theta} \, \mathrm{e}^{-\mathrm{i} \Phi} \\ \sqrt{1-(+) \cos \Theta} \end{matrix} \right] .
\end{equation}
The unknown scattering~coefficients~$ a^{(1)} $ and $ b^{(1)} $ in the given scattering~states indicate normal~reflection of the incoming electronlike~quasiparticle at the left interface without and with a spin~flip, respectively, while $ c^{(1)} $ and $ d^{(1)} $ are the corresponding spin-resolved Andreev~reflection~coefficients. Accordingly, transmission into the right superconductor as an electronlike or holelike~quasiparticle with spin~up or spin~down is incorporated in the amplitudes~$ m^{(1)} $, $ n^{(1)} $, $ o^{(1)} $, and $ p^{(1)} $. To attain these scattering~coefficients, we apply the boundary~conditions
\begin{widetext}
	\begin{eqnarray}
		\Psi^{(1)} (z) \big|_{z=0_-} &&= \Psi^{(1)}(z) \big|_{z=0_+} , \label{EqBound1} \\
		\Psi^{(1)} (z) \big|_{z=d_-} &&= \Psi^{(1)}(z) \big|_{z=d_+} , \label{EqBound2} \\
		\left[ -\frac{\hbar^2}{2 m_\mathrm{F}} \frac{\mathrm{d}}{\mathrm{d} z} + V_\mathrm{L} d_\mathrm{L} \right] \boldsymbol{\eta} \Psi^{(1)}(z) \big|_{z=0_+} + \left[ \begin{matrix} \mathbf{\Omega_L} \cdot \hat{\boldsymbol{\sigma}} & 0 \\ 0 & -\mathbf{\Omega_L} \cdot \hat{\boldsymbol{\sigma}} \end{matrix} \right] \Psi^{(1)} (z) \big|_{z=0_+} &&= -\frac{\hbar^2}{2m_\mathrm{S}} \frac{\mathrm{d}}{\mathrm{d} z} \boldsymbol{\eta} \Psi^{(1)}(z) \big|_{z=0_-} ,\label{EqBound3} \\
		\left[ \frac{\hbar^2}{2m_\mathrm{F}} \frac{\mathrm{d}}{\mathrm{d} z} + V_\mathrm{R} d_\mathrm{R} \right] \boldsymbol{\eta} \Psi^{(1)} (z) \big|_{z=d_-} + \left[ \begin{matrix} \mathbf{\Omega_R} \cdot \hat{\boldsymbol{\sigma}} & 0 \\ 0 & -\mathbf{\Omega_R} \cdot \hat{\boldsymbol{\sigma}} \end{matrix} \right] \Psi^{(1)} (z)\big|_{z=d_-} &&= \frac{\hbar^2}{2m_\mathrm{S}} \frac{\mathrm{d}}{\mathrm{d} z} \boldsymbol{\eta} \Psi^{(1)}(z) \big|_{z=d_+} , \label{EqBound4} 
	\end{eqnarray}
\end{widetext}
with
\begin{equation}
	\boldsymbol{\eta} = \left[ \begin{matrix} \mathbbm{1}_{2 \times 2} & 0 \\ 0 & -\mathbbm{1}_{2 \times 2} \end{matrix} \right] ,
\end{equation}
to the obtained scattering~states and numerically solve the resulting linear system of equations for the scattering~coefficients.

The Bogoljubov--de~Gennes scattering~states for the injection of an electronlike~quasiparticle with spin~down, as well as for incoming holelike~quasiparticles with spin~up or spin~down from the left superconducting electrode are constructed in the same way.

\newpage
\begin{widetext}
Following the Furusaki-Tsukada~technique~\cite{Furusaki1991}, the (dc)~Josephson~current is given by 
	\begin{eqnarray}
		\label{EqJosephsonCurrent}
		I_\mathrm{J} = \frac{e k_\mathrm{B} T}{4 \hbar} \left| \Delta_\mathrm{S}(T) \right| \frac{A}{4 \pi^2} \int \mathrm{d}^2 \mathbf{k_\parallel} \sum_{\omega_n} \frac{1}{\sqrt{\omega_n^2 + \left| \Delta_\mathrm{S}(T) \right|^2}} \left[ q_\mathrm{ez}(\mathrm{i} \omega_n) + q_\mathrm{hz}(\mathrm{i} \omega_n) \right] &&\left[ \frac{d^{(1)}(\mathrm{i} \omega_n) + d^{(2)}(\mathrm{i} \omega_n)}{q_\mathrm{ez}(\mathrm{i} \omega_n)} \right. \nonumber \\
		&&- \left. \frac{d^{(3)}(\mathrm{i} \omega_n)+d^{(4)}(\mathrm{i} \omega_n)}{q_\mathrm{hz}(\mathrm{i} \omega_n)} \right] ,
	\end{eqnarray}
\end{widetext}
where $ e $ is the (positive)~elementary~charge, $ k_\mathrm{B} $ is the Boltzmann~constant, $ A $ denotes the contact~area, and $ q_\mathrm{ez \, (hz)} (\mathrm{i} \omega_n) $ are the $ \hat{z} $-components of the wave~vectors for electronlike~(holelike)~quasiparticles in the superconductors~(see above) after analytically continuing $ E $ to $ \mathrm{i} \omega_n $~[$ \omega_n = (2n+1) \pi k_\mathrm{B} T $ with $ n=0,\pm 1, \pm 2, \ldots $ are the fermionic Matsubara~frequencies]. As explained before, the scattering~coefficient~$ d^{(1)} $ refers to the situation that an incoming electronlike~quasiparticle with spin~up is Andreev~reflected as a holelike~quasiparticle with the same spin at the left junction~interface. Analogously, $ d^{(2)} $, $ d^{(3)} $, and $ d^{(4)} $ are the Andreev~reflection amplitudes for the other involved quasiparticle~injection~processes. The temperature~dependence of the superconducting~energy~gap within the BCS~theory is $ | \Delta_\mathrm{S}(T) | = | \Delta_\mathrm{S}(0) | \tanh ( 1.74 \sqrt{ \slfrac{T_\mathrm{C}}{T} - 1} ) $, with $ T_\mathrm{C} $ being the critical~temperature of the superconductor and $ | \Delta_\mathrm{S} (0) | $ its energy~gap at zero temperature. Finally, the two-dimensional integration over the in-plane~wave~vector~$ \mathbf{k_\parallel} $ is introduced to average over all possible directions of incoming quasiparticles.

For a numerical evaluation of Eq.~\eqref{EqJosephsonCurrent}, we use realistic values for the superconducting~energy~gap and the critical~temperature of conventional superconductors, i.e., $ \left| \Delta_\mathrm{S}(0) \right| \sim 2.5 \, \mathrm{meV} $ and $ T_\mathrm{C} \sim 16 \, \mathrm{K} $~[for~instance~\cite{Carbotte1990}, $ \mathrm{V}_3 \mathrm{Ga} $~alloy has $ \left| \Delta_\mathrm{S}(0) \right|^{\mathrm{V_3 Ga}} \approx 2.7 \, \mathrm{meV} $ and $ T_\mathrm{C}^{\mathrm{V_3 Ga}} \approx 15 \, \mathrm{K} $]. For the Fermi~level in the ferromagnet, we take a typical value of $ \mu_\mathrm{F} = 1000 \, \left| \Delta_\mathrm{S} (0) \right| $. To compactify the analysis, we define dimensionless parameters: $ Z_\mathrm{L \; (R)}=\slfrac{V_\mathrm{L \; (R)} d_\mathrm{L \; (R)} \sqrt{m_\mathrm{F} m_\mathrm{S}}}{(\hbar^2 \sqrt{ k_\mathrm{F} q_\mathrm{F}})} $, where $ k_\mathrm{F} $ and $ q_\mathrm{F} $ are the Fermi~wave~vectors in the ferromagnetic and superconducting regions, determines the strength of the potential~barrier at the left~(right)~interface, $ P=\slfrac{(\Delta_\mathrm{XC}/2)}{\mu_\mathrm{F}} $ quantifies the spin~polarization in the ferromagnet, and $ \lambda^\alpha_\mathrm{L \; (R)}=\slfrac{2 \alpha_\mathrm{L \; (R)} \sqrt{m_\mathrm{F} m_\mathrm{S}}}{\hbar^2} $ as well as $ \lambda^\beta_\mathrm{L \; (R)}=\slfrac{2 \beta_\mathrm{L \; (R)} \sqrt{m_\mathrm{F} m_\mathrm{S}}}{\hbar^2} $ parametrize the interfacial Rashba and Dresselhaus~SOC at the left~(right)~interface. Since the Rashba~SOC~strengths~$ \alpha_\mathrm{L \; (R)} $ in the Josephson~junctions depend not only on the concrete combinations of ferromagnetic and superconducting materials, but also on the bands contributing to electrical~transport, these parameters are typically unknown and have to be extracted from \textit{ab initio} calculations~\cite{MatosAbiague2009}. Therefore, we will treat~$ \lambda^\alpha_\mathrm{L \; (R)} $ as phenomenological parameters throughout this paper. Mismatches of the effective~masses and the Fermi~wave~vectors in the superconducting and ferromagnetic constituents of the junction---the latter result from different charge~carrier~densities in the materials~\cite{Zutic2000}---are included in the dimensionless parameters $ F_\mathrm{K} = q_\mathrm{F} / k_\mathrm{F} $ and $ F_\mathrm{M} = m_\mathrm{S} / m_\mathrm{F} $, respectively. Table~\ref{TabSystemParameters} summarizes all used system~parameters in a compact way.
\renewcommand{\arraystretch}{2.5}
\begin{table}[h]
	\caption{Dimensionless system parameters.  \label{TabSystemParameters}}
	\begin{ruledtabular}
		\begin{tabular}{l l}
		$ P=\slfrac{(\Delta_\mathrm{XC}/2)}{\mu_\mathrm{F}} $ & spin polarization in ferromagnet \\
		$ Z_\mathrm{L}=\dfrac{V_\mathrm{L}d_\mathrm{L} \sqrt{m_\mathrm{F} m_\mathrm{S}}}{\hbar^2 \sqrt{k_\mathrm{F} q_\mathrm{F}}} $ & barrier strength at left interface \\
		$ Z_\mathrm{R}=\dfrac{V_\mathrm{R}d_\mathrm{R} \sqrt{m_\mathrm{F} m_\mathrm{S}}}{\hbar^2 \sqrt{k_\mathrm{F} q_\mathrm{F}}} $ & barrier strength at right interface \\
		$ \lambda^\alpha_\mathrm{L}=\slfrac{2\alpha_\mathrm{L} \sqrt{m_\mathrm{F} m_\mathrm{S}}}{\hbar^2} $ & Rashba SOC at left interface \\
		$ \lambda^\alpha_\mathrm{R}=\slfrac{2\alpha_\mathrm{R} \sqrt{m_\mathrm{F} m_\mathrm{S}}}{\hbar^2} $ & Rashba SOC at right interface \\
		$ \lambda^\beta_\mathrm{L}=\slfrac{2\beta_\mathrm{L} \sqrt{m_\mathrm{F} m_\mathrm{S}}}{\hbar^2} $ & Dresselhaus SOC at left interface \\
		$ \lambda^\beta_\mathrm{R}=\slfrac{2\beta_\mathrm{R} \sqrt{m_\mathrm{F} m_\mathrm{S}}}{\hbar^2} $ & Dresselhaus SOC at right interface \\
		$ F_\mathrm{K} = \slfrac{q_\mathrm{F}}{k_\mathrm{F}} $ & Fermi~wave~vector mismatch \\
		$ F_\mathrm{M} = \slfrac{m_\mathrm{S}}{m_\mathrm{F}} $ & mismatch of effective~masses
		\end{tabular}
	\end{ruledtabular}
\end{table}

In the following, we present numerical results for the Josephson~current at low temperature~$ T = 0.1 T_\mathrm{C} $. To simplify the discussion and illustrate the main points, we mostly suppose equal effective~masses and Fermi~wave~vectors in all components of the Josephson~junction, i.e., $ F_\mathrm{K} = F_\mathrm{M} = 1 $. The impact of Fermi~wave~vector or mass mismatch on the outcomes is analyzed in Sec.~\ref{SectionV}.

\section{Reversal of Josephson~current induced by changing the interlayer~thickness\label{SectionIII}}

\begin{figure}
	\includegraphics[width=0.47\textwidth]{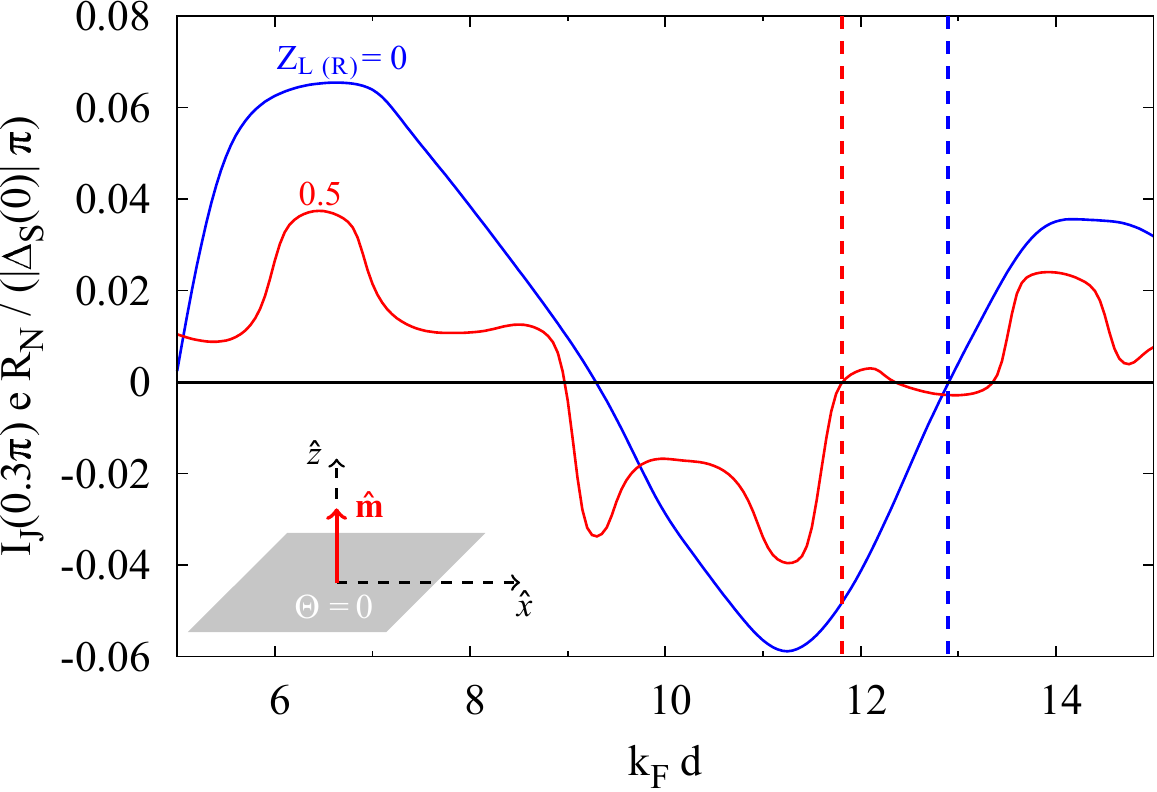}
	\colorcaption{\label{FigkfdWithoutSOC} Calculated dependence of the (normalized) Josephson~current~$ I_\mathrm{J} $~[normalization~constant~$ R_\mathrm{N}=\slfrac{2 \pi^2 \hbar}{(A e^2 k_\mathrm{F}^2)} $ refers to the resistance of a perfectly transparent N/N/N~tunnel~junction] on the effective interlayer~thickness~$ k_\mathrm{F} d $ for S/F/S~Josephson~junctions with transparent interfaces~($ Z_\mathrm{L}=Z_\mathrm{R}=0 $) or weak interfacial barriers~($ Z_\mathrm{L} = Z_\mathrm{R}=0.5 $) and without Fermi~wave~vector or mass~mismatch~($ F_\mathrm{K}=F_\mathrm{M}=1 $) at a fixed superconducting~phase~difference of $ \phi_\mathrm{S}=0.3\pi $. The spin~polarization in the ferromagnetic part is $ P=0.7 $, neither Rashba nor Dresselhaus~SOC are present~($ \lambda^\alpha_\mathrm{L}=\lambda^\alpha_\mathrm{R}=\lambda^\beta_\mathrm{L}=\lambda^\beta_\mathrm{R}=0 $), and the magnetization~direction is oriented perpendicular to the ferromagnetic layer~($ \Theta = 0 $ and $ \Phi = 0 $; see illustration).}
\end{figure}
At first, we investigate the pure influence of changing the effective thickness~$ k_\mathrm{F} d $ of the metallic interlayer on the Josephson~current~flow across the regarded Josephson~junctions. Figure~\ref{FigkfdWithoutSOC} illustrates the calculated dependence of the Josephson~current on~$ k_\mathrm{F} d $ for S/F/S~model~junctions in which neither interfacial Rashba nor Dresselhaus~SOC are present, i.e., $ \lambda^\alpha_\mathrm{L}=\lambda^\alpha_\mathrm{R}=\lambda^\beta_\mathrm{L}=\lambda^\beta_\mathrm{R}=0 $. For the spin~polarization in the metallic~interlayer, we choose a realistic value of $ P=0.7 $, which would correspond in experiments to an iron~layer, and the magnetization~direction is aligned perpendicular to the junction~interfaces. To evaluate the Josephson~current numerically, the superconducting~phase~difference across the junction is set to a fixed value, for~instance, $ \phi_\mathrm{S}=0.3\pi $. The qualitative results occurring at other superconducting~phase~differences~$ 0 < \phi_\mathrm{S} < \pi $ are analog and not explicitly presented. Since present-day microfabrication~techniques enable the experimental realization of ballistic metal/superconductor~multilayer~structures with highly transparent interfaces~\cite{DeFranceschi1998, Vasko1998, Wan2015}, we concentrate on the cases of perfectly transparent junctions as well as weak symmetric tunneling barriers at the interfaces, modeled by $ Z_\mathrm{L}=Z_\mathrm{R}=0 $ and $ Z_\mathrm{L}=Z_\mathrm{R}=0.5 $, respectively.

Even in junctions with perfectly transparent interfaces~($ Z_\mathrm{L}=Z_\mathrm{R}=0 $), the Josephson~current exhibits an oscillatory dependence on the effective interlayer~thickness~$ k_\mathrm{F} d $. Our model~calculations show that, owing to these oscillations, the direction~(sign) of the Josephson~current~flow can be reversed for certain values of $ k_\mathrm{F} d $, indicating transitions between $ 0 $ and $ \pi $~states. Since the oscillations in the $ I_\mathrm{J} $-$ k_\mathrm{F} d $~relation are solely caused by the exchange~interaction in the central layer of the junctions, they characteristically appear only in S/F/S~Josephson~junctions and are absent in S/N/S~Josephson~junctions with a normal~metal~interlayer. As a consequence, $ \pi $~states can emerge exclusively in S/F/S and not in S/N/S~Josephson~junctions.

\begin{figure}
	\includegraphics[width=0.45\textwidth]{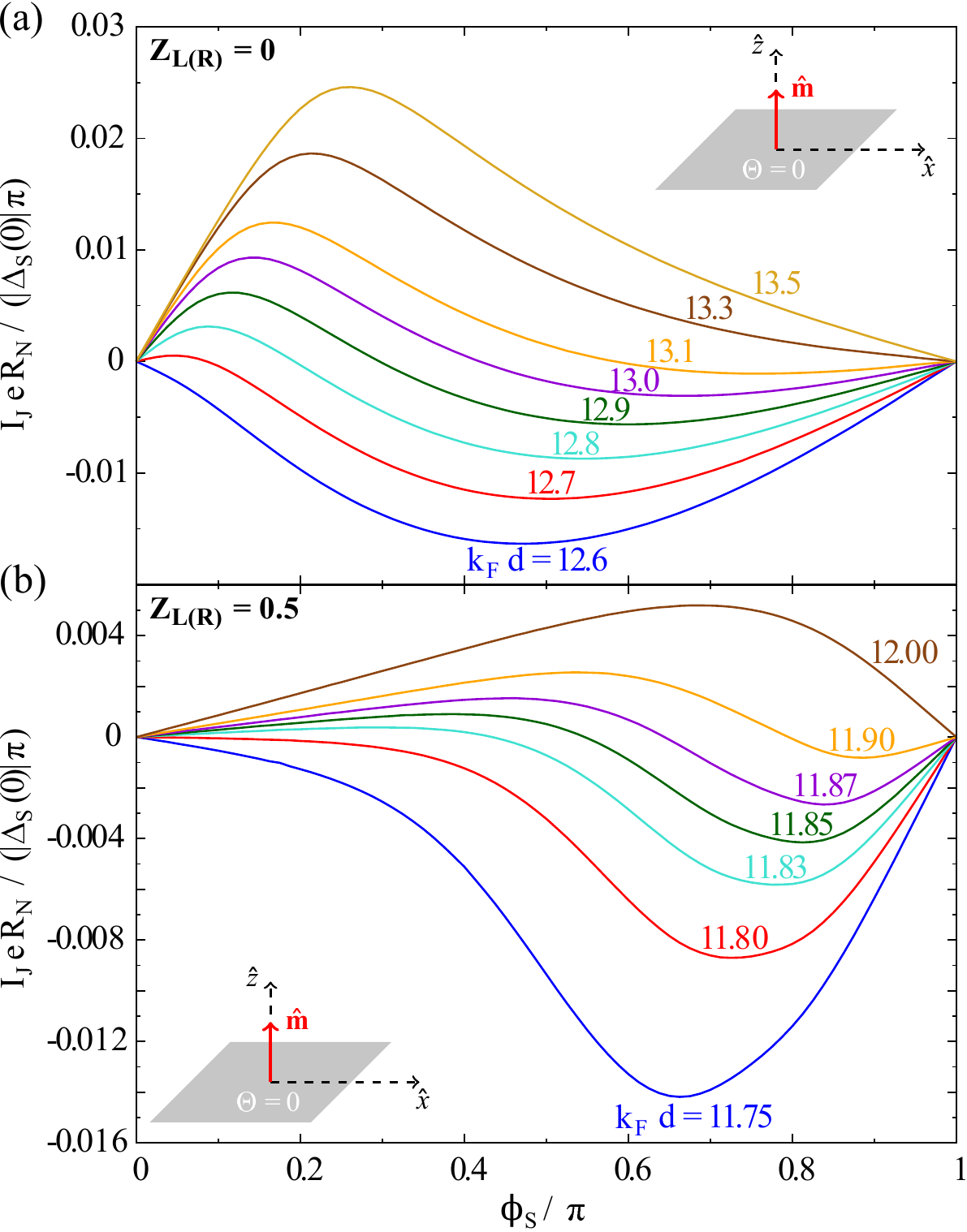}
	\colorcaption{\label{FigCPRKFD} (a)~Calculated (normalized) Josephson~current~$ I_\mathrm{J} $ as a function of the superconducting~phase~difference~$ \phi_\mathrm{S} $ (current-phase~relation) for S/F/S~Josephson~junction with transparent interfaces~($ Z_\mathrm{L}=Z_\mathrm{R}=0 $), spin~polarization~$ P=0.7 $ (the chosen parameters correspond, for~instance, to an iron~interlayer), and for various values of the effective interlayer~thickness~$ k_\mathrm{F} d $ in the vicinity of one $ \pi $-$ 0 $~transition. Rashba and Dresselhaus~SOC are absent~($ \lambda^\alpha_\mathrm{L}=\lambda^\alpha_\mathrm{R}=\lambda^\beta_\mathrm{L}=\lambda^\beta_\mathrm{R}=0 $), the Fermi~wave~vectors and effective~masses in the superconducting and ferromagnetic regions are equal~($ F_\mathrm{K}=F_\mathrm{M}=1 $), and the magnetization~direction~$ \hat{\mathbf{m}} $ is perpendicular to the ferromagnetic layer~($ \Theta=0 $ and $ \Phi=0 $; see illustration). (b)~Calculated current-phase~relation for the same junction as in~(a), but in the presence of weak interfacial barriers~($ Z_\mathrm{L}=Z_\mathrm{R}=0.5 $).}
\end{figure}
To become more familiar with the physical concepts behind the $ 0 $-$ \pi $~transitions~\cite{Andreev1991, Demler1997}, we have a closer look at the intermediate metallic region. Due to the proximity~effect, spin-singlet Cooper~pairs can leak from one superconductor across the interlayer into the other one, inducing superconducting~correlations in the metal and generating a net supercurrent~flow across the system. However, if the central region consists of a ferromagnet, the exchange~interaction in the material opens an exchange~energy~gap between the spin~up and spin~down subbands. Consequently, the majority~spin electron of the penetrating Cooper~pair lowers its potential~energy in the metallic interlayer, whereas that of the minority~spin electron increases. Since the total energy of the electrons needs to be conserved, the kinetic~part must compensate the changes of the potential~energies and the Cooper~pair acquires a finite center-of-mass~momentum. This response of the transferred Cooper~pairs to the spin-dependent potentials in the ferromagnet gives rise to spatial oscillations of the proximity-induced superconducting~order~parameter in the ferromagnetic~layer~\cite{Andreev1991, Demler1997}. If the thickness of the metallic region is now comparable to half of the period of these oscillations, the superconducting~order~parameter may differ in sign at both junction~interfaces and an additional intrinsic $ \pi $~shift to the superconducting~phase~difference, entailing a transition from $ 0 $ to $ \pi $~states, may arise.

In the presence of weak interfacial barriers~($ Z_\mathrm{L}=Z_\mathrm{R}=0.5 $), additional oscillations due to quasiparticle~resonances~(so-called \textit{geometrical~oscillations}~\cite{Brinkman2000}) are superimposed on the oscillations originating from the exchange~interaction. Nevertheless, the previous arguments concerning $ 0 $-$ \pi $~transitions are still valid and crossovers between $ 0 $ and $ \pi $~states are again possible for certain values of effective interlayer~thickness~$ k_\mathrm{F} d $.

To characterize the transitions between $ 0 $ and $ \pi $~states further, the calculated dependence of the Josephson~current on the superconducting~phase~difference (current-phase~relation) is shown for different effective interlayer~thicknesses~$ k_\mathrm{F} d $ close to one $ 0 $-$ \pi $~transition in Fig.~\ref{FigCPRKFD}. In particular, the presented values of $ k_\mathrm{F} d $ are chosen in the vicinity of the first $ \pi $ to $ 0 $~transitions in Fig.~\ref{FigkfdWithoutSOC}, indicated by the second sign~changes of the Josephson~current (see dashed lines in Fig.~\ref{FigkfdWithoutSOC}). All other system~parameters are the same as before. Without interfacial tunneling~barriers~[$ Z_\mathrm{L}=Z_\mathrm{R}=0 $; see Fig.~\ref{FigCPRKFD}(a)], increasing the effective interlayer~thickness from $ k_\mathrm{F} d = 12.6 $ to $ k_\mathrm{F} d = 13.1 $ gives rise to a crossover from $ \pi $ to $ 0 $~states. In the transition~region in between~($ k_\mathrm{F} d \approx 12.7 \ldots 13.1 $), the coexistence of $ 0 $ and $ \pi $~states leads to nonsinusoidal variations of the current-phase~relation. These outcomes are fully consistent with earlier obtained results of Radovi{\'{c}} and co-workers~\cite{Radovic2003}. Similar characteristics were also predicted for $ 0 $-$ \pi $~transitions controlled by changing the temperature~\cite{Radovic2003} or in the presence of inhomogeneous magnetization~\cite{Bergeret2001a} in S/F/S~junctions, as well as in S/F/c/F/S~junctions with geometrical constrictions~\cite{Golubov2002,*Golubov2002alt}. Moreover, a crossover between $ 0 $ and $ \pi $~states may also be achieved in dirty S/F/S~Josephson~junctions by reducing the interfacial transparency~\cite{Buzdin2003, Golubov2004}.

If moderate barriers~[$ Z_\mathrm{L}=Z_\mathrm{R}=0.5 $; see Fig.~\ref{FigCPRKFD}(b)] are present at both junction~interfaces, the crossover~points between $ 0 $ and $ \pi $~states are shifted to lower interlayer~thicknesses~(compare also to Fig.~\ref{FigkfdWithoutSOC}) and the region of coexisting $ 0 $ and $ \pi $~states~($ k_\mathrm{F} d \approx 11.8 \ldots 11.9 $) is nearly four times narrower than in the case of the perfectly transparent junction. The $ 0 $-$ \pi $~transitions appearing at other values of $ k_\mathrm{F} d $ lead to analog characteristics and are not explicitly analyzed.

\section{Spin-orbit~coupling induced 0-$ \pmb{\pi} $~transitions and increase in critical~current\label{SectionIV}}

\begin{figure}
	\includegraphics[width=0.45\textwidth]{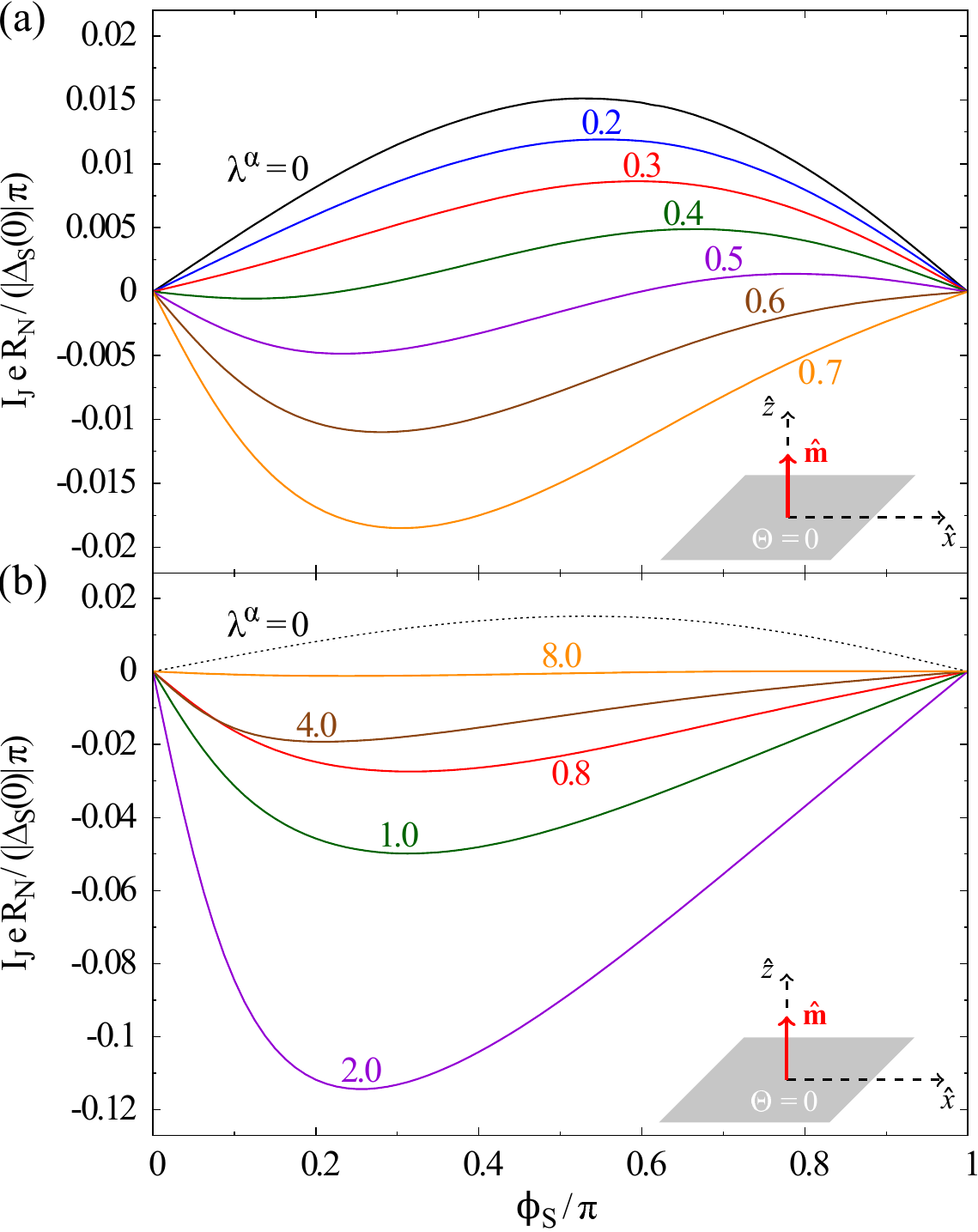}
	\colorcaption{\label{FigCPROutofplane}  (a)~Calculated current-phase~relation for S/F/S~Josephson~junction with weak interfacial barriers~($ Z_\mathrm{L}=Z_\mathrm{R}=0.5 $), spin~polarization~$ P=0.7 $, effective interlayer~thickness~$ k_\mathrm{F} d = 8.2 $ (the chosen parameters correspond, for~instance, to an iron~interlayer with thickness~$ d \approx 1 \, \mathrm{nm} $), various moderate Rashba~SOC~strengths~$ \lambda^\alpha_\mathrm{L}=\lambda^\alpha_\mathrm{R}=\lambda^\alpha $, and without Fermi~wave~vector or mass~mismatch~($ F_\mathrm{K}=F_\mathrm{M}=1 $). Magnetization~$ \hat{\mathbf{m}} $ is perpendicular to the ferromagnetic layer~($ \Theta=0 $ and $ \Phi=0 $; see illustration) and Dresselhaus~SOC is not present~($ \lambda^\beta_\mathrm{L}=\lambda^\beta_\mathrm{R}=0 $). (b)~Calculated current-phase~relation for the same junction as in~(a), but larger Rashba~SOC~strengths are considered~($ \lambda^\alpha=0 $ is again shown for orientation).}
\end{figure}
To study the impact of interfacial SOC on the Josephson~current~flow, the current-phase-relation for a realistic S/F/S~Josephson~junction is shown for various strengths of symmetric Rashba~spin-orbit~fields at the junction~interfaces~($ \lambda^\alpha_\mathrm{L}=\lambda^\alpha_\mathrm{R}=\lambda^\alpha $) in Fig.~\ref{FigCPROutofplane}. The situation of antisymmetric Rashba~spin-orbit~fields~($ \lambda^\alpha_\mathrm{L}=-\lambda^\alpha_\mathrm{R} $) is rather unrealistic in real junctions and therefore not considered in this paper. The spin~polarization in the ferromagnet is again chosen to be $ P=0.7 $, its effective thickness is $ k_\mathrm{F} d = 8.2 $ (these parameters would correspond in experiments to an iron~interlayer with thickness~$ d \approx 1 \, \mathrm{nm} $ as $ k_\mathrm{F} \approx 8.05 \times 10^7 \, \mathrm{cm}^{-1} $ in iron~\cite{Wang2003}), and the magnetization~direction is oriented perpendicular to the ferromagnetic layer. To simplify the discussion, Dresselhaus~SOC is absent~($ \lambda^\beta_\mathrm{L}=\lambda^\beta_\mathrm{R}=0 $) in all calculations throughout this section. Since reducing the interfacial transparency would not significantly change the qualitative features, we solely discuss the calculations for a junction with moderate interfacial barriers~($ Z_\mathrm{L}=Z_\mathrm{R}=0.5 $). 

Without interfacial Rashba~SOC~($ \lambda^\alpha=0 $), the junction is in the $ 0 $~state, in which the Josephson~current approaches a sinusoidal dependence on $ \phi_\mathrm{S} $. Increasing the strengths of the Rashba~fields slightly, reverses the direction~(sign) of the Josephson~current~flow~[see Fig.~\ref{FigCPROutofplane}(a)] and leads to a crossover from $ 0 $ to $ \pi $~states. Characteristic for the transition~region between pure $ 0 $ and $ \pi $~states~($ \lambda^\alpha \approx 0.40  \ldots 0.55 $) is again a nonsinusoidal variation of the current-phase~relation due to the coexistence of $ 0 $ and $ \pi $~states as we already explained for the $ 0 $-$ \pi $~transitions controlled by altering the interlayer~thickness in Sec.~\ref{SectionIII}. However, our calculations suggest that analogous physical effects may also be caused by modulating the strength of the interfacial Rashba~fields, e.g., by means of an applied gate~voltage~\cite{Nitta1997, Nitta2002}, without the need to change the interlayer~thickness of the Josephson~junction.

Regarding the amplitudes of the Josephson~current, the critical~current becomes already at a moderate Rashba~SOC strength of $ \lambda^\alpha=0.7 $ greater than in a junction without SOC. This is due to additional contributions to the Josephson~current from spin-flip~processes at the interfaces, enabled by the presence of spin-orbit~fields. Increasing the Rashba~SOC~strength~$ \lambda^\alpha $ further~[see~Fig.~\ref{FigCPROutofplane}(b)], the critical~current first increases again, reaching its maximal value for $ \lambda^\alpha \approx 2.0 $; the critical~current there is one order of magnitude greater than the critical~current in the absence of SOC. Nevertheless, we need to mention that SOC also introduces more scattering at the interfaces. This scattering starts to dominate at~$ \lambda^\alpha \gtrsim 4.0 $ and reverses the increasing trend in the critical~current.

\begin{figure}
	\includegraphics[width=0.45\textwidth]{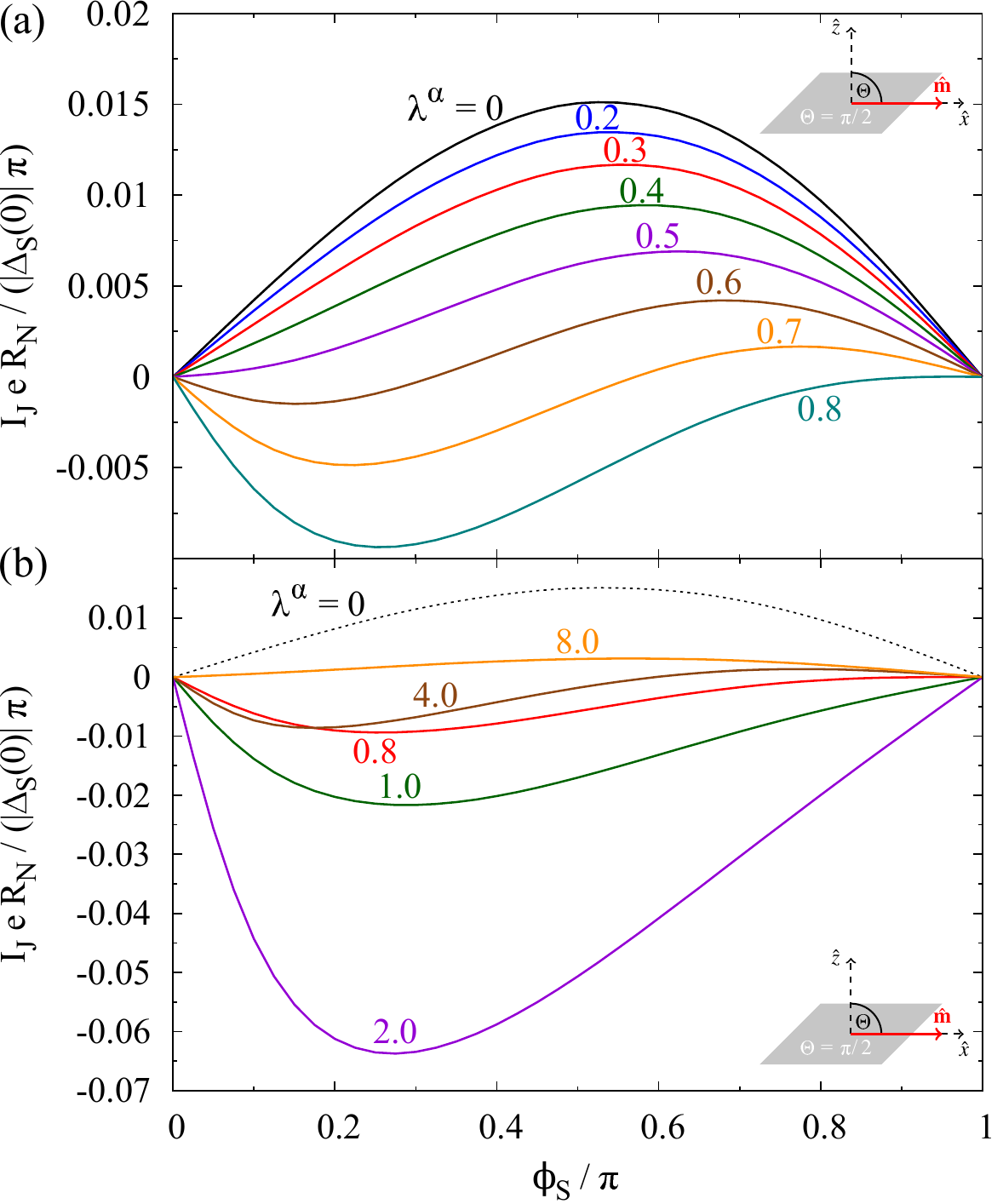}
	\colorcaption{\label{FigCPRInplane} (a)~Calculated current-phase~relation for S/F/S~Josephson~junction with weak interfacial barriers~($ Z_\mathrm{L}=Z_\mathrm{R}=0.5 $), spin~polarization~$ P=0.7 $, effective interlayer~thickness~$ k_\mathrm{F} d = 8.2 $, various moderate Rashba~SOC~strengths~$ \lambda^\alpha_\mathrm{L}=\lambda^\alpha_\mathrm{R}=\lambda^\alpha $, and without Fermi~wave~vector or mass~mismatch~($ F_\mathrm{K}=F_\mathrm{M}=1 $). Magnetization~$ \hat{\mathbf{m}} $ is parallel to the ferromagnetic layer~($ \Theta=\pi / 2 $ and $ \Phi=0 $; see illustration) and Dresselhaus~SOC is not present~($ \lambda^\beta_\mathrm{L}=\lambda^\beta_\mathrm{R}=0 $). (b)~Calculated current-phase~relation for the same junction as in~(a), but larger Rashba~SOC~strengths are considered~($ \lambda^\alpha=0 $ is again shown for orientation).}
\end{figure}

If the magnetization~direction in the ferromagnetic layer is aligned parallel to the layer, the main qualitative physical characteristics do not change~(see Fig.~\ref{FigCPRInplane}). 
Increasing the Rashba~SOC~strength from $ \lambda^\alpha=0 $ still gives rise to a crossover from $ 0 $ to $ \pi $~states. Compared to the previously discussed case, in which the magnetization was oriented perpendicular to the ferromagnet, slightly stronger Rashba spin-orbit~fields are required to induce $ 0 $-$ \pi $~transitions in the junction with in-plane magnetization and also the region of coexisting $ 0 $ and $ \pi $~states is somewhat larger~($ \lambda^\alpha \approx 0.55 \ldots 0.75 $) than before~[$ \lambda^\alpha \approx 0.40 \ldots 0.55 $; see Fig.~\ref{FigCPROutofplane}(a)]. A further increase of the Rashba~SOC parameter again reflects the nonmonotonic dependence of the Josephson~current on the SOC~strength as its amplitudes first increase due to the formation of spin-triplet Cooper~pairs, but finally decrease owing to the additional scattering introduced by SOC. However, when comparing the outcomes to the ones for perpendicular magnetization in Fig.~\ref{FigCPROutofplane}, we observe that the critical~current is remarkably smaller for all considered strengths of Rashba~SOC and in-plane magnetization, suggesting that the generation of spin-triplet~Cooper~pairs becomes suppressed if the magnetization is parallel to the ferromagnetic layer. Nonetheless, it is important to stress here that the spin-triplet contribution to the Josephson~current can still be dominant for certain strengths of the Rashba~spin-orbit~fields~[e.g., for $ \lambda^\alpha=2.0 $; see Fig.~\ref{FigCPRInplane}(b)], significantly enhancing the critical~current compared to the case without Rashba~SOC~($ \lambda^\alpha=0 $). This finding differs from earlier studies of \textit{diffusive} lateral S/F/S~Josephson~junctions with SOC in the ferromagnetic region and in-plane magnetization~\cite{Bergeret2014}, in which a long-range spin-triplet supercurrent component can only exist if both Rashba and Dresselhaus~SOC are present.

Similar results are also obtained in S/F/S~Josephson~junctions with other values of interlayer thickness and spin~polarization in the ferromagnet as we show in detail in Appendix~\ref{AppA}.

\section{Impact of Fermi~wave~vector or mass~mismatch on the Josephson~current\label{SectionV}}

In this section we want to illustrate the influence of different effective~masses or Fermi~wave~vectors in the superconducting and ferromagnetic constituents of the considered S/F/S~Josephson~junctions on the Josephson~current~flow. To quantify mismatches of the effective~masses or Fermi~wave~vectors, we have introduced the dimensionless parameters $ F_\mathrm{M}=\slfrac{m_\mathrm{S}}{m_\mathrm{F}} $ and $ F_\mathrm{K}=\slfrac{q_\mathrm{F}}{k_\mathrm{F}} $ in the theoretical model presented in Sec.~\ref{SectionII}.

\begin{figure}
	\centering
	\includegraphics[width=0.47\textwidth]{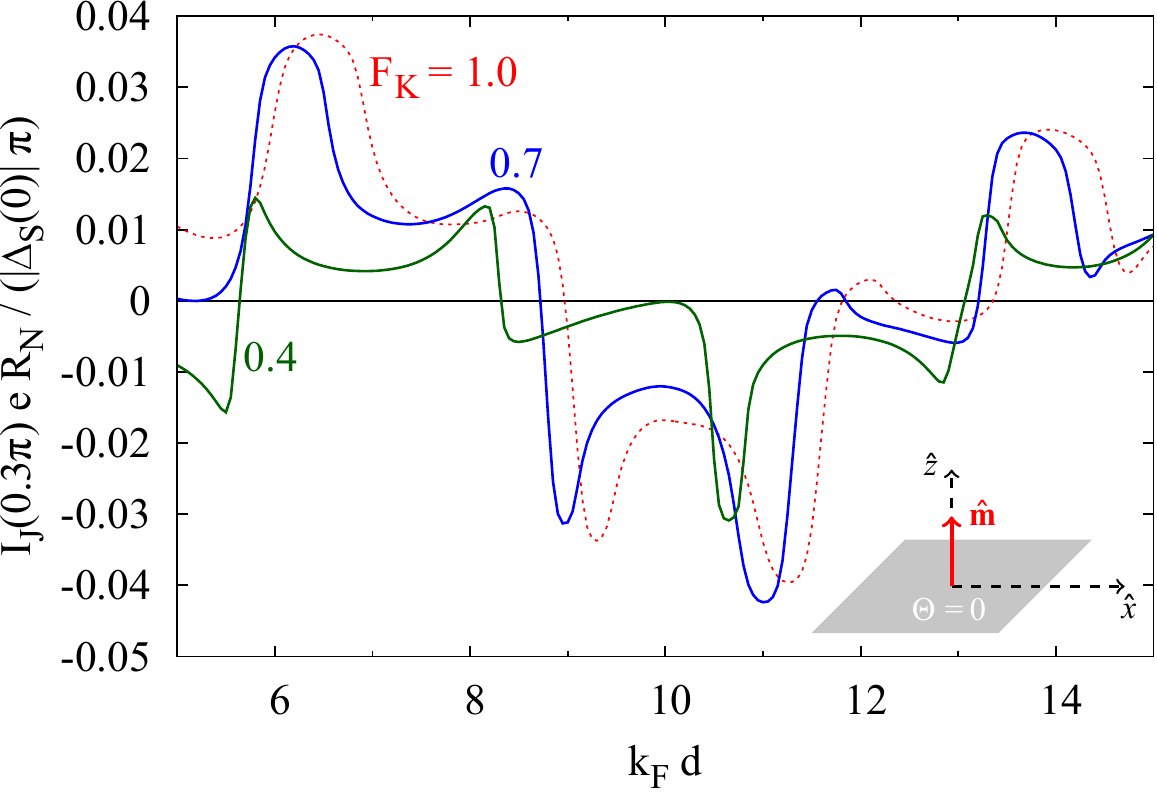}
	\colorcaption{\label{FigkfdFWVM} Calculated dependence of the (normalized) Josephson~current~$ I_\mathrm{J} $ on the effective interlayer~thickness~$ k_\mathrm{F} d $ for S/F/S~Josephson~junctions with weak interfacial barriers~($ Z_\mathrm{L}=Z_\mathrm{R}=0.5 $), spin~polarization~$ P=0.7 $, without mass~mismatch~($ F_\mathrm{M}=1 $), and for different values of Fermi~wave~vector mismatch~$ F_\mathrm{K} $ at a fixed superconducting phase~difference of $ \phi_\mathrm{S}=0.3 \pi $. Rashba and Dresselhaus~SOC are absent~($ \lambda^\alpha_\mathrm{L}=\lambda^\alpha_\mathrm{R}=\lambda^\beta_\mathrm{L}=\lambda^\beta_\mathrm{R}=0 $) and the magnetization~direction is oriented perpendicular to the ferromagnetic layer~($ \Theta=0 $ and $ \Phi=0 $; see illustration).}
\end{figure}
First, we suppose equal effective~masses of quasiparticles in all components of the Josephson~junctions~($ F_\mathrm{M}=1 $) and study the consequences of different Fermi~wave~vectors in the superconducting and ferromagnetic regions, which originate in real junctions from differing charge~carrier~densities in the materials~\cite{Zutic2000}. Figure~\ref{FigkfdFWVM} shows the dependence of the Josephson~current on the effective interlayer~thickness~$ k_\mathrm{F} d $ for Josephson~junctions with weak interfacial barriers~($ Z_\mathrm{L}=Z_\mathrm{R}=0.5 $) and spin~polarization~$ P=0.7 $ in the absence of both Rashba as well as Dresselhaus spin-orbit~fields~($ \lambda^\alpha_\mathrm{L}=\lambda^\alpha_\mathrm{R}=\lambda^\beta_\mathrm{L}=\lambda^\beta_\mathrm{R}=0 $). For the parameter~$ F_\mathrm{K} $, incorporating Fermi~wave~vector~mismatch, we choose the values $ F_\mathrm{K}=1 $~(no mismatch), $ F_\mathrm{K}=0.7 $~(moderate mismatch), and $ F_\mathrm{K}=0.4 $~(large mismatch), respectively. It is important to observe that an increase of the mismatch between the Fermi~wave~vectors in the superconducting and ferromagnetic materials of the Josephson~junctions (which means decreasing~$ F_\mathrm{K} $) does not change the oscillatory dependence of the Josephson~current~flow on the effective interlayer~thickness~$ k_\mathrm{F} d $ qualitatively. Therefore, altering the interlayer~thickness in such Josephson~junctions can still provide a practicable way to reverse the direction~(sign) of the Josephson~current and induce a crossover between $ 0 $ and $ \pi $~states as already mentioned in Sec.~\ref{SectionIII}. Nevertheless, we observe here that Fermi~wave~vector~mismatch shifts the transition~points as a function of the effective~interlayer~thickness~$ k_\mathrm{F} d $.

\begin{figure}
	\centering
	\includegraphics[width=0.47\textwidth]{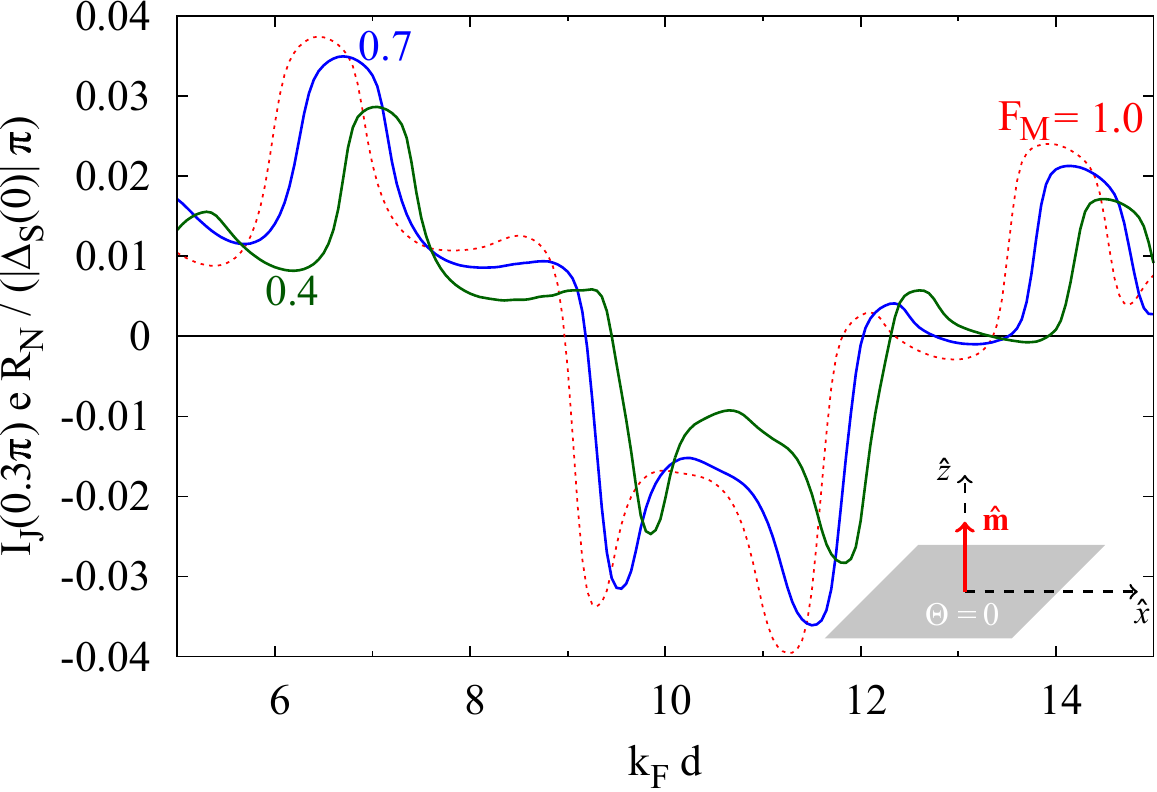}
	\colorcaption{\label{FigkfdMM} Calculated dependence of the (normalized) Josephson~current~$ I_\mathrm{J} $ on the effective interlayer~thickness~$ k_\mathrm{F} d $ for S/F/S~Josephson~junctions with weak interfacial barriers~($ Z_\mathrm{L}=Z_\mathrm{R}=0.5 $), spin~polarization~$ P=0.7 $, without Fermi~wave~vector~mismatch~($ F_\mathrm{K}=1 $), and for different values of mass~mismatch~$ F_\mathrm{M} $ at a fixed superconducting phase~difference of $ \phi_\mathrm{S}=0.3 \pi $. Rashba and Dresselhaus~SOC are absent~($ \lambda^\alpha_\mathrm{L}=\lambda^\alpha_\mathrm{R}=\lambda^\beta_\mathrm{L}=\lambda^\beta_\mathrm{R}=0 $) and the magnetization~direction is oriented perpendicular to the ferromagnetic layer~($ \Theta=0 $ and $ \Phi=0 $; see illustration).}
\end{figure}
To investigate the effects of different effective~masses in the superconductors and ferromagnet, the $ I_\mathrm{J} $-$ k_\mathrm{F} d $~relation is presented for the same junction as before, but with equal Fermi~wave~vectors~($ F_\mathrm{K}=1 $) and differing effective~masses in the superconducting and metallic parts in Fig.~\ref{FigkfdMM}. Similarly to the discussion of Fermi~wave~vector~mismatch, we also distinguish the situations of no mass~mismatch~($ F_\mathrm{M}=1 $), moderate mismatch~($ F_\mathrm{M}=0.7 $), as well as strongly differing effective~masses~($ F_\mathrm{M}=0.4 $). Again, the oscillatory dependence of the Josephson~current on the effective interlayer~thickness~$ k_\mathrm{F} d $ is still clearly visible, even at large mass~mismatch. As we have already asserted for different Fermi~wave~vectors in Fig.~\ref{FigkfdFWVM}, also an increase of mass~mismatch (which means decreasing~$ F_\mathrm{M} $) shifts the transition~points, separating $ 0 $ and $ \pi $~states, to other values of the effective interlayer~thickness~$ k_\mathrm{F} d $.

From the presented calculations, we can conclude that the influence of Fermi~wave~vector or mass~mismatch on the $ I_\mathrm{J} $-$ k_\mathrm{F} d $~relation is comparable to the effects of reduced interfacial~transparency~(see Sec.~\ref{SectionIII}). Similar effects of Fermi~wave~vector mismatch were predicted earlier by Radovi{\'{c}} and co-workers~\cite{Radovic2003}.

\begin{figure}
	\centering
	\includegraphics[width=0.47\textwidth]{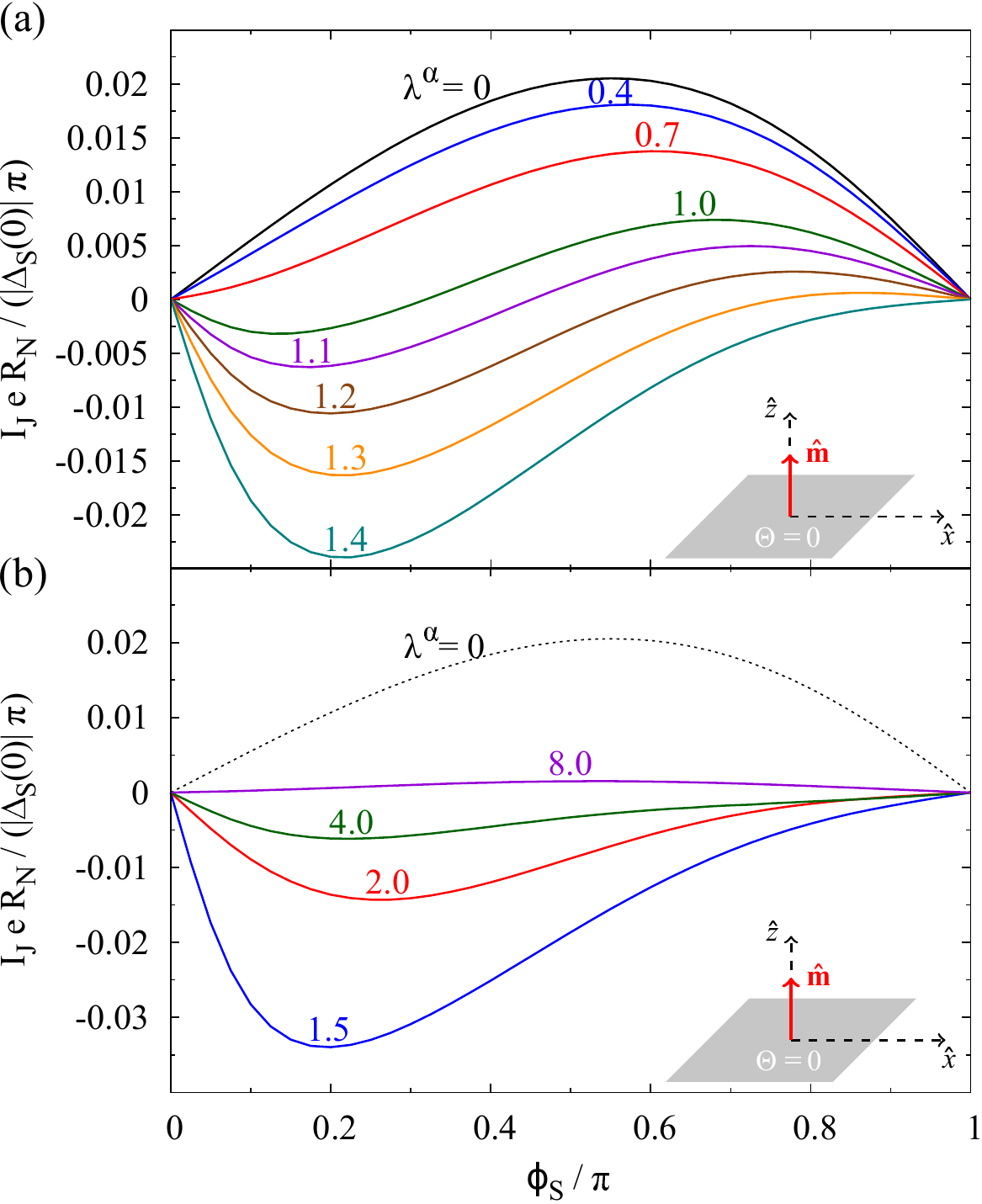}
	\colorcaption{\label{FigCPRFWVM} (a)~Calculated current-phase~relation for S/F/S~Josephson~junction with weak interfacial barriers~($ Z_\mathrm{L}=Z_\mathrm{R}=0.5 $), spin~polarization~$ P=0.7 $, effective interlayer~thickness~$ k_\mathrm{F} d = 8.2 $, different moderate Rashba~SOC~strengths~$ \lambda^\alpha_\mathrm{L}=\lambda^\alpha_\mathrm{R}=\lambda^\alpha $, and with moderate mismatch of Fermi~wave~vectors~$ F_\mathrm{K}=0.7 $. Dresselhaus~SOC is absent~($ \lambda^\beta_\mathrm{L}=\lambda^\beta_\mathrm{R}=0 $), the effective~masses in the superconducting and ferromagnetic parts are equal~($ F_\mathrm{M}=1 $), and the magnetization direction is oriented perpendicular to the ferromagnetic layer~($ \Theta = 0 $ and $ \Phi=0 $; see illustration). (b)~Calculated current-phase~relation for the same junction as in~(a), but larger Rashba~SOC~strengths are considered~($ \lambda^\alpha=0 $ is again shown for orientation).}
\end{figure}
\begin{figure}
	\centering
	\includegraphics[width=0.47\textwidth]{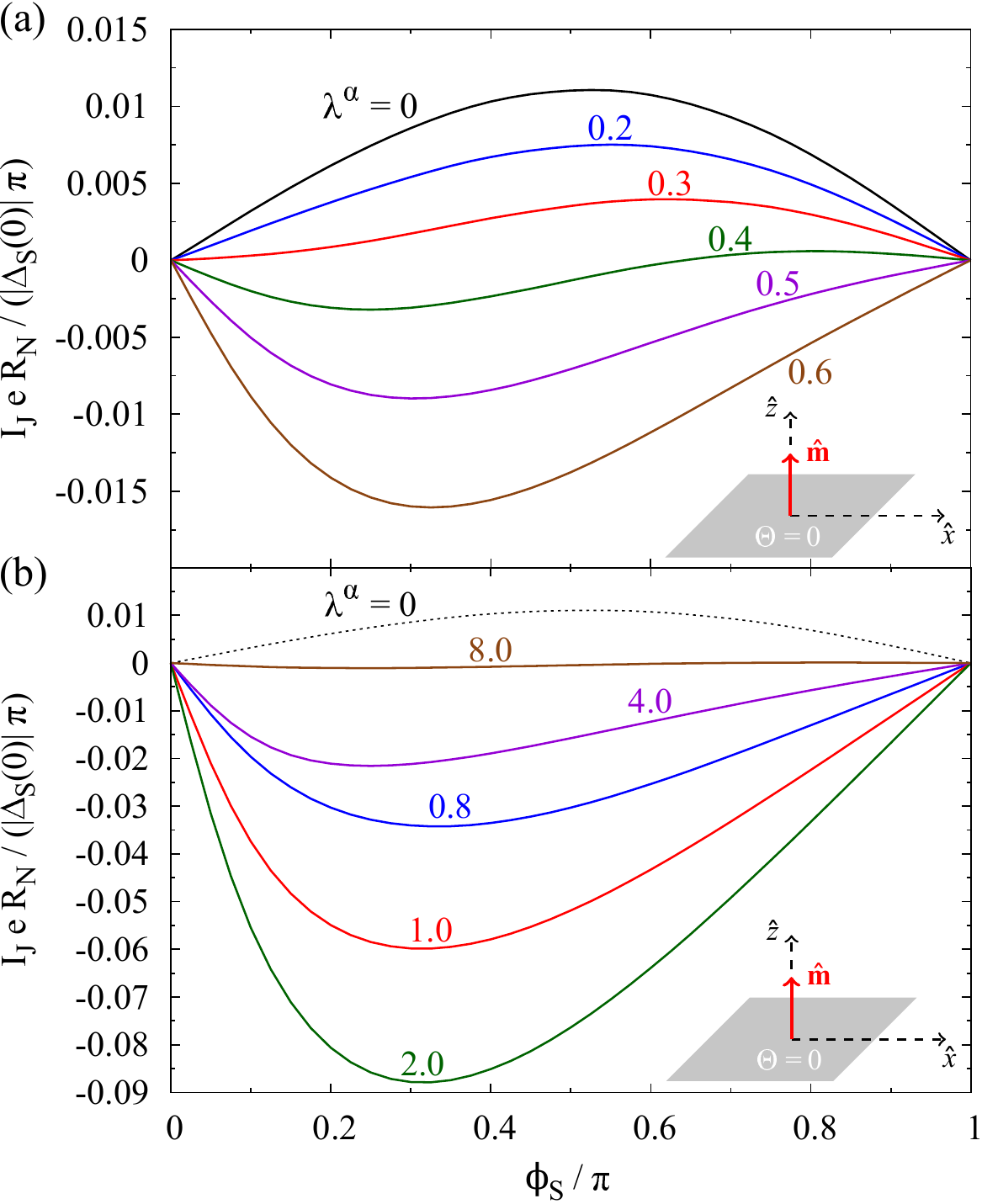}
	\colorcaption{\label{FigCPRMM} (a)~Calculated current-phase~relation for S/F/S~Josephson~junction with weak interfacial barriers~($ Z_\mathrm{L}=Z_\mathrm{R}=0.5 $), spin~polarization~$ P=0.7 $, effective interlayer~thickness~$ k_\mathrm{F} d = 8.2 $, different moderate Rashba~SOC~strengths~$ \lambda^\alpha_\mathrm{L}=\lambda^\alpha_\mathrm{R}=\lambda^\alpha $, and with moderate mismatch of effective~masses~$ F_\mathrm{M}=0.7 $. Dresselhaus~SOC is absent~($ \lambda^\beta_\mathrm{L}=\lambda^\beta_\mathrm{R}=0 $), the Fermi~wave~vectors in the superconducting and ferromagnetic parts are equal~($ F_\mathrm{K}=1 $), and the magnetization~direction is oriented perpendicular to the ferromagnetic layer~($ \Theta=0 $ and $ \Phi=0 $; see illustration). (b)~Calculated current-phase~relation for the same junction as in~(a), but larger Rashba~SOC~strengths are considered~($ \lambda^\alpha=0 $ is again shown for orientation).}
\end{figure}

At the end of this section, we briefly analyze the effects of Fermi~wave~vector or mass~mismatch on the spin-orbit~coupling induced $ 0 $-$ \pi $~transitions predicted in Sec.~\ref{SectionIV}. To find the most general features, we  focus on two cases: in the first one~(see~Fig.~\ref{FigCPRFWVM}), we assume equal effective~masses~($ F_\mathrm{M}=1 $) and moderate Fermi~wave~vector~mismatch~($ F_\mathrm{K}=0.7 $), whereas in the second one~(see~Fig.~\ref{FigCPRMM}), the Fermi~wave~vectors are equal~($ F_\mathrm{K}=1 $) and moderate mismatch of the effective~masses~($ F_\mathrm{M}=0.7 $) is present. The other junction~parameters are the same as for the junction discussed in Sec.~\ref{SectionIV}, i.e., weak interfacial barrier strengths~($ Z_\mathrm{L}=Z_\mathrm{R}=0.5 $), spin~polarization~$ P=0.7 $, and effective interlayer~thickness~$ k_\mathrm{F} d = 8.2 $. In order to compare the results to the situation without any mismatches~(see Fig.~\ref{FigCPROutofplane}), we show the current-phase~relation for weak Rashba~SOC strengths in the vicinity of the induced $ 0 $-$ \pi $~transitions~[see Figs.~\ref{FigCPRFWVM}(a) and \ref{FigCPRMM}(a)] and the one at stronger Rashba~SOC~[see Figs.~\ref{FigCPRFWVM}(b) and \ref{FigCPRMM}(b)] separately~(Dresselhaus~SOC is always absent, $ \lambda^\beta_\mathrm{L}=\lambda^\beta_\mathrm{R}=0 $). 
Interestingly, already rather weak Fermi~wave~vector mismatch~($ F_\mathrm{K}=0.7 $) is sufficient to shift the transition~point between $ 0 $ and $ \pi $~states to remarkably larger values of the Rashba~SOC~strength. Moreover, the region of coexisting $ 0 $ and $ \pi $~states is significantly larger in the presence of moderate Fermi~wave~vector mismatch~[$ \lambda^\alpha \approx 0.80 \ldots 1.30 $; see Fig.~\ref{FigCPRFWVM}(a)] than in the junction without mismatches~[$ \lambda^\alpha \approx 0.40 \ldots 0.55 $; see Fig.~\ref{FigCPROutofplane}(a)]. Regarding the amplitudes of the Josephson~current, the maximal critical~current occurs in the junction with Fermi~wave~vector mismatch at a slightly smaller Rashba~SOC~strength~[$ \lambda^\alpha \approx 1.5 $; see Fig.~\ref{FigCPRFWVM}(b)], but is by far not as large as in the junction without mismatches~[compare to Fig.~\ref{FigCPROutofplane}(b)]. Contrarily, in the case of moderate mass~mismatch~($ F_\mathrm{M}=0.7 $), the spin-orbit~coupling induced crossover between $ 0 $ and $ \pi $~states already emerges at extremely weak Rashba~SOC~strengths and the transition~region with coexisting $ 0 $ and $ \pi $~states is quite narrow~[$ \lambda^\alpha \approx 0.30 \ldots 0.40 $; see Fig.~\ref{FigCPRMM}(a)]. The maximal critical~current in the presence of mass~mismatch can flow across the junction at a Rashba~SOC~strength~$ \lambda^\alpha \approx 2.0 $~[see Fig.~\ref{FigCPRMM}(b)] and is notably larger than in the junction with Fermi~wave~vector~mismatch, but still not as large as in junctions without mismatches. Therefore, our calculations suggest that the enhancement of the Josephson~current owing to the generation of spin-triplet Cooper~pairs becomes maximal in junctions without Fermi~wave~vector or mass~mismatch. Nevertheless, since the condition of perfectly matching Fermi~wave~vectors or effective masses in the superconducting and ferromagnetic materials of the Josephson~junction is practically not achievable, the discussed effects associated with mismatch might play a quantitative role in experiments, although they do not change the qualitative characteristics.

\section{Magnetoanisotropic~Josephson~current\label{SectionVI}}

\begin{figure}
	\includegraphics[width=0.475\textwidth]{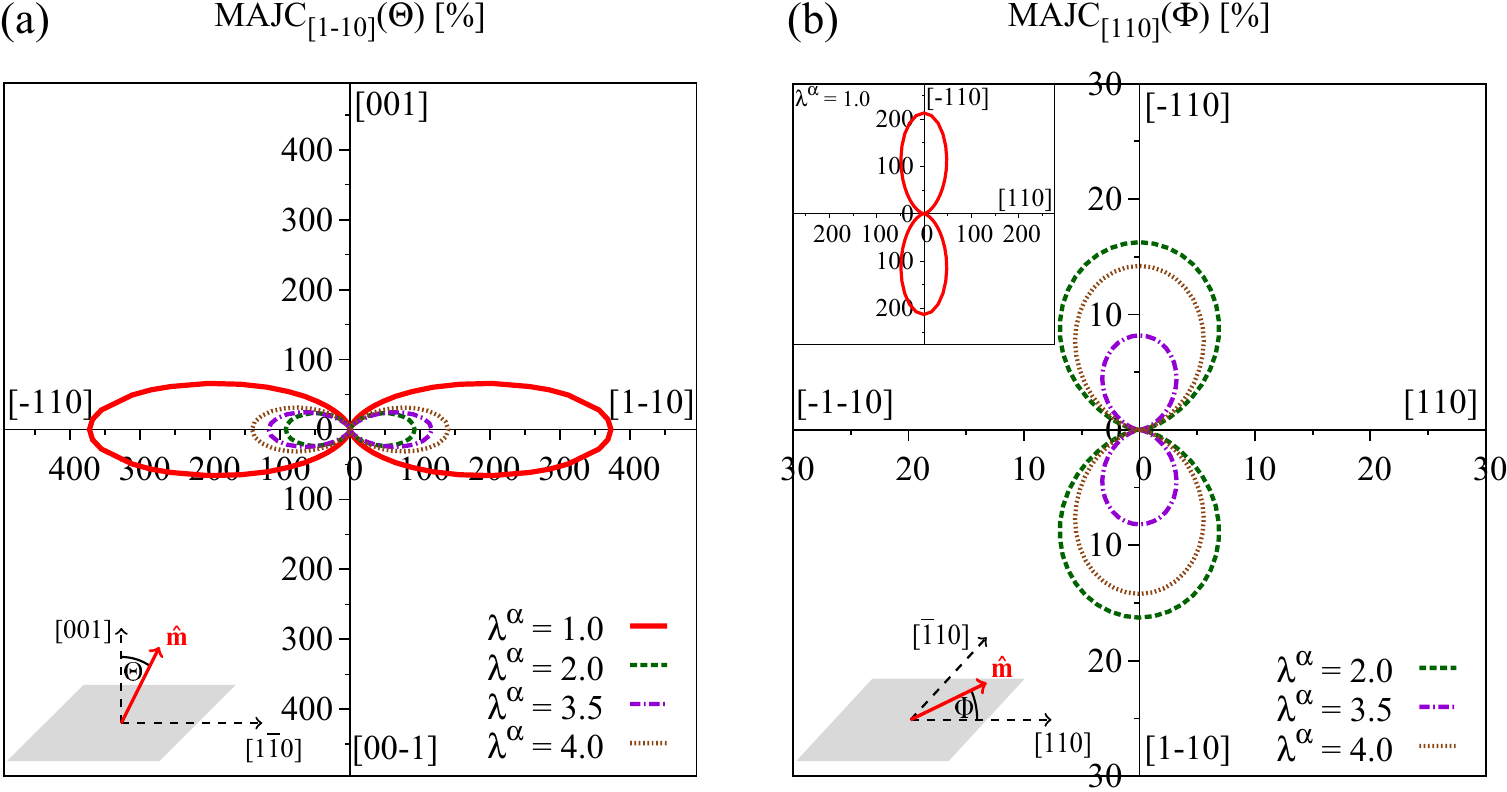}
	\colorcaption{\label{FigMAJC} (a)~Calculated angular~dependence of the out-of-plane~magnetoanisotropic~Josephson~current~(MAJC) with $ [1 \overline{1} 0] $ crystallographic reference axis for S/F/S~Josephson~junction with weak interfacial barriers~($ Z_\mathrm{L}=Z_\mathrm{R}=0.5 $), spin~polarization~$ P=0.7 $, effective interlayer~thickness~$ k_\mathrm{F} d = 8.2 $, moderate Dresselhaus~SOC~$ \lambda^\beta_\mathrm{L}=\lambda^\beta_\mathrm{R}=0.2 $, various Rashba~SOC~strengths~$ \lambda^\alpha_\mathrm{L}=\lambda^\alpha_\mathrm{R}=\lambda^\alpha $, and without Fermi~wave~vector or mass~mismatch~($ F_\mathrm{K}=F_\mathrm{M}=1 $). (b)~Calculated angular~dependence of the in-plane~MAJC with $ [110] $ crystallographic reference axis for the same junction~parameters as in~(a). The in-plane~MAJC for Rashba~SOC~strength~$ \lambda^\alpha=1.0 $ is shown in the inset.}
\end{figure}

While the significant enhancement of the Josephson~current due to the generation of spin-triplet Cooper~pairs is already a first precursor for the presence of interfacial spin-orbit~fields, magnetoanisotropic transport properties are a clear indication. Two configurations are important for investigating transport~anisotropies in vertical junctions: out-of-plane, in which the magnetization~direction~$ \hat{\mathbf{m}} $ is rotated along the polar~angle~$ \Theta $ in a plane perpendicular to the ferromagnetic layer, and in-plane, with changes of the azimuthal~angle~$ \Phi $ of~$ \hat{\mathbf{m}} $ in a plane parallel to the ferromagnetic layer. To quantify the anisotropies in both cases, we define the \textit{out-of-plane~magnetoanisotropic~Josephson~current~(MAJC)} as
\begin{equation}
	\label{EqAJCOut}
	\mathrm{MAJC}_{[1 \overline{1}0]} (\Theta) = \left. \frac{I_\mathrm{C} (0, \Phi) - I_\mathrm{C} (\Theta, \Phi)}{I_\mathrm{C} (\Theta, \Phi)} \right|_{\Phi=-90^\circ} ,
\end{equation}
and the \textit{in-plane~MAJC} as
\begin{equation}
	\label{EqAJCIn}
	\mathrm{MAJC}_{[110]} (\Phi) = \left. \frac{I_\mathrm{C} (\Theta,0) - I_\mathrm{C} (\Theta,\Phi)}{I_\mathrm{C} (\Theta,\Phi)} \right|_{\Theta=90^\circ} ,
\end{equation}
with $ I_\mathrm{C} $ being the critical~current. In general, the out-of-plane~MAJC depends on the azimuthal~angle~$ \Phi $, but we choose $ \Phi=-90^\circ $ as its reference.

Numerical results for the angular~dependencies of the out-of-plane and in-plane~MAJC in a realistic Josephson~junction with weak interfacial barriers~($ Z_\mathrm{L}=Z_\mathrm{R}=0.5 $), spin~polarization~$ P=0.7 $, and effective interlayer~thickness~$ k_\mathrm{F} d = 8.2 $ are presented in Fig.~\ref{FigMAJC}. The strengths of the Dresselhaus~spin-orbit~fields are chosen to be rather moderate~($ \lambda^\beta_\mathrm{L}=\lambda^\beta_\mathrm{R}=0.2 $), whereas the Rashba~SOC~parameters are varied from $ \lambda^\alpha_\mathrm{L}=\lambda^\alpha_\mathrm{R}=1.0 $ up to $ \lambda^\alpha_\mathrm{L}=\lambda^\alpha_\mathrm{R}=4.0 $. To simplify the analysis, the Fermi~wave~vectors and effective~masses in the superconducting and ferromagnetic parts are again assumed to be equal~($ F_\mathrm{K}=F_\mathrm{M}=1 $) in the following.

Similarly to earlier investigated TAMR~\cite{Moser2007, MatosAbiague2009} and MAAR~\cite{Hoegl2015} effects, the interplay of ferromagnetism and the interfacial spin-orbit~fields gives rise to marked magnetoanisotropies in the Josephson~current~flow. As a clear indication that the MAJC originates from this interplay, the characteristic $ C_{2v} $~symmetry of the spin-orbit~fields at the interfaces is transferred to the angular~dependencies of both the out-of-plane and in-plane~MAJC.

Since the in-plane~anisotropy stems from the interference of the interfacial Rashba and Dresselhaus~spin-orbit~fields~\cite{Fabian2004, Moser2007, MatosAbiague2009}, the in-plane~MAJC vanishes if one of the two fields is absent. In contrast, the out-of-plane~anisotropy arises from the Rashba or Dresselhaus~fields alone and is finite even in the presence of only one of the fields~(see Appendix~\ref{AppB}), making out-of-plane~MAJC measurements a robust probe for the presence of interfacial SOC, while the in-plane anisotropy is a sensitive probe of the interfacial symmetry. Owing to the complex interplay of the spin-orbit~fields and ferromagnetism, the amplitudes of the out-of-plane and in-plane~MAJC are very sensitive to changes of the Rashba~SOC~strength~$ \lambda^\alpha $ and vary nonmonotonically with respect to an increase of $ \lambda^\alpha $. Nevertheless, all calculated maximal MAJC~values are giant compared to TAMR in similar junctions~(for~example, TAMR in Fe/GaAs/Au~junctions is less than $ 1 \, \% $~\cite{Moser2007}). The Josephson~current flowing across the S/F/S~junctions is extremely sensitive to rotations of the magnetization~direction relative to the Rashba~spin-orbit~fields in the vicinity of $ 0 $-$ \pi $~transitions. Therefore, especially giant MAJC~amplitudes occur close to $ 0 $-$ \pi $~transitions~[e.g., $ \mathrm{MAJC}_{[1 \overline{1} 0]} (\Theta = \pi/2) \approx 373 \, \% $ in the out-of-plane and $ \mathrm{MAJC}_{[110]} (\Phi=\pi/2) \approx 213 \, \% $ in the in-plane~case at Rashba~SOC~strength~$ \lambda^\alpha=1.0 $], allowing one to identify the vicinity of a $ 0 $-$ \pi $~transition from MAJC~measurements, without
changing the thickness of the ferromagnetic layer in the controlling $ k_\mathrm{F} d $~parameter. At greater Rashba~SOC~magnitudes, the junctions are in stable $ \pi $~states and both the out-of-plane as well as the in-plane~MAJC are remarkably suppressed.

In Appendix~\ref{AppB} we study the dependence of the MAJC on the spin~polarization in the ferromagnet, covering both the Zeeman and exchange~coupling magnitudes. It is worth to mention here that the predicted anisotropies are negligible in the Zeeman~limit, which would correspond to exchange~fields less than $ 1 \, \mathrm{meV} $.

To investigate MAJC~effects experimentally, a junction~geometry comparable to the one of earlier TAMR~experiments~\cite{Moser2007} could be used. In such experiments, the magnetization~direction of the ferromagnetic layer is typically rotated by an external magnetic field. In order to realize this in the regarded S/F/S~Josephson~junction, two aspects are of great importance. First, the thickness of the superconducting electrodes has to be in a range so that the applied magnetic field can penetrate into the central ferromagnetic region of the junction. On the other hand, one needs to ensure that the magnetic field does not suppress the superconducting properties in the left and right electrode of the junction. One possible way would be the usage of dysprosium~magnets~\cite{Betthausen2012}, in which the magnetization can be oriented by an external field, but does not need the presence of the external field to remain in this position. As a consequence, the dysprosium~layer in the Josephson~junction could be premagnetized by applying an external magnetic~field that rotates the magnetization into a certain direction. After switching off the field, the magnetic layer still shows a permanent magnetization in the chosen direction and the MAJC~amplitudes can be measured without suppressing superconductivity. Moreover, transport through heterojunctions with $ \mathrm{NbN} $~contacts has been investigated very recently~\cite{Wan2015}. Owing to the large critical~fields of these superconducting electrodes~($ > 16 \, \mathrm{T} $) compared to the external magnetic~fields which are typically required to tilt the magnetization in the ferromagnet~($ \sim 0.1 \, \mathrm{T} $ up to a few T), Josephson~junctions with $ \mathrm{NbN} $~electrodes could also be of special interest to study the predicted magnetoanisotropic effects.

\section{Reversal of Josephson~current induced by changing the magnetization~orientation\label{SectionVII}}

\begin{figure}
	\includegraphics[width=0.45\textwidth]{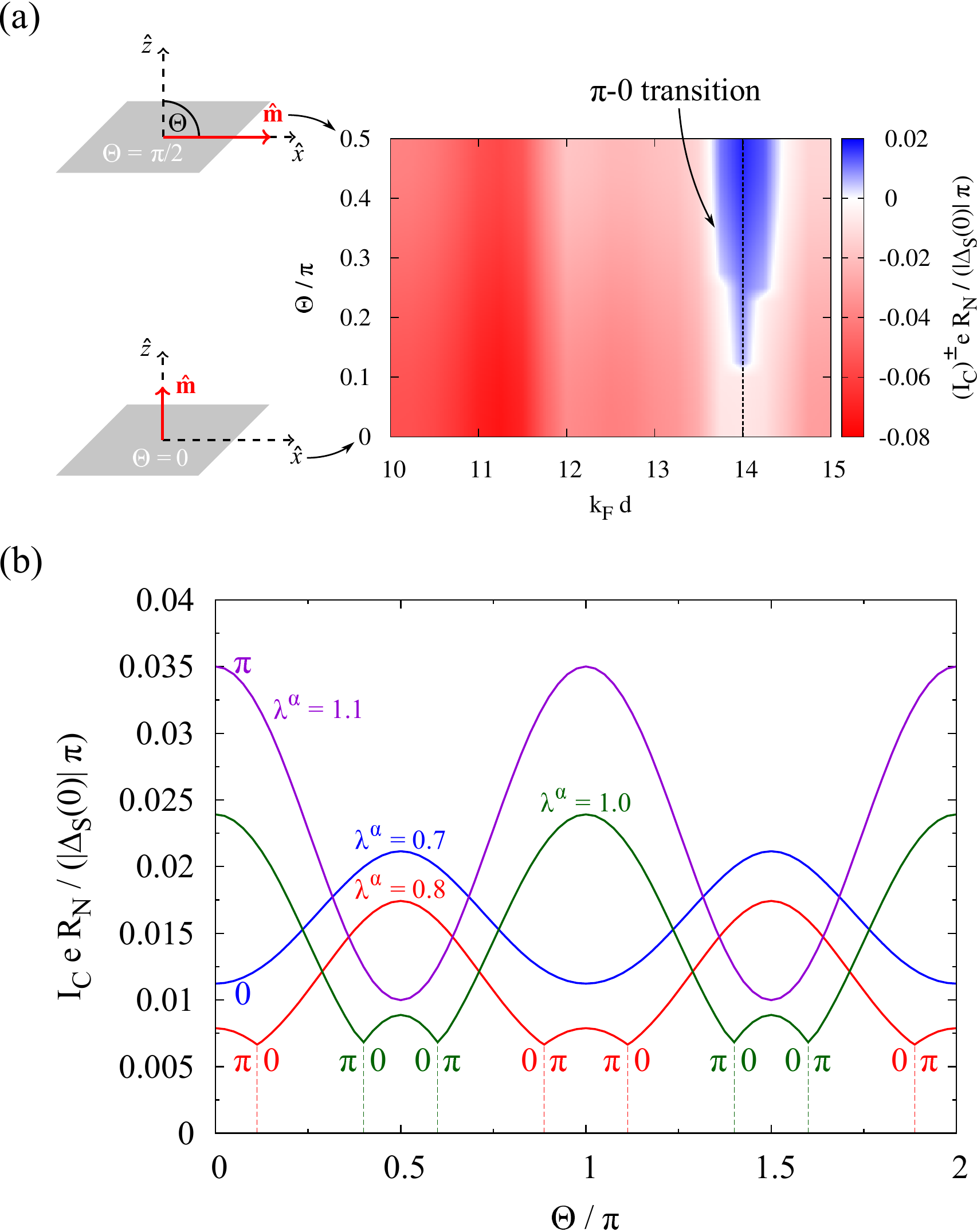}
	\colorcaption{\label{FigMagnetization}  (a)~Calculated (normalized) oriented~critical~current~$ \left( I_\mathrm{C} \right)^\pm $~(amplitude of the critical~current with positive sign for $ 0 $ and negative sign for $ \pi $~states) as a function of the magnetization~polar~angle~$ \Theta $ (azimuthal~angle~$ \Phi=0 $ is fixed) and effective interlayer~thickness~$ k_\mathrm{F} d $ for S/F/S~Josephson~junctions with weak interfacial barriers~($ Z_\mathrm{L}=Z_\mathrm{R}=0.5 $), spin~polarization~$ P=0.7 $, Rashba~SOC~strength~$ \lambda^\alpha_\mathrm{L}=\lambda^\alpha_\mathrm{R}=0.8 $ (for the choice of these parameters, see Appendix~\ref{AppC}), and without Fermi~wave~vector or mass~mismatch~($ F_\mathrm{K}=F_\mathrm{M}=1 $). The sign~change of the oriented~critical~current in the vicinity of $ k_\mathrm{F} d = 14.0 $~(white regions) indicates a transition from $ \pi $ to $ 0 $~states. (b)~Calculated dependence of the (normalized) critical~current~$ I_\mathrm{C} $ on the magnetization~polar~angle~$ \Theta $ (azimuthal~angle~$ \Phi=0 $ is fixed) for a specific S/F/S~Josephson~junction with weak interfacial barriers~($ Z_\mathrm{L}=Z_\mathrm{R}=0.5 $), spin~polarization~$ P=0.7 $, effective interlayer~thickness~$ k_\mathrm{F} d = 14.0 $~[compare to dashed line in~(a)], various Rashba~SOC~strengths~$ \lambda^\alpha_\mathrm{L}=\lambda^\alpha_\mathrm{R}=\lambda^\alpha $, and without Fermi~wave~vector or mass~mismatch~($ F_\mathrm{K}=F_\mathrm{M}=1 $). Transitions between $ 0 $ and $ \pi $~states are indicated by the dips in the~$ I_\mathrm{C} $-$ \Theta $~relation. In all calculations, Dresselhaus~SOC is not present~($ \lambda^\beta_\mathrm{L}=\lambda^\beta_\mathrm{R}=0 $).}
\end{figure}

Following our above discussion on the huge magnetoanisotropies close to the $ 0 $-$ \pi $~transitions, we now focus
on the transitions themselves. We introduce the oriented~critical~current~$ \left( I_\mathrm{C} \right)^\pm $, given by the amplitude of the critical~current~$ I_\mathrm{C} $ with a positive sign for $ 0 $ and a negative sign for $ \pi $~states. The calculated dependence of~$ \left( I_\mathrm{C} \right)^\pm $ on the polar~magnetization~angle~$ \Theta $ and the effective interlayer~thickness~$ k_\mathrm{F} d $ is depicted for realistic S/F/S~Josephson~junctions with weak interfacial barriers~($ Z_\mathrm{L}=Z_\mathrm{R}=0.5 $), spin~polarization~$ P=0.7 $, and moderate Rashba~SOC~strengths~$ \lambda^\alpha_\mathrm{L}=\lambda^\alpha_\mathrm{R}=0.8 $~(Dresselhaus~SOC is again absent) in Fig.~\ref{FigMagnetization}(a).

As long as the magnetization is aligned perpendicular to the ferromagnetic layer, the Josephson~junctions are in stable $ \pi $~states for all investigated values of $ k_\mathrm{F} d $~[$ \left( I_\mathrm{C} \right)^\pm < 0 $]. However, close to $ k_\mathrm{F} d =14.0 $, rotating the magnetization towards the plane reverses the supercurrent~direction~[$ \left( I_\mathrm{C} \right)^\pm > 0 $], signifying $\pi $ to $ 0 $~transitions. The impact of the Rashba~SOC~strength on these transitions is studied for a specific junction with $ k_\mathrm{F} d = 14.0 $ in Fig.~\ref{FigMagnetization}(b), which plots the dependence of the critical~current~$ I_\mathrm{C} $ on the magnetization~polar~angle~$ \Theta $ for various Rashba~SOC~strengths~$ \lambda^\alpha $. All other parameters are the same as in Fig.~\ref{FigMagnetization}(a). The results in Fig.~\ref{FigMagnetization}(a) correspond to the case of~$ \lambda^\alpha=0.8 $ in Fig.~\ref{FigMagnetization}(b), where the transition between $ \pi $ and $ 0 $~states now occurs as a dip in the~$ I_\mathrm{C} $-$ \Theta $~relation. The transition~points are very sensitive to the Rashba~SOC~strength. For~example, increasing the Rashba~parameter from $ \lambda^\alpha=0.8 $ to $ \lambda^\alpha=1.0 $ shifts the first crossover between $ \pi $ and $ 0 $~states from $ \Theta \approx 0.11 \pi $ to $ \Theta \approx 0.40 \pi $. Already for slightly weaker~($ \lambda^\alpha = 0.7 $) or stronger~($ \lambda^\alpha = 1.1 $) Rashba~SOC, magnetization~orientation controlled $ 0 $-$ \pi $~transitions are absent. Instead, the junctions are in stable $ 0 $~states in the former and stable $ \pi $~states in the latter case for all regarded magnetization~orientations. Similar phenomena are predicted to occur in lateral S/N/S~junctions with Zeeman~splitting and uniform Rashba~coupling~\cite{Yokoyama2014, Yokoyama2014a, Arjoranta2016}. In Appendix~\ref{AppC} we provide a detailed analysis of the interlayer~thickness dependence of the transitions.

Experimental realization of these predictions could follow the same way as suggested for MAJC~measurements in the previous section.

\section{Summary\label{SectionVIII}}

In this paper we studied the interplay of the Josephson~effect, interfacial SOC, and exchange~coupling in ballistic vertical S/F/S~junctions. Our presented numerical calculations for realistic model~junctions confirmed previous findings that changing the thickness of the metallic interlayer offers one practical way to manipulate $ 0 $-$ \pi $~transitions. In the presence of interfacial Rashba~SOC, we found that modulating the strength of these spin-orbit~fields can not only significantly enhance the critical~current due to the formation of spin-triplet Cooper~pairs, but may also facilitate a crossover from $ 0 $ to $ \pi $~states, without the need to alter the thickness of the ferromagnetic layer. As a clear signature for the interfacial spin-orbit~fields, we propose to investigate out-of-plane~(Rashba) as well as in-plane~(Rashba and Dresselhaus) magnetoanisotropies in the Josephson~current~flow. These anisotropies are giant compared to normal-state TAMR in magnetic tunnel~junctions and even MAAR in single F/S~junctions, especially in the vicinity of $ 0 $ to $ \pi $ transitions. Vice versa, we showed that these $ 0 $-$ \pi $~transitions can also be controlled by solely rotating the magnetization direction in junctions with certain interlayer~thicknesses and Rashba~SOC~strengths.

\begin{acknowledgments}
	This work was supported by DFG~SFB Grant~No.~689 and by the International Doctorate Program Topological~Insulators of the Elite~Network of Bavaria. This project has received funding from the European~Union's Horizon~2020 research and innovation~programme under Grant~agreement No.~696656. The authors gratefully acknowledge useful discussions with Farkhad~Aliev and Christoph~Strunk on experimental realizations of the presented results. 
\end{acknowledgments}

\appendix

\section{Impact of interlayer thickness and spin~polarization on the Josephson current\label{AppA}}

In Sec.~\ref{SectionIV} we analyze the spin-orbit~coupling induced $ 0 $-$ \pi $~transitions in a realistic model~S/F/S~Josephson~junction. Here we qualitatively discuss the influence of the effective interlayer~thickness~$ k_\mathrm{F} d $ and the spin~polarization~$ P $ on these transitions. To simplify the analysis, we again assume equal effective~masses~($ F_\mathrm{M}=1 $) and Fermi~wave~vectors~($ F_\mathrm{K} = 1 $) in the superconducting and metallic regions of the regarded Josephson~junctions. If not specifically indicated, the magnetization~direction in the ferromagnet is aligned perpendicular to the junction~interfaces. 

\begin{figure}
	\includegraphics[width=0.47\textwidth]{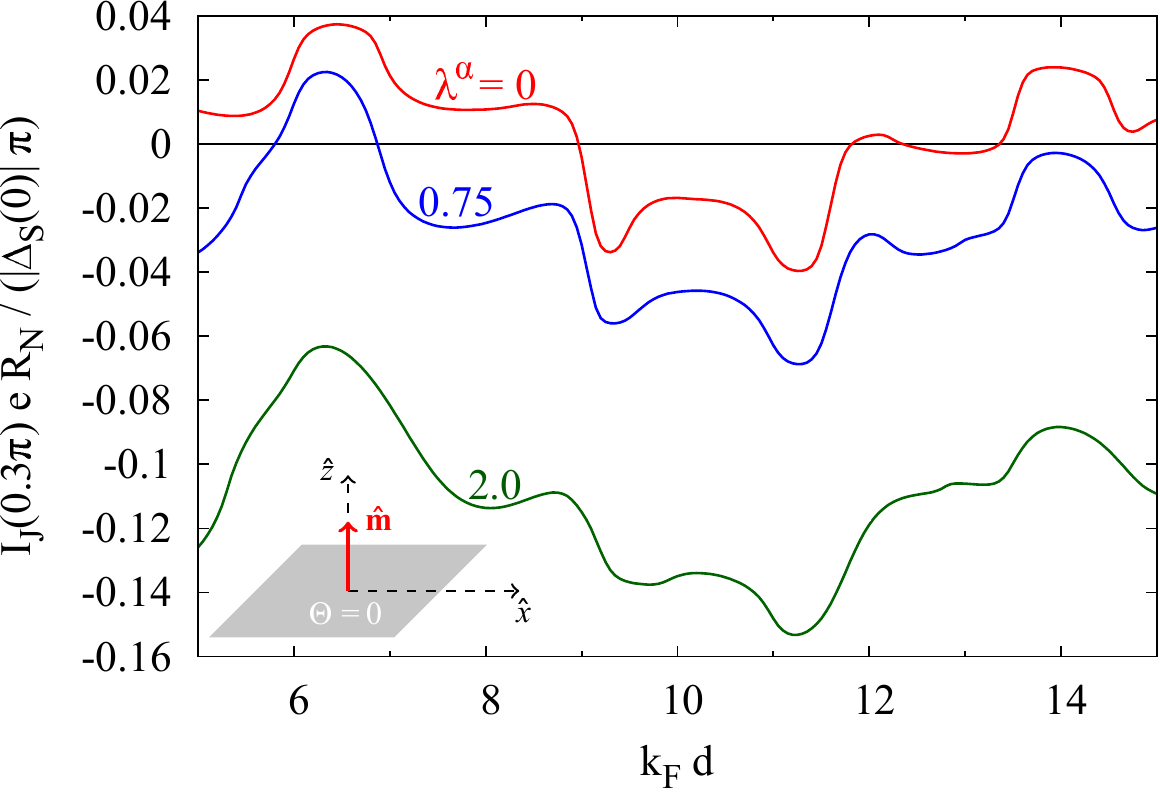}
	\colorcaption{\label{FigkfdWithSOC} Calculated dependence of the (normalized) Josephson~current~$ I_\mathrm{J} $ on the effective interlayer~thickness~$ k_\mathrm{F} d $ for S/F/S~Josephson~junctions with weak interfacial barriers~($ Z_\mathrm{L}=Z_\mathrm{R}=0.5 $), spin~polarization~$ P=0.7 $, different Rashba~SOC~strengths~$ \lambda^\alpha_\mathrm{L}=\lambda^\alpha_\mathrm{R}=\lambda^\alpha $, and without Fermi~wave~vector or mass~mismatch~($ F_\mathrm{K}=F_\mathrm{M}=1 $) at a fixed superconducting phase~difference of $ \phi_\mathrm{S}=0.3 \pi $. Dresselhaus~SOC is absent~($ \lambda^\beta_\mathrm{L}=\lambda^\beta_\mathrm{R}=0 $) and the magnetization~direction is oriented perpendicular to the ferromagnetic layer~($ \Theta=0 $ and $ \Phi=0 $; see illustration).}
\end{figure}
\begin{figure}
	\includegraphics[width=0.47\textwidth]{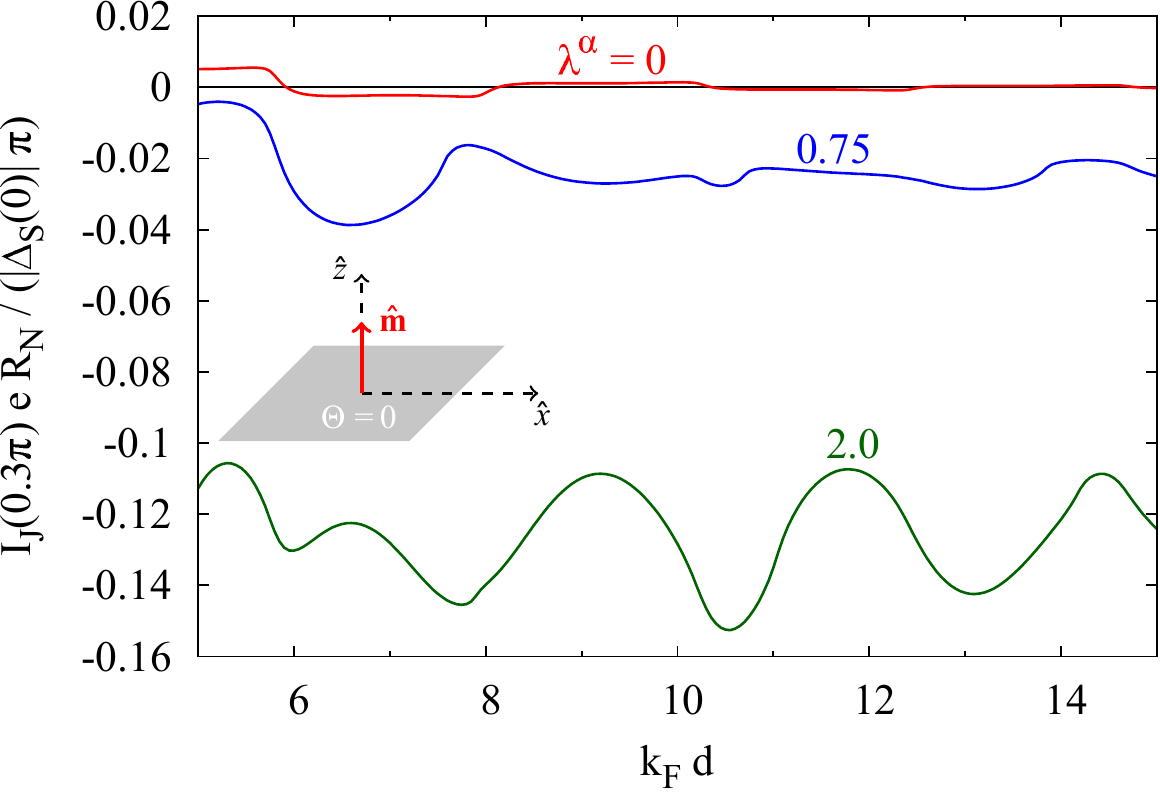}
	\colorcaption{\label{FigkfdWithSOCHalfMetal} Calculated dependence of the (normalized) Josephson~current~$ I_\mathrm{J} $ on the effective interlayer~thickness~$ k_\mathrm{F} d $ for S/F/S~Josephson~junctions with weak interfacial barriers~($ Z_\mathrm{L}=Z_\mathrm{R}=0.5 $), spin~polarization~$ P=1.0 $, different Rashba~SOC~strengths~$ \lambda^\alpha_\mathrm{L}=\lambda^\alpha_\mathrm{R}=\lambda^\alpha $, and without Fermi~wave~vector or mass~mismatch~($ F_\mathrm{K}=F_\mathrm{M}=1 $) at a fixed superconducting phase~difference of $ \phi_\mathrm{S}=0.3 \pi $. Dresselhaus~SOC is absent~($ \lambda^\beta_\mathrm{L}=\lambda^\beta_\mathrm{R}=0 $) and the magnetization~direction is oriented perpendicular to the ferromagnetic layer~($ \Theta=0 $ and $ \Phi=0 $; see illustration).}
\end{figure}
Figure~\ref{FigkfdWithSOC} shows the calculated dependence of the Josephson~current on the effective interlayer~thickness~$ k_\mathrm{F} d $ for S/F/S~Josephson~junctions with weak interfacial barriers~($ Z_\mathrm{L}=Z_\mathrm{R}=0.5 $), spin~polarization~$ P=0.7 $, and for three different strengths of symmetric Rashba~spin-orbit~fields~$ \lambda^\alpha_\mathrm{L}=\lambda^\alpha_\mathrm{R}=\lambda^\alpha $ with $ \lambda^\alpha=0 $, $ \lambda^\alpha=0.75 $, and $ \lambda^\alpha=2.0 $ at a fixed superconducting~phase~difference of $ \phi_\mathrm{S}=0.3\pi $. The findings at other phase~differences~$ 0<\phi_\mathrm{S}<\pi $ are comparable. As for the calculations discussed in Sec.~\ref{SectionIV}, Dresselhaus~SOC is absent~($ \lambda^\beta_\mathrm{L}=\lambda^\beta_\mathrm{R}=0 $). The situation of perfectly transparent interfaces is again not explicitly considered since the qualitative tendencies are analog. Without SOC, we recover the results explained in detail in Sec.~\ref{SectionIII}. Owing to the exchange~interaction in the ferromagnet, the Josephson~current exhibits an oscillatory dependence on the effective interlayer~thickness~$ k_\mathrm{F} d $. For certain values of $ k_\mathrm{F} d $, these oscillations reverse the direction~(sign) of the Josephson~current, signifying $ 0 $-$ \pi $~transitions. Also in the presence of interfacial Rashba~SOC, the oscillatory dependence of the Josephson~current on the effective interlayer~thickness still clearly appears. Regarding the amplitudes of the Josephson~current, we assert that an increase of the Rashba~SOC~strength to $ \lambda^\alpha = 0.75 $ decreases the Josephson~current for all regarded values of effective interlayer~thickness and narrows the regions in which the junctions realize $ 0 $~states~(positive Josephson~current) drastically. If the Rashba~SOC~strength gets increased to $ \lambda^\alpha=2.0 $, the amplitudes of the Josephson~current are lowered further and finally, the junctions are in $ \pi $~states for all presented values of $ k_\mathrm{F} d $. This finding is quite notable since it suggests that the crossover from $ 0 $ to $ \pi $~states, caused exclusively by modulating the Rashba~SOC~strength as we have already analyzed in Sec.~\ref{SectionIV} for one realistic model~junction with constant thickness of the metallic link, is quite general and emerges also in junctions with other values of interlayer~thickness. For completeness, we want to mention that a further enlargement of the Rashba~SOC~parameter~$ \lambda^\alpha $ again suppresses the absolute amplitudes of the Josephson~current remarkably because of the additional scattering introduced by SOC at the junction~interfaces (see also previous explanations). 

\begin{figure}
	\includegraphics[width=0.47\textwidth]{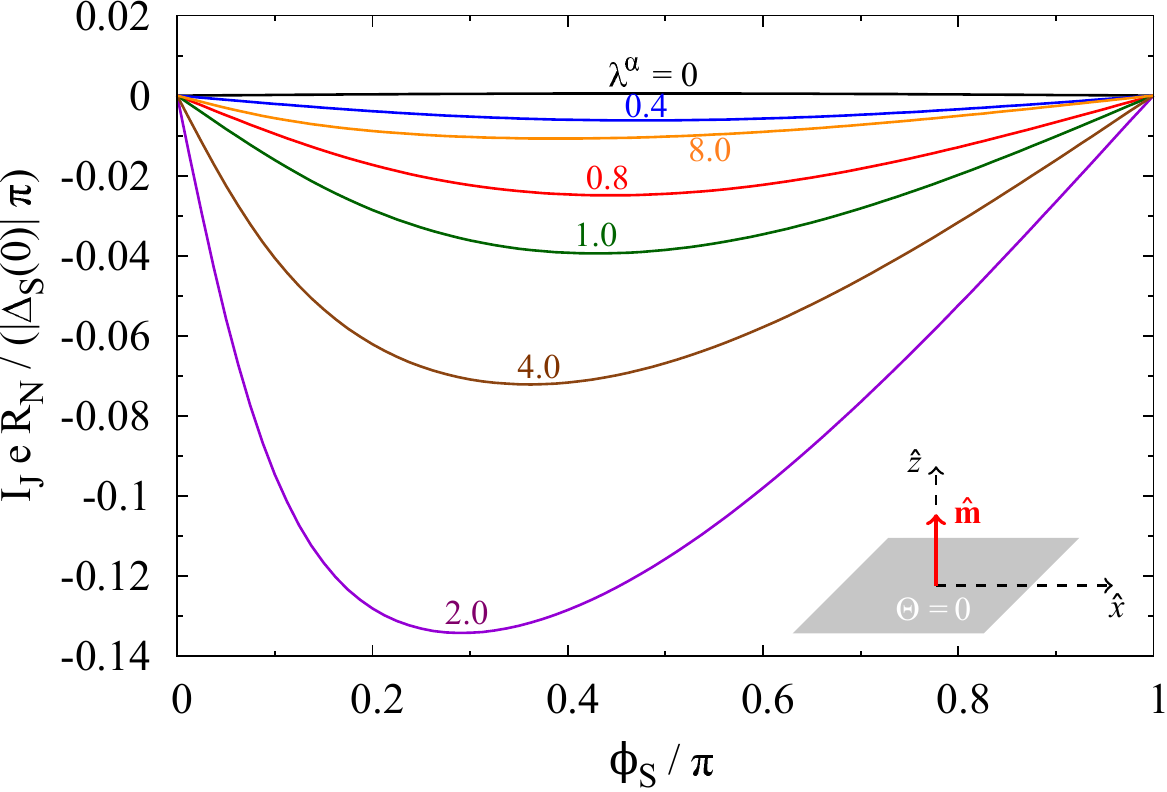}
	\colorcaption{\label{FigCPRSupplemental}  Calculated current-phase~relation for S/F/S~Josephson~junction with weak interfacial barriers~($ Z_\mathrm{L}=Z_\mathrm{R}=0.5 $), spin~polarization~$ P=1.0 $, effective interlayer~thickness~$ k_\mathrm{F} d = 8.2 $, different Rashba~SOC~strengths~$ \lambda^\alpha_\mathrm{L}=\lambda^\alpha_\mathrm{R}=\lambda^\alpha $, and without Fermi~wave~vector or mass~mismatch~($ F_\mathrm{K}=F_\mathrm{M}=1 $). Dresselhaus~SOC is absent~($ \lambda^\beta_\mathrm{L}=\lambda^\beta_\mathrm{R}=0 $) and the magnetization~direction is oriented perpendicular to the ferromagnetic layer~($ \Theta=0 $ and $ \Phi=0 $; see illustration).}
\end{figure}
In S/F/S~Josephson~junctions, in which the interlayer consists of a half-metallic~ferromagnet~(spin~polarization~$ P=1.0 $), the qualitative outcomes~(see Fig.~\ref{FigkfdWithSOCHalfMetal}) are similar to the preceding situation and hence, we only address a few interesting properties specifically. In junctions without interfacial Rashba~SOC~($ \lambda^\alpha = 0 $), the amplitudes of the Josephson~current become drastically damped with an increase of the effective interlayer~thickness~$ k_\mathrm{F} d $. This observation is a consequence of the density~of~states in half-metallic~ferromagnets, in which only one spin~subband is occupied at the Fermi~level. The resulting insulating character for the other spin~subband leads to a strong suppression of the transfer of spin-singlet Cooper~pairs, consisting of two correlated electrons with \textit{opposite} spin, across the metallic link and reduces the amount of supercurrent flowing in the system significantly. However, Keizer~\textit{et~al.} reported one experiment~\cite{Keizer2006} which predicts the existence of a measurable supercurrent~component in the half-metallic link of diffusive $ \mathrm{NbTiN/CrO_2/NbTiN} $~Josephson~junctions even over long length~scales~($ d \sim 1 \, \upmu \mathrm{m} $) compared to the coherence~length in half-metallic~ferromagnets. Since spin-singlet~Cooper~pairs cannot pass through the half-metallic $ \mathrm{CrO_2} $~layer, the observed Josephson~current~flow must be attributed to the generation of spin-triplet~Cooper~pairs via interfacial spin-flip~processes at the junction~interfaces~\cite{Keizer2006, Zheng2009}, carrying the supercurrent across the metallic region. Within our model, spin-flip~scattering at the superconductor/metal~interfaces is enabled by the presence of spin-orbit~fields. Indeed, the presented calculations reveal that increasing the Rashba~SOC~strength from $ \lambda^\alpha=0 $ to $ \lambda^\alpha=0.75 $ or $ \lambda^\alpha=2.0 $ enlarges the probability for the conversion of spin-singlet into spin-triplet~Cooper~pairs at the left interface and enhances the supercurrent~flow remarkably. These tendencies are also visible in the current-phase~relation for a superconductor/half-metallic~ferromagnet/superconductor~model~junction with an effective interlayer~thickness of $ k_\mathrm{F} d = 8.2 $, which is presented in Fig.~\ref{FigCPRSupplemental}. As in the junction with lower spin~polarization~(see Sec.~\ref{SectionIV}), the maximal critical~current appears for Rashba~SOC~strength~$ \lambda^\alpha \approx 2.0 $. Its amplitude is now more than two orders of magnitude larger than in the absence of SOC, which again reflects the dominant role of spin-triplet~Cooper~pairs in Josephson~junctions with half-metallic~links. Moreover, modulating the Rashba~parameter can also be identified as one possible way to control transitions between $ 0 $ and $ \pi $~states in superconductor/half-metallic~ferromagnet/superconductor Josephson~junctions. In comparison with the system in Sec.~\ref{SectionIV}~(spin~polarization~$ P=0.7 $), the crossover from $ 0 $ to $ \pi $~states is shifted to rather small SOC~strengths~($ \lambda^\alpha \approx 0.12 $) and the transition~region, in which $ 0 $ and $ \pi $~states coexist, is significantly narrowed~($ \lambda^\alpha \approx 0.11 \ldots 0.13 $).

\begin{figure}
	\includegraphics[width=0.47\textwidth]{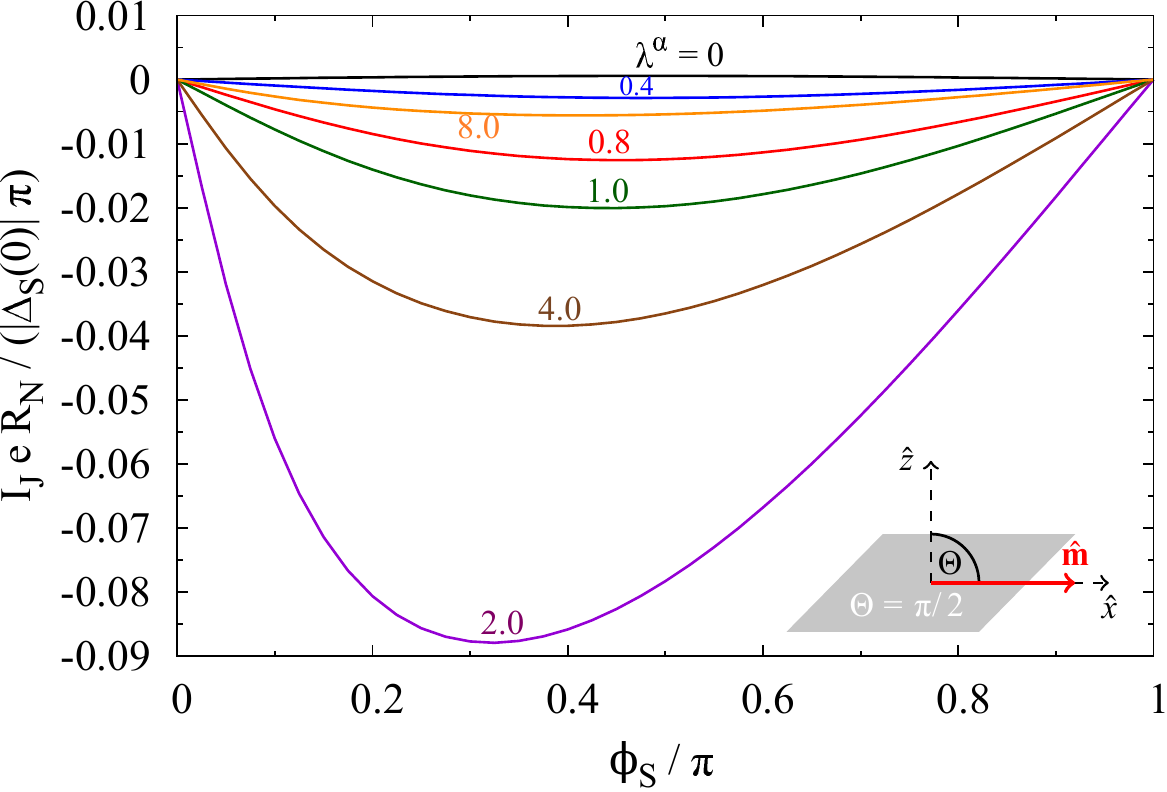}
	\colorcaption{\label{FigCPRSupplementalInPlane}  Calculated current-phase~relation for S/F/S~Josephson~junction with weak interfacial barriers~($ Z_\mathrm{L}=Z_\mathrm{R}=0.5 $), spin~polarization~$ P=1.0 $, effective interlayer~thickness~$ k_\mathrm{F} d = 8.2 $, different Rashba~SOC~strengths~$ \lambda^\alpha_\mathrm{L}=\lambda^\alpha_\mathrm{R}=\lambda^\alpha $, and without Fermi~wave~vector or mass~mismatch~($ F_\mathrm{K}=F_\mathrm{M}=1 $). Dresselhaus~SOC is absent~($ \lambda^\beta_\mathrm{L}=\lambda^\beta_\mathrm{R}=0 $) and the magnetization~direction is oriented parallel to the ferromagnetic layer~($ \Theta=\pi/2 $ and $ \Phi=0 $; see illustration).}
\end{figure}
Before finishing this part, we briefly want to discuss the calculated current-phase~relation for the same junction as before, but with the magnetization~direction aligned parallel to the ferromagnetic layer instead of perpendicular orientation~(see~Fig.~\ref{FigCPRSupplementalInPlane}). As we have already mentioned for the junction with lower spin~polarization in Sec.~\ref{SectionIV}, the contribution of the long-range spin-triplet~supercurrent to the total Josephson~current is remarkably smaller for in-plane~magnetization than for perpendicular magnetization. Therefore, the critical~current in the present junction is suppressed for all regarded strengths of the interfacial Rashba~spin-orbit~fields compared to the junction with perpendicular magnetization~(see~Fig.~\ref{FigCPRSupplemental}). However, increasing the Rashba~parameter still gives rise to a transition from $ 0 $ to $ \pi $~states. The region of coexisting $ 0 $ and $ \pi $~states is shifted to slightly larger Rashba~SOC~strengths~($ \lambda^\alpha \approx 0.16 \ldots 0.18 $) than in the case of perpendicular magnetization~($ \lambda^\alpha \approx 0.11 \ldots 0.13 $).

\section{Impact of interlayer~thickness and spin~polarization on magnetoanisotropic Josephson~current\label{AppB}}

\begin{figure}
	\centering
	\includegraphics[width=0.317\textwidth]{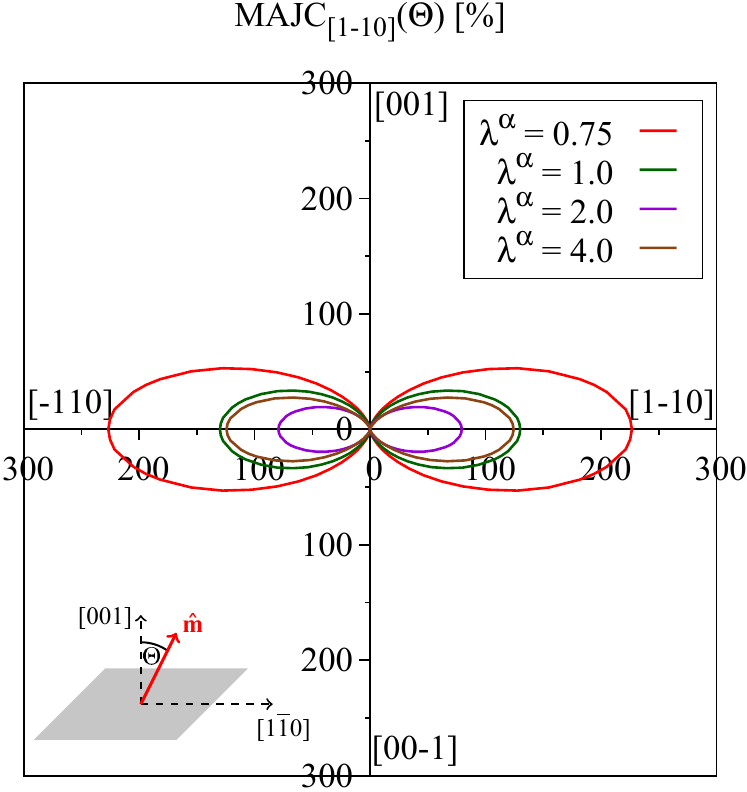}
	\colorcaption{\label{FigAJCRashba}  Calculated angular dependence of the out-of-plane~MAJC for S/F/S~Josephson~junction with weak interfacial barriers~($ Z_\mathrm{L}=Z_\mathrm{R}=0.5 $), spin~polarization~$ P=0.7 $, effective interlayer~thickness~$ k_\mathrm{F} d = 8.2 $, various Rashba~SOC~strengths~$ \lambda^\alpha_\mathrm{L}=\lambda^\alpha_\mathrm{R}=\lambda^\alpha $, and without Fermi~wave~vector or mass~mismatch~($ F_\mathrm{K}=F_\mathrm{M}=1 $). Dresselhaus~SOC is absent~($ \lambda^\alpha_\mathrm{L}=\lambda^\beta_\mathrm{R}=0 $).}
\end{figure}
\begin{figure}
	\includegraphics[width=0.3\textwidth]{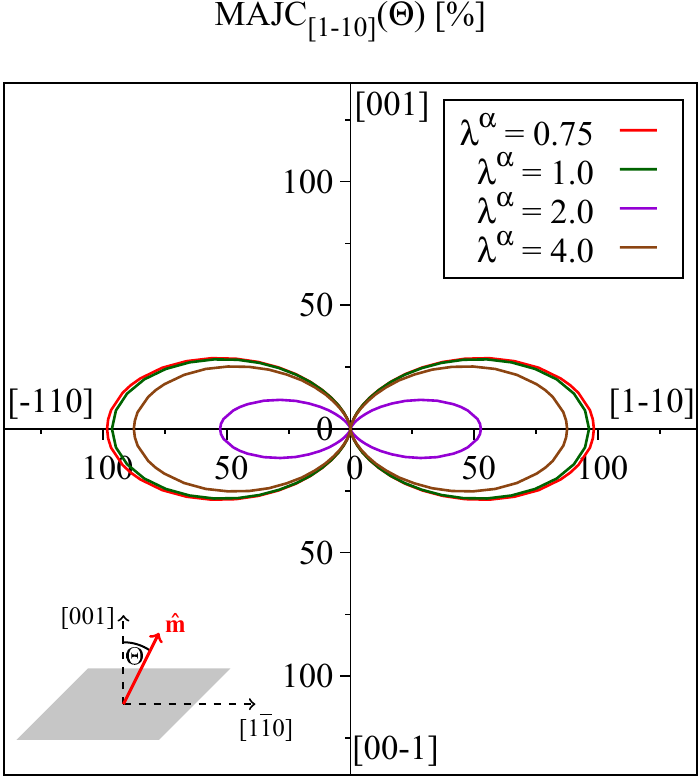}
	\colorcaption{\label{FigAJCPaperHalfMetal}  Calculated angular dependence of the out-of-plane~MAJC for S/F/S~Josephson~junction with weak interfacial barriers~($ Z_\mathrm{L}=Z_\mathrm{R}=0.5 $), spin~polarization~$ P=1.0 $, effective interlayer~thickness~$ k_\mathrm{F} d = 8.2 $, various Rashba~SOC~strengths~$ \lambda^\alpha_\mathrm{L}=\lambda^\alpha_\mathrm{R}=\lambda^\alpha $, and without Fermi~wave~vector or mass~mismatch~($ F_\mathrm{K}=F_\mathrm{M}=1 $). Dresselhaus~SOC is absent~($ \lambda^\beta_\mathrm{L}=\lambda^\beta_\mathrm{R}=0 $).}
\end{figure}

As explained in Sec.~\ref{SectionVI}, magnetic anisotropy of the Josephson~current~flow across S/F/S~junctions is a clear indication for the presence of interfacial spin-orbit~fields. While the in-plane~MAJC vanishes if only Rashba or Dresselhaus~spin-orbit~fields are present at the junction interfaces, the out-of-plane~MAJC arises from Rashba or Dresselhaus~SOC alone and can also be observed in the absence of one of the two spin-orbit~fields, providing a reliable way to identify the presence of SOC. To stress this finding, we show the angular~dependence of the out-of-plane~MAJC for a model S/F/S~Josephson~junction with weak interfacial barriers~($ Z_\mathrm{L}=Z_\mathrm{R}=0.5 $), spin~polarization~$ P=0.7 $, effective interlayer~thickness~$ k_\mathrm{F} d = 8.2 $, and for various values of Rashba~SOC~strengths~$ \lambda^\alpha_\mathrm{L}=\lambda^\alpha_\mathrm{R}=\lambda^\alpha $ in  Fig.~\ref{FigAJCRashba}~(Dresselhaus~SOC is absent, i.e., $ \lambda^\beta_\mathrm{L}=\lambda^\beta_\mathrm{R}=0 $). In all calculations analyzed in the following, we assume equal effective~masses and Fermi~wave~vectors in the superconducting and ferromagnetic materials~($ F_\mathrm{K}=F_\mathrm{M}=1 $) to simplify discussions. As expected, the out-of-plane~MAJC is finite for all considered strengths of interfacial Rashba~spin-orbit~fields and reflects $ C_{2v} $~symmetry as a clear indication for the presence of interfacial SOC. Similarly to the junction considered in Sec.~\ref{SectionVI}, the amplitudes of the out-of-plane~MAJC depend sensitively on the Rashba~SOC~parameters and change nonmonotonically with increasing SOC~strength. The maximal amplitudes of the out-of-plane~MAJC can again reveal huge values---especially in the vicinity of $ 0 $-$ \pi $~transitions---such as $ \mathrm{MAJC}_{[1 \overline{1} 0]}(\Theta = \pi/2) \approx 227 \, \% $ for $ \lambda^\alpha=0.75 $. If the metallic link is composed of a half-metallic~ferromagnet~(spin~polarization $ P=1.0 $; see Fig.~\ref{FigAJCPaperHalfMetal}), the nonmonotonic dependence of the out-of-plane~MAJC on the strength of the interfacial Rashba~spin-orbit~fields still appears.
For all presented Rashba~SOC~parameters, the (maximal) amplitudes of the out-of-plane~MAJC in the half-metallic~case are smaller than in the junction with spin~polarization~$ P=0.7 $. Furthermore, an increase of the Rashba~SOC~strength impacts the MAJC~amplitudes in superconductor/half-metallic~ferromagnet/superconductor~Josephson~junctions significantly only at stronger SOC~($ \lambda^\alpha > 1.0 $), whereas those in the previously discussed junction with lower spin~polarization are extremely sensitive to a change of the SOC~parameters even at rather moderate strengths of SOC.

\begin{figure}
	\includegraphics[width=0.47\textwidth]{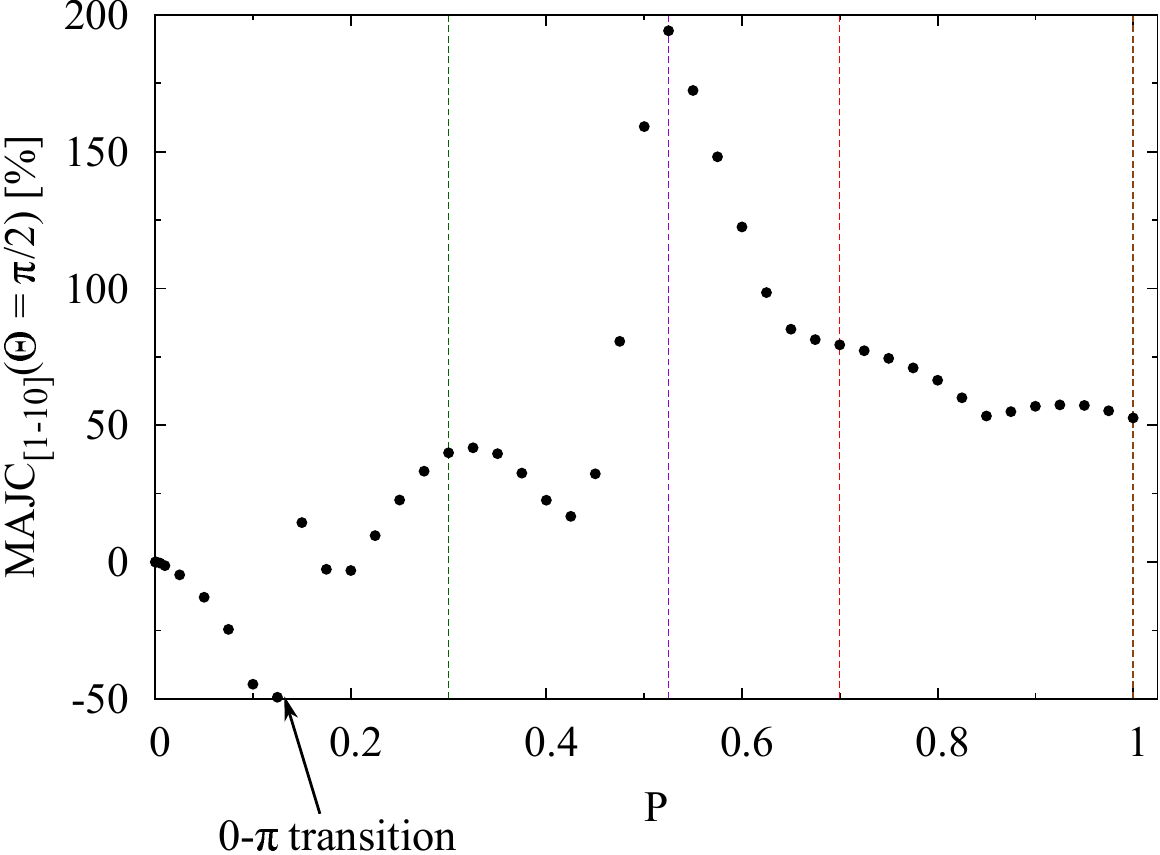}
	\colorcaption{\label{FigAJCDiffSpinPol}  Calculated dependence of the maximal amplitude of the out-of-plane~MAJC~[i.e., $ \mathrm{MAJC}_{[1\overline{1}0]} (\Theta=\pi/2) $] on the spin~polarization~$ P $ in S/F/S~Josephson~junctions with weak interfacial barriers~($ Z_\mathrm{L}=Z_\mathrm{R}=0.5 $), effective interlayer~thickness~$ k_\mathrm{F} d = 8.2 $, Rashba~SOC~strength~$ \lambda^\alpha_\mathrm{L}=\lambda^\alpha_\mathrm{R}=2.0 $, and without Fermi~wave~vector or mass~mismatch~($ F_\mathrm{K}=F_\mathrm{M}=1 $). Dresselhaus~SOC is absent~($ \lambda^\beta_\mathrm{L}=\lambda^\beta_\mathrm{R}=0 $).}
\end{figure}

\begin{figure}
	\includegraphics[width=0.3\textwidth]{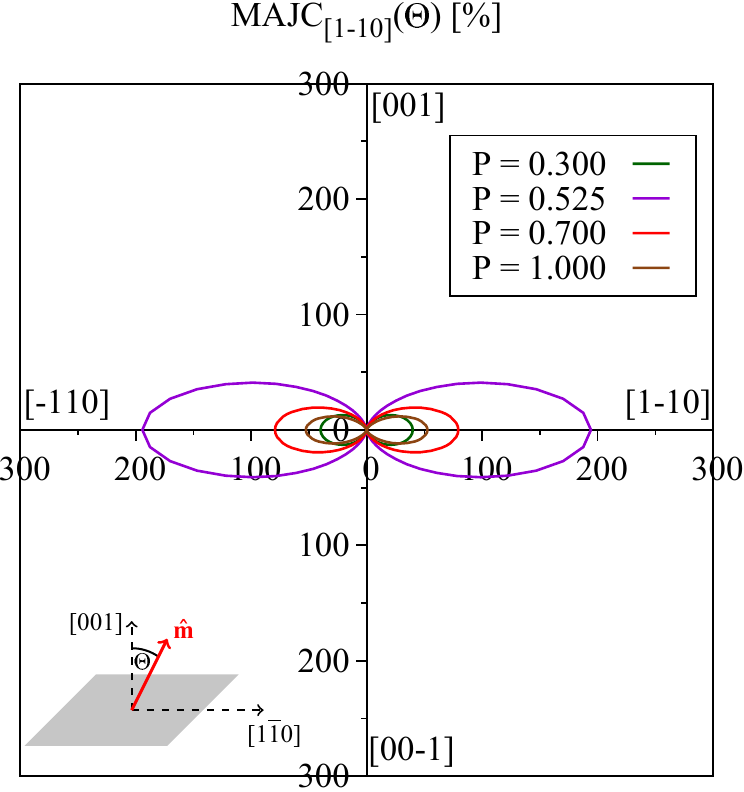}
	\colorcaption{\label{FigAJCSpinPol}  Calculated angular dependence of the out-of-plane~MAJC for S/F/S~Josephson~junction with weak interfacial barriers~($ Z_\mathrm{L}=Z_\mathrm{R}=0.5 $), effective interlayer~thickness~$ k_\mathrm{F} d = 8.2 $, Rashba~SOC~strength~$ \lambda^\alpha_\mathrm{L}=\lambda^\alpha_\mathrm{R}=2.0 $, various values of the spin~polarization~$ P $ in the ferromagnetic layer, and without Fermi~wave~vector or mass~mismatch~($ F_\mathrm{K}=F_\mathrm{M}=1 $). Dresselhaus~SOC is absent~($ \lambda^\beta_\mathrm{L}=\lambda^\beta_\mathrm{R}=0 $).}
\end{figure}

\begin{figure}
	\includegraphics[width=0.48\textwidth]{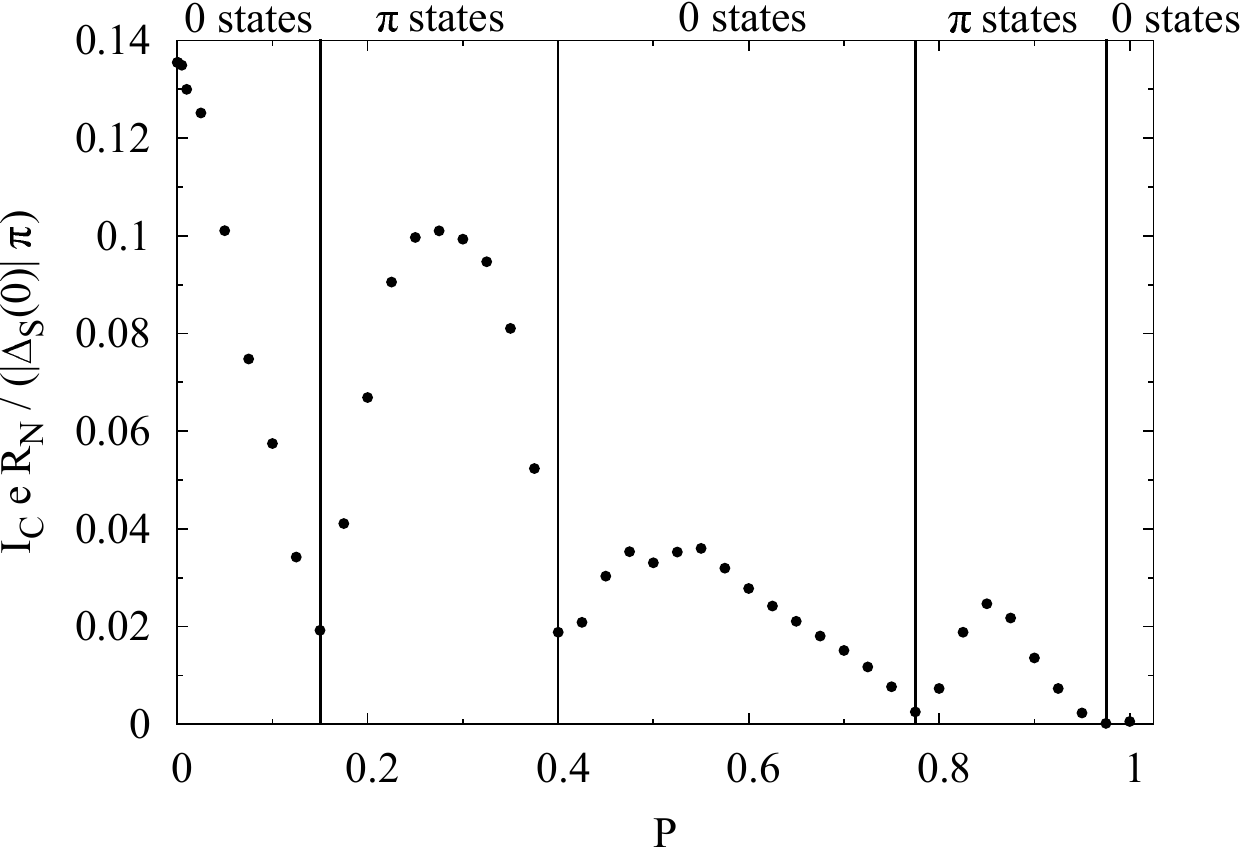}
	\colorcaption{\label{FigCritCurrentWithoutSOC}  Calculated dependence of the (normalized) critical~current~$ I_\mathrm{C} $ on the spin~polarization~$ P $ in S/F/S~Josephson~junctions with weak interfacial barriers~($ Z_\mathrm{L}=Z_\mathrm{R}=0.5 $) and effective interlayer~thickness~$ k_\mathrm{F} d = 8.2 $ in the absence of interfacial Rashba and Dresselhaus~SOC~($ \lambda^\alpha_\mathrm{L}=\lambda^\alpha_\mathrm{R}=\lambda^\beta_\mathrm{L}=\lambda^\beta_\mathrm{R}=0 $). The Fermi~wave~vectors and effective~masses in the superconductors and the ferromagnet are assumed to be equal~($ F_\mathrm{K}=F_\mathrm{M}=1 $).}
\end{figure}

\begin{figure}
	\includegraphics[width=0.47\textwidth]{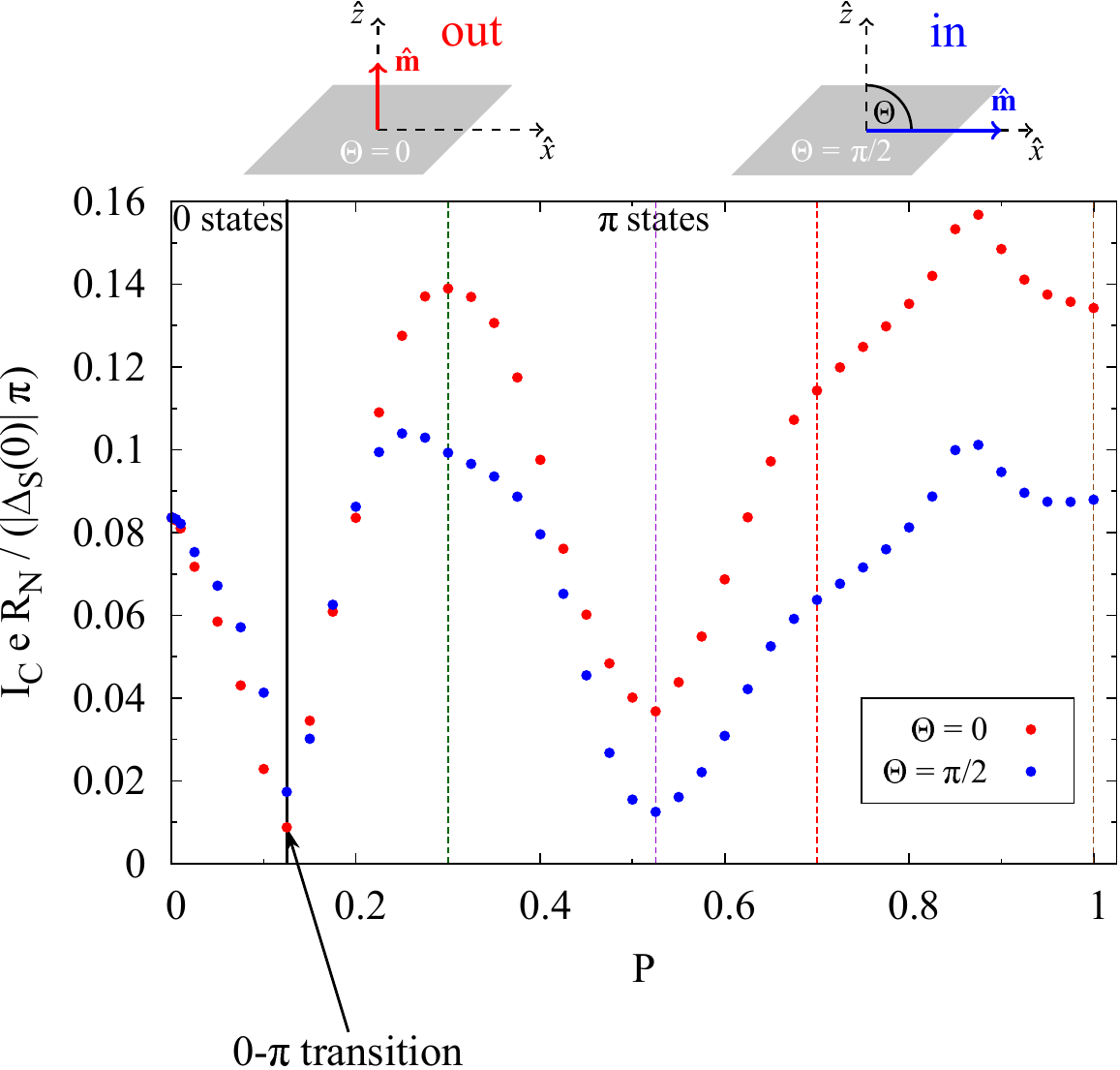}
	\colorcaption{\label{FigCritCurrent}  Calculated dependence of the (normalized) critical~current~$ I_\mathrm{C} $ on the spin polarization~$ P $ in S/F/S~Josephson~junctions with weak interfacial barriers~($ Z_\mathrm{L}=Z_\mathrm{R}=0.5 $), effective interlayer~thickness~$ k_\mathrm{F} d = 8.2 $, Rashba~SOC~strength~$ \lambda^\alpha_\mathrm{L}=\lambda^\alpha_\mathrm{R}=2.0 $, and without Fermi~wave~vector or mass~mismatch~($ F_\mathrm{K}=F_\mathrm{M}=1 $). Dresselhaus~SOC is absent~($ \lambda^\beta_\mathrm{L}=\lambda^\beta_\mathrm{R}=0 $) and the magnetization~direction in the ferromagnet can be oriented either perpendicular~($ \Theta=0 $ and $ \Phi=0 $; abbreviated with ``out'') or parallel~($ \Theta=\pi/2 $ and $ \Phi = 0 $; abbreviated with ``in'') to the ferromagnetic layer~(see illustration).}
\end{figure}

To get a deeper insight, how the strength of the exchange~splitting in the ferromagnetic region of the junctions impacts the magnetoanisotropy in the Josephson~current~flow, we present the maximal amplitudes of the out-of-plane~MAJC~[i.e., $ \mathrm{MAJC}_{[1\overline{1}0]}(\Theta = \pi/2) $] for interfacial Rashba~SOC~strength~$ \lambda^\alpha_\mathrm{L}=\lambda^\alpha_\mathrm{R}=2.0 $ and different values of the spin~polarization~$ P $ in the ferromagnetic layer in Fig.~\ref{FigAJCDiffSpinPol}. 
The other system~parameters are the same as in the previous calculations. The maxima of the out-of-plane~MAJC are extremely sensitive to a change of the spin~polarization~$ P $ in the interlayer and vary nonmonotonically with increasing~$ P $. The calculated MAJC~values switch sign from negative to positive values at a spin~polarization of $ P \sim 0.15 $, which is related to a crossover from $ 0 $ to $ \pi $~states. Noticeably huge MAJC~ratios appear for rather large spin~polarizations of $ P \approx 0.525 $~(see dashed violet line in Fig.~\ref{FigAJCDiffSpinPol}), whereas the anisotropy is nearly not measurable in junctions in which the spin~polarization in the intermediate~layer is extremely small~(e.g., in S/N/S~Josephson~junctions in which an applied magnetic~field gives rise to a comparatively small Zeeman~splitting). The angular~dependence of the out-of-plane~MAJC in the regarded Josephson~junctions is shown for four different spin~polarizations, i.e., $ P = 0.3 $, $ P=0.525 $, $ P=0.7 $, and $ P=1.0 $ in Fig.~\ref{FigAJCSpinPol}. 
The outcomes again reveal $ C_{2v} $~symmetry, originating from the omnipresent interfacial Rashba~spin-orbit~fields, and confirm the nonmonotonic variation of the maximal MAJC~amplitudes with an increase of the exchange~splitting~(spin~polarization) in the intermediate region of the Josephson~junctions~(compare to dashed colored lines in Fig.~\ref{FigAJCDiffSpinPol}). To understand the huge MAJC~amplitudes at spin~polarizations in the vicinity of $ P \approx 0.525 $, which signify huge magnetoanisotropies of the critical~current, we have a closer look at the dependence of the critical~current flowing across the Josephson~junctions on the spin~polarization in the ferromagnetic layer. At first, we regard the junctions in the absence of interfacial Rashba and Dresselhaus~SOC, $ \lambda^\alpha_\mathrm{L}=\lambda^\alpha_\mathrm{R}=\lambda^\beta_\mathrm{L}=\lambda^\beta_\mathrm{R}=0 $~(see Fig.~\ref{FigCritCurrentWithoutSOC}). The oscillatory dependence of the Josephson~current, and therefore also the critical~current, on the exchange~splitting~(spin~polarization) in the interlayer is an important property of S/F/S~Josephson~junctions and was studied in detail earlier~\cite{Radovic2003, Yokoyama2014, Yokoyama2014a}. Similarly to altering the interlayer~thickness, the oscillations of the Josephson~current induced by increasing the spin~polarization in the ferromagnetic layer can result in $ 0 $-$ \pi $~transitions for certain combinations of spin~polarization, effective interlayer~thickness, and barrier~strengths. The transition~points, separating $ 0 $ and $ \pi $~states, are indicated by the sharp dips in the~$ I_\mathrm{C} $-$ P $~relation in Fig.~\ref{FigCritCurrentWithoutSOC}. In particular for the chosen parameters, we predict the existence of four transitions between $ 0 $~ and $ \pi $~states at spin~polarizations of $ P \approx 0.15 $, $ P \approx 0.4 $, $ P \approx 0.775 $, and $ P \approx 0.975 $, respectively. The amplitudes of the critical~current remarkably decrease at large spin~polarizations since the transfer of spin-singlet~Cooper~pairs, which can solely carry the supercurrent in junctions without interfacial SOC, from one superconductor into the other one across the ferromagnetic part becomes drastically suppressed with an increase of the spin~polarization. As the magnetoanisotropies in the Josephson~current~flow stem from the interplay of the interfacial spin-orbit~fields and ferromagnetism, rotating the magnetization~direction in the ferromagnet has no influence on the Josephson~current~flow as long as SOC is absent. 
The situation becomes different if interfacial Rashba~spin-orbit~fields are present. Figure~\ref{FigCritCurrent} shows the dependence of the critical~current on the spin~polarization in the interlayer for the considered Josephson~junctions in the presence of interfacial Rashba~spin-orbit~fields~$ \lambda^\alpha_\mathrm{L}=\lambda^\alpha_\mathrm{R}=2.0 $~(Dresselhaus~SOC is still absent, i.e., $ \lambda^\beta_\mathrm{L}=\lambda^\beta_\mathrm{R}=0 $). We distinguish the situations in which the magnetization is aligned either perpendicular~($ \Theta = 0 $ and $ \Phi = 0 $; abbreviated with ``out'') or parallel~($ \Theta = \pi/2 $ and $ \Phi = 0 $; abbreviated with ``in'') to the ferromagnetic layer. Compared to the junctions without interfacial SOC, increasing the exchange~splitting~(spin~polarization) leads to only one crossover from $ 0 $ to $ \pi $~states at $ P \approx 0.125 $. The other three $ 0 $-$ \pi $~transitions, which we predict by changing the spin~polarization in the absence of Rashba~SOC~(see Fig.~\ref{FigCritCurrentWithoutSOC}), are suppressed if Rashba~spin-orbit~fields with strength $ \lambda^\alpha_\mathrm{L}=\lambda^\alpha_\mathrm{R}=2.0 $ are present at the junction~interfaces. Consequently, the junctions are now in stable $ \pi $~states for all spin~polarizations~$ 0.125 \lesssim P \leq 1 $. Particularly for $ 0.4 \leq P \leq 0.775 $, this means that the presence of sufficiently strong Rashba~SOC facilitates transitions from $ 0 $ to $ \pi $~states, confirming our previous findings that interfacial Rashba~SOC might be used to manipulate $ 0 $-$ \pi $~transitions effectively. By comparing the~$ I_\mathrm{C} $-$ P $~relation to the calculated maximal MAJC~amplitudes in Fig.~\ref{FigAJCDiffSpinPol}, we assert that the huge MAJC~ratios also mainly appear in that range of spin~polarizations~($ 0.4 \leq P \leq 0.775 $), for which the Rashba~spin-orbit~fields cause transitions from $ 0 $~states~(see Fig.~\ref{FigCritCurrentWithoutSOC}) to $ \pi $~states~(see vicinity of dashed violet line in Fig.~\ref{FigCritCurrent}). This observation suggests that the $ \pi $~states, induced by interfacial Rashba~SOC, are extremely sensitive to a change of the magnetization~direction in the ferromagnetic layer so that rotating the magnetization from the ``out'' to the ``in''~configuration can give rise to huge relative anisotropies in the Josephson~current~flow close to the spin-orbit~coupling induced $ 0 $-$\pi $~transitions as already mentioned in Sec.~\ref{SectionVI}.

\begin{figure}[t]
	\includegraphics[width=0.285\textwidth]{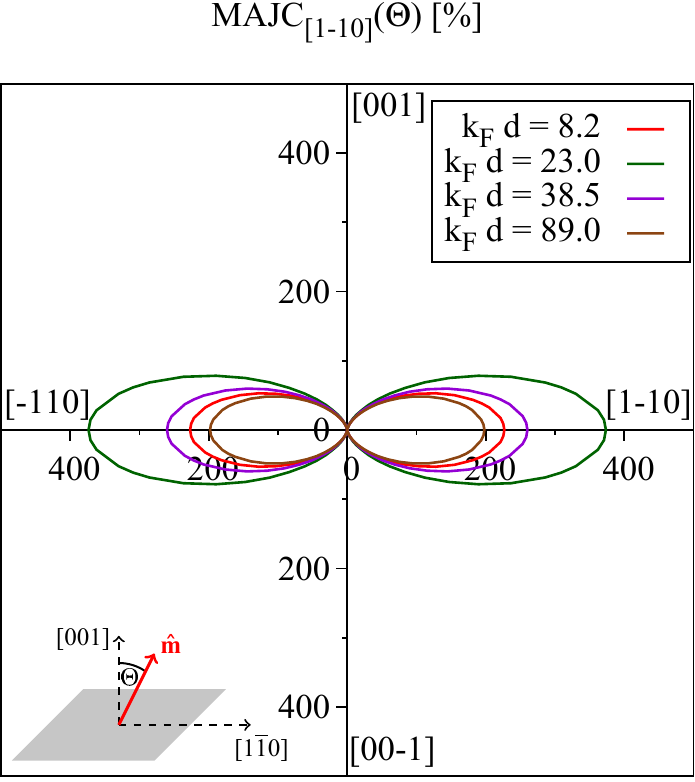}
	\colorcaption{\label{FigAJCkfd}  Calculated angular dependence of the out-of-plane~MAJC for S/F/S~Josephson~junction with weak interfacial barriers~($ Z_\mathrm{L}=Z_\mathrm{R}=0.5 $), spin~polarization~$ P=0.7 $, moderate Rashba~SOC~strength~$ \lambda^\alpha_\mathrm{L}=\lambda^\alpha_\mathrm{R}=0.75 $, various values of effective interlayer~thickness~$ k_\mathrm{F}  d $, and without Fermi~wave~vector or mass~mismatch~($ F_\mathrm{K}=F_\mathrm{M}=1 $). Dresselhaus~SOC is absent~($ \lambda^\beta_\mathrm{L}=\lambda^\beta_\mathrm{R}=0) $.}
\end{figure}
\begin{figure}[b]
	\includegraphics[width=0.285\textwidth]{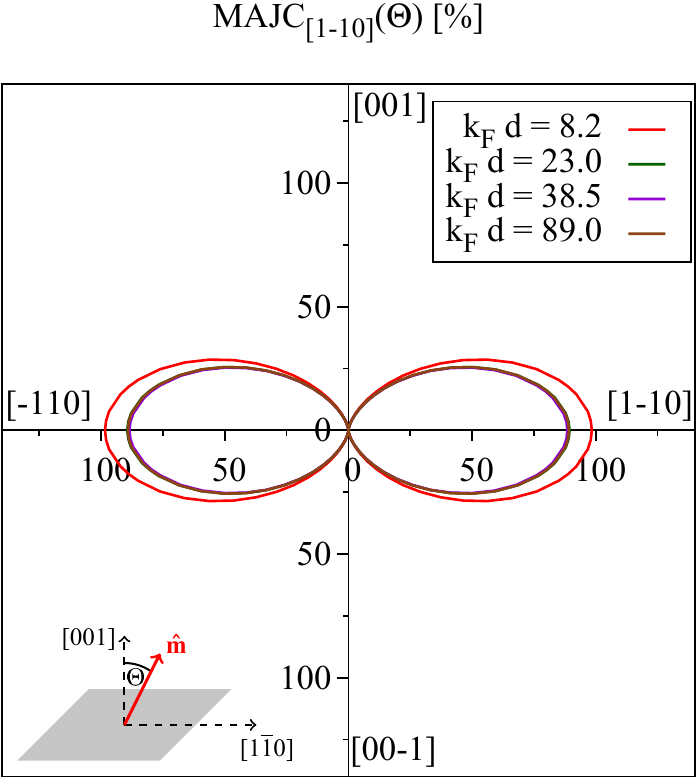}
	\colorcaption{\label{FigAJCkfdHalfMetal}  Calculated angular dependence of the out-of-plane~MAJC for S/F/S~Josephson~junction with weak interfacial barriers~($ Z_\mathrm{L}=Z_\mathrm{R}=0.5 $), spin~polarization~$ P=1.0 $, moderate Rashba~SOC~strength~$ \lambda^\alpha_\mathrm{L}=\lambda^\alpha_\mathrm{R}=0.75 $, various values of effective interlayer~thickness~$ k_\mathrm{F} d $, and without Fermi~wave~vector or mass~mismatch~($ F_\mathrm{K}=F_\mathrm{M}=1 $). Dresselhaus~SOC is absent~($ \lambda^\beta_\mathrm{L}=\lambda^\beta_\mathrm{R}=0 $).}
\end{figure}

In order to analyze the role of the interlayer~thickness, the angular~dependence of the out-of-plane~MAJC is shown for junctions with weak interfacial barriers~($ Z_\mathrm{L}=Z_\mathrm{R}=0.5 $), spin polarization~$ P=0.7 $, and different values of the effective interlayer~thickness~$ k_\mathrm{F} d $ in Fig.~\ref{FigAJCkfd}. The strengths of the Rashba~spin-orbit~fields at the junction~interfaces are set to rather moderate values of~$ \lambda^\alpha_\mathrm{L}=\lambda^\alpha_\mathrm{R}=0.75 $ and Dresselhaus~SOC is again not present~($ \lambda^\beta_\mathrm{L}=\lambda^\beta_\mathrm{R}=0 $). Similar considerations at other Rashba~SOC~parameters reveal analog characteristics. For all regarded parameter combinations, the out-of-plane~MAJC shows a nonmonotonic behavior with respect to an increase of the interlayer~thickness. Analogously to the preceding situations, the maximal amplitudes of the out-of-plane~MAJC can again exhibit huge values as $ \mathrm{MAJC}_{[110]} (\Theta = \slfrac{\pi}{2}) \approx 374 \, \% $ in junctions with an effective interlayer~thickness of $ k_\mathrm{F} d = 23.0 $. The chosen parameters correspond to realistic junctions---for~instance, to Josephson~junctions with an iron interlayer, which has a thickness of $ d \approx 2.9 \, \mathrm{nm} $~(as $ k_\mathrm{F} \approx 8.05 \times 10^7 \, \mathrm{cm}^{-1} $ in iron~\cite{Wang2003})---so that the predicted huge MAJC~ratios should also be observable in future experiments. If the metallic link is replaced by a half-metallic~ferromagnet with spin~polarization~$ P=1.0 $~(see Fig.~\ref{FigAJCkfdHalfMetal}), the qualitative features of the out-of-plane~MAJC do not change. Nevertheless, we observe that the maximal amplitudes of the MAJC get drastically suppressed compared to the previous case with spin~polarization~$ P=0.7 $. Moreover, the MAJC~amplitudes are less sensitive to changes of the interlayer~thickness than in the preceding systems and are nearly independent of $ k_\mathrm{F} d $ at $ k_\mathrm{F} d \gtrsim 23.0 $.

The impact of changing the spin~polarization in the ferromagnet or the interlayer~thickness on the in-plane~MAJC is similar to the one on the out-of-plane~MAJC. Therefore, we do not present explicit calculations for the in-plane~MAJC here.

\newpage

\begin{widetext}
\section{Impact of interlayer~thickness on magnetization~orientation controlled 0-$ \pmb{\pi} $~transitions\label{AppC}}

\begin{figure}[h]
		\centering
		\includegraphics[width=0.95\textwidth]{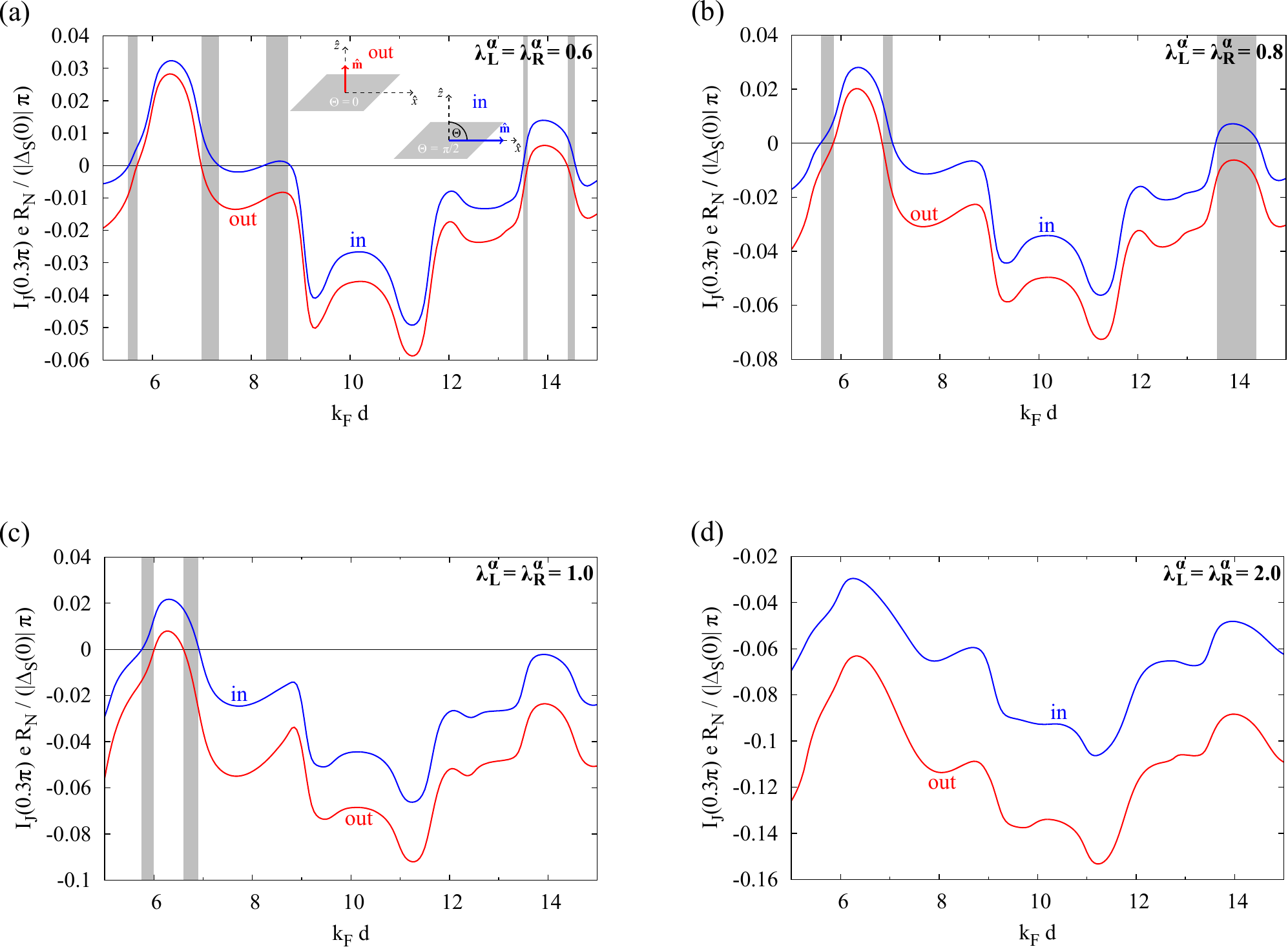}
		\colorcaption{\label{FigRotMagn}  Calculated dependence of the (normalized) Josephson~current~$ I_\mathrm{J} $ on the effective interlayer~thickness~$ k_\mathrm{F}  d $ for S/F/S~Josephson~junctions with weak interfacial barriers~($ Z_\mathrm{L}=Z_\mathrm{R}=0.5 $), spin~polarization~$ P=0.7 $, and without Fermi~wave~vector or mass~mismatch~($ F_\mathrm{K}=F_\mathrm{M}=1 $) at a fixed superconducting phase~difference of $ \phi_\mathrm{S}=0.3 \pi $. The magnetization~direction in the ferromagnet can be oriented either perpendicular~($ \Theta=0 $ and $ \Phi=0 $; abbreviated with ``out'') or parallel~($ \Theta=\slfrac{\pi}{2} $ and $ \Phi=0 $; abbreviated with ``in'') to the ferromagnetic layer as shown in the illustration in~(a). The Rashba~spin-orbit~fields at both interfaces have the effective strengths (a)~$ \lambda^\alpha_\mathrm{L}=\lambda^\alpha_\mathrm{R}=0.6 $, (b)~$ \lambda^\alpha_\mathrm{L}=\lambda^\alpha_\mathrm{R}=0.8 $, (c)~$ \lambda^\alpha_\mathrm{L}=\lambda^\alpha_\mathrm{R}=1.0 $, and (d)~$ \lambda^\alpha_\mathrm{L}=\lambda^\alpha_\mathrm{R}=2.0 $, respectively, whereas Dresselhaus~SOC is not present ($ \lambda^\beta_\mathrm{L}=\lambda^\beta_\mathrm{R}=0 $). The regions of interlayer~thicknesses, for which rotating the magnetization from the ``out'' to the ``in''~configuration switches the sign of the Josephson~current and induces potentially transitions between $ 0 $ and $ \pi $~states, are shaded.}
\end{figure}
\end{widetext}

We also examine in Sec.~\ref{SectionVII}, whether the interplay of ferromagnetism and the present spin-orbit~fields could offer another practical possibility to switch between $ 0 $ and $ \pi $~states by solely rotating the magnetization~direction in the ferromagnet. From an experimental point of view, the magnetization~direction could be changed by using Josephson~junctions in which the interlayer consists of dysprosium~magnets together with the application of an external magnetic~field as explained in detail in Sec.~\ref{SectionVI}. To get a first theoretical impression, we regard the dependence of the Josephson~current on the effective interlayer~thickness~$ k_\mathrm{F} d $ for a realistic system with spin~polarization~$ P=0.7 $, moderate interfacial barriers~($ Z_\mathrm{L}=Z_\mathrm{R}=0.5 $), and various strengths of Rashba~SOC in the cases that the magnetization is aligned either perpendicular to the ferromagnetic layer~($ \Theta=0 $ and $ \Phi=0 $; abbreviated with ``out'') or in a plane parallel to the ferromagnetic layer~($ \Theta=\slfrac{\pi}{2} $ and $ \Phi=0 $; abbreviated with ``in'') in Fig.~\ref{FigRotMagn}. As before, the superconducting phase~difference across the junction is set to a fixed value of $ \phi_\mathrm{S}=0.3 \pi $, which is sufficient to deduce qualitative trends. Indeed, we assert that rotating the magnetization in the ferromagnetic layer from the ``out'' to the ``in''~configuration may reverse the direction~(sign) of the Josephson~current~flow for certain combinations of the effective interlayer~thickness~$ k_\mathrm{F} d $ and Rashba~SOC~strengths~$ \lambda^\alpha $~(see shaded regions in Fig.~\ref{FigRotMagn}), which might be a first indication for the emergence of $ 0 $-$ \pi $~transitions. It is important to stress that the intervals of $ k_\mathrm{F} d $, for which the Josephson~current can switch its sign, probably signifying $ 0 $-$ \pi $~transitions, depend very sensitively on the Rashba~SOC~strength~$ \lambda^\alpha $ as one can see by comparing the shaded areas in the different panels of Fig.~\ref{FigRotMagn}, referring to different~$ \lambda^\alpha $, respectively. Also the strength of the exchange~splitting~(spin~polarization) in the ferromagnet may have a substantial impact on the magnetization~orientation controlled $ 0 $-$ \pi $~transitions. In Sec.~\ref{SectionVII}, we present a more detailed investigation focusing on the Rashba~SOC~strength~$ \lambda^\alpha_\mathrm{L}=\lambda^\alpha_\mathrm{R}=0.8 $, for which changing the magnetization~direction from the ``out'' to the ``in''~configuration can reverse the direction of the Josephson~current~flow in a comparatively wide range of interlayer~thicknesses in the vicinity of $ k_\mathrm{F} d = 14.0 $~[see shaded region around $ k_\mathrm{F} d = 14.0 $ in Fig.~\ref{FigRotMagn}(b)]. However, the results in Fig.~\ref{FigRotMagn} suggest that analog features can occur in junctions with other interlayer~thickness or Rashba~SOC~strength likewise so that an experimental verification of our predictions is not necessarily restricted to the presented junction~parameters.

\bibliography{paper}

\begin{thebibliography}{71}%
\makeatletter
\providecommand \@ifxundefined [1]{%
 \@ifx{#1\undefined}
}%
\providecommand \@ifnum [1]{%
 \ifnum #1\expandafter \@firstoftwo
 \else \expandafter \@secondoftwo
 \fi
}%
\providecommand \@ifx [1]{%
 \ifx #1\expandafter \@firstoftwo
 \else \expandafter \@secondoftwo
 \fi
}%
\providecommand \natexlab [1]{#1}%
\providecommand \enquote  [1]{``#1''}%
\providecommand \bibnamefont  [1]{#1}%
\providecommand \bibfnamefont [1]{#1}%
\providecommand \citenamefont [1]{#1}%
\providecommand \href@noop [0]{\@secondoftwo}%
\providecommand \href [0]{\begingroup \@sanitize@url \@href}%
\providecommand \@href[1]{\@@startlink{#1}\@@href}%
\providecommand \@@href[1]{\endgroup#1\@@endlink}%
\providecommand \@sanitize@url [0]{\catcode `\\12\catcode `\$12\catcode
  `\&12\catcode `\#12\catcode `\^12\catcode `\_12\catcode `\%12\relax}%
\providecommand \@@startlink[1]{}%
\providecommand \@@endlink[0]{}%
\providecommand \url  [0]{\begingroup\@sanitize@url \@url }%
\providecommand \@url [1]{\endgroup\@href {#1}{\urlprefix }}%
\providecommand \urlprefix  [0]{URL }%
\providecommand \Eprint [0]{\href }%
\providecommand \doibase [0]{http://dx.doi.org/}%
\providecommand \selectlanguage [0]{\@gobble}%
\providecommand \bibinfo  [0]{\@secondoftwo}%
\providecommand \bibfield  [0]{\@secondoftwo}%
\providecommand \translation [1]{[#1]}%
\providecommand \BibitemOpen [0]{}%
\providecommand \bibitemStop [0]{}%
\providecommand \bibitemNoStop [0]{.\EOS\space}%
\providecommand \EOS [0]{\spacefactor3000\relax}%
\providecommand \BibitemShut  [1]{\csname bibitem#1\endcsname}%
\let\auto@bib@innerbib\@empty
\bibitem [{\citenamefont {Eschrig}(2011)}]{Eschrig2011}%
  \BibitemOpen
  \bibfield  {author} {\bibinfo {author} {\bibfnamefont {M.}~\bibnamefont
  {Eschrig}},\ }\href
  {http://scitation.aip.org/content/aip/magazine/physicstoday/article/64/1/10.1063/1.3541944}
  {\bibfield  {journal} {\bibinfo  {journal} {Phys. Today}\ }\textbf {\bibinfo
  {volume} {64}},\ \bibinfo {pages} {43} (\bibinfo {year} {2011})}\BibitemShut
  {NoStop}%
\bibitem [{\citenamefont {Linder}\ and\ \citenamefont
  {Robinson}(2015)}]{Linder2015}%
  \BibitemOpen
  \bibfield  {author} {\bibinfo {author} {\bibfnamefont {J.}~\bibnamefont
  {Linder}}\ and\ \bibinfo {author} {\bibfnamefont {J.~W.~A.}\ \bibnamefont
  {Robinson}},\ }\href
  {http://www.nature.com/nphys/journal/v11/n4/abs/nphys3242.html} {\bibfield
  {journal} {\bibinfo  {journal} {Nature Phys.}\ }\textbf {\bibinfo {volume}
  {11}},\ \bibinfo {pages} {307} (\bibinfo {year} {2015})}\BibitemShut
  {NoStop}%
\bibitem [{\citenamefont {Gingrich}\ \emph {et~al.}(2016)\citenamefont
  {Gingrich}, \citenamefont {Niedzielski}, \citenamefont {Glick}, \citenamefont
  {Wang}, \citenamefont {Miller}, \citenamefont {Loloee}, \citenamefont {{Pratt
  Jr}},\ and\ \citenamefont {Birge}}]{Gingrich2016}%
  \BibitemOpen
  \bibfield  {author} {\bibinfo {author} {\bibfnamefont {E.~C.}\ \bibnamefont
  {Gingrich}}, \bibinfo {author} {\bibfnamefont {B.~M.}\ \bibnamefont
  {Niedzielski}}, \bibinfo {author} {\bibfnamefont {J.~A.}\ \bibnamefont
  {Glick}}, \bibinfo {author} {\bibfnamefont {Y.}~\bibnamefont {Wang}},
  \bibinfo {author} {\bibfnamefont {D.~L.}\ \bibnamefont {Miller}}, \bibinfo
  {author} {\bibfnamefont {R.}~\bibnamefont {Loloee}}, \bibinfo {author}
  {\bibfnamefont {W.~P.}\ \bibnamefont {{Pratt Jr}}}, \ and\ \bibinfo {author}
  {\bibfnamefont {N.~O.}\ \bibnamefont {Birge}},\ }\href
  {http://dx.doi.org/10.1038/nphys3681} {\bibfield  {journal} {\bibinfo
  {journal} {Nat. Phys.}\ }\textbf {\bibinfo {volume} {12}},\ \bibinfo {pages}
  {564} (\bibinfo {year} {2016})}\BibitemShut {NoStop}%
\bibitem [{\citenamefont {Golubov}\ \emph {et~al.}(2004)\citenamefont
  {Golubov}, \citenamefont {Kupriyanov},\ and\ \citenamefont
  {Il'ichev}}]{Golubov2004}%
  \BibitemOpen
  \bibfield  {author} {\bibinfo {author} {\bibfnamefont {A.~A.}\ \bibnamefont
  {Golubov}}, \bibinfo {author} {\bibfnamefont {M.~Y.}\ \bibnamefont
  {Kupriyanov}}, \ and\ \bibinfo {author} {\bibfnamefont {E.}~\bibnamefont
  {Il'ichev}},\ }\href {http://link.aps.org/doi/10.1103/RevModPhys.76.411}
  {\bibfield  {journal} {\bibinfo  {journal} {Rev. Mod. Phys.}\ }\textbf
  {\bibinfo {volume} {76}},\ \bibinfo {pages} {411} (\bibinfo {year}
  {2004})}\BibitemShut {NoStop}%
\bibitem [{\citenamefont {Buzdin}(2005)}]{Buzdin2005}%
  \BibitemOpen
  \bibfield  {author} {\bibinfo {author} {\bibfnamefont {A.~I.}\ \bibnamefont
  {Buzdin}},\ }\href {http://link.aps.org/doi/10.1103/RevModPhys.77.935}
  {\bibfield  {journal} {\bibinfo  {journal} {Rev. Mod. Phys.}\ }\textbf
  {\bibinfo {volume} {77}},\ \bibinfo {pages} {935} (\bibinfo {year}
  {2005})}\BibitemShut {NoStop}%
\bibitem [{\citenamefont {Bergeret}\ \emph {et~al.}(2005)\citenamefont
  {Bergeret}, \citenamefont {Volkov},\ and\ \citenamefont
  {Efetov}}]{Bergeret2005}%
  \BibitemOpen
  \bibfield  {author} {\bibinfo {author} {\bibfnamefont {F.~S.}\ \bibnamefont
  {Bergeret}}, \bibinfo {author} {\bibfnamefont {A.~F.}\ \bibnamefont
  {Volkov}}, \ and\ \bibinfo {author} {\bibfnamefont {K.~B.}\ \bibnamefont
  {Efetov}},\ }\href {http://link.aps.org/doi/10.1103/RevModPhys.77.1321}
  {\bibfield  {journal} {\bibinfo  {journal} {Rev. Mod. Phys.}\ }\textbf
  {\bibinfo {volume} {77}},\ \bibinfo {pages} {1321} (\bibinfo {year}
  {2005})}\BibitemShut {NoStop}%
\bibitem [{\citenamefont {Bulaevskii}\ \emph
  {et~al.}(1977{\natexlab{a}})\citenamefont {Bulaevskii}, \citenamefont
  {Kuzii},\ and\ \citenamefont {Sobyanin}}]{Bulaevskii1977}%
  \BibitemOpen
  \bibfield  {author} {\bibinfo {author} {\bibfnamefont {L.~N.}\ \bibnamefont
  {Bulaevskii}}, \bibinfo {author} {\bibfnamefont {V.~V.}\ \bibnamefont
  {Kuzii}}, \ and\ \bibinfo {author} {\bibfnamefont {A.~A.}\ \bibnamefont
  {Sobyanin}},\ }\href@noop {} {\bibfield  {journal} {\bibinfo  {journal}
  {Pis'ma Zh. Eksp. Teor. Fiz.}\ }\textbf {\bibinfo {volume} {25}},\ \bibinfo
  {pages} {314} (\bibinfo {year} {1977}{\natexlab{a}})}\BibitemShut {NoStop}%
\bibitem [{\citenamefont {Bulaevskii}\ \emph
  {et~al.}(1977{\natexlab{b}})\citenamefont {Bulaevskii}, \citenamefont
  {Kuzii},\ and\ \citenamefont {Sobyanin}}]{Bulaevskii1977alt}%
  \BibitemOpen
  \bibfield  {author} {\bibinfo {author} {\bibfnamefont {L.~N.}\ \bibnamefont
  {Bulaevskii}}, \bibinfo {author} {\bibfnamefont {V.~V.}\ \bibnamefont
  {Kuzii}}, \ and\ \bibinfo {author} {\bibfnamefont {A.~A.}\ \bibnamefont
  {Sobyanin}},\ }\href
  {http://www.jetpletters.ac.ru/ps/1410/article_21163.shtml} {\bibfield
  {journal} {\bibinfo  {journal} {JETP Lett.}\ }\textbf {\bibinfo {volume}
  {25}},\ \bibinfo {pages} {290} (\bibinfo {year}
  {1977}{\natexlab{b}})}\BibitemShut {NoStop}%
\bibitem [{\citenamefont {Buzdin}\ \emph
  {et~al.}(1982{\natexlab{a}})\citenamefont {Buzdin}, \citenamefont
  {Bulaevskii},\ and\ \citenamefont {Panyukov}}]{Buzdin1982}%
  \BibitemOpen
  \bibfield  {author} {\bibinfo {author} {\bibfnamefont {A.~I.}\ \bibnamefont
  {Buzdin}}, \bibinfo {author} {\bibfnamefont {L.~N.}\ \bibnamefont
  {Bulaevskii}}, \ and\ \bibinfo {author} {\bibfnamefont {S.~V.}\ \bibnamefont
  {Panyukov}},\ }\href@noop {} {\bibfield  {journal} {\bibinfo  {journal}
  {Pis'ma Zh. Eksp. Teor. Fiz.}\ }\textbf {\bibinfo {volume} {35}},\ \bibinfo
  {pages} {147} (\bibinfo {year} {1982}{\natexlab{a}})}\BibitemShut {NoStop}%
\bibitem [{\citenamefont {Buzdin}\ \emph
  {et~al.}(1982{\natexlab{b}})\citenamefont {Buzdin}, \citenamefont
  {Bulaevskii},\ and\ \citenamefont {Panyukov}}]{Buzdin1982alt}%
  \BibitemOpen
  \bibfield  {author} {\bibinfo {author} {\bibfnamefont {A.~I.}\ \bibnamefont
  {Buzdin}}, \bibinfo {author} {\bibfnamefont {L.~N.}\ \bibnamefont
  {Bulaevskii}}, \ and\ \bibinfo {author} {\bibfnamefont {S.~V.}\ \bibnamefont
  {Panyukov}},\ }\href
  {http://www.jetpletters.ac.ru/ps/1314/article_19853.shtml} {\bibfield
  {journal} {\bibinfo  {journal} {JETP Lett.}\ }\textbf {\bibinfo {volume}
  {35}},\ \bibinfo {pages} {178} (\bibinfo {year}
  {1982}{\natexlab{b}})}\BibitemShut {NoStop}%
\bibitem [{\citenamefont {Ryazanov}\ \emph {et~al.}(2001)\citenamefont
  {Ryazanov}, \citenamefont {Oboznov}, \citenamefont {Rusanov}, \citenamefont
  {Veretennikov}, \citenamefont {Golubov},\ and\ \citenamefont
  {Aarts}}]{Ryazanov2001}%
  \BibitemOpen
  \bibfield  {author} {\bibinfo {author} {\bibfnamefont {V.~V.}\ \bibnamefont
  {Ryazanov}}, \bibinfo {author} {\bibfnamefont {V.~A.}\ \bibnamefont
  {Oboznov}}, \bibinfo {author} {\bibfnamefont {A.~Y.}\ \bibnamefont
  {Rusanov}}, \bibinfo {author} {\bibfnamefont {A.~V.}\ \bibnamefont
  {Veretennikov}}, \bibinfo {author} {\bibfnamefont {A.~A.}\ \bibnamefont
  {Golubov}}, \ and\ \bibinfo {author} {\bibfnamefont {J.}~\bibnamefont
  {Aarts}},\ }\href {http://link.aps.org/doi/10.1103/PhysRevLett.86.2427}
  {\bibfield  {journal} {\bibinfo  {journal} {Phys. Rev. Lett.}\ }\textbf
  {\bibinfo {volume} {86}},\ \bibinfo {pages} {2427} (\bibinfo {year}
  {2001})}\BibitemShut {NoStop}%
\bibitem [{\citenamefont {Kontos}\ \emph {et~al.}(2002)\citenamefont {Kontos},
  \citenamefont {Aprili}, \citenamefont {Lesueur}, \citenamefont {Gen{\^{e}}t},
  \citenamefont {Stephanidis},\ and\ \citenamefont {Boursier}}]{Kontos2002}%
  \BibitemOpen
  \bibfield  {author} {\bibinfo {author} {\bibfnamefont {T.}~\bibnamefont
  {Kontos}}, \bibinfo {author} {\bibfnamefont {M.}~\bibnamefont {Aprili}},
  \bibinfo {author} {\bibfnamefont {J.}~\bibnamefont {Lesueur}}, \bibinfo
  {author} {\bibfnamefont {F.}~\bibnamefont {Gen{\^{e}}t}}, \bibinfo {author}
  {\bibfnamefont {B.}~\bibnamefont {Stephanidis}}, \ and\ \bibinfo {author}
  {\bibfnamefont {R.}~\bibnamefont {Boursier}},\ }\href
  {http://link.aps.org/doi/10.1103/PhysRevLett.89.137007} {\bibfield  {journal}
  {\bibinfo  {journal} {Phys. Rev. Lett.}\ }\textbf {\bibinfo {volume} {89}},\
  \bibinfo {pages} {137007} (\bibinfo {year} {2002})}\BibitemShut {NoStop}%
\bibitem [{\citenamefont {Robinson}\ \emph {et~al.}(2006)\citenamefont
  {Robinson}, \citenamefont {Piano}, \citenamefont {Burnell}, \citenamefont
  {Bell},\ and\ \citenamefont {Blamire}}]{Robinson2006}%
  \BibitemOpen
  \bibfield  {author} {\bibinfo {author} {\bibfnamefont {J.~W.~A.}\
  \bibnamefont {Robinson}}, \bibinfo {author} {\bibfnamefont {S.}~\bibnamefont
  {Piano}}, \bibinfo {author} {\bibfnamefont {G.}~\bibnamefont {Burnell}},
  \bibinfo {author} {\bibfnamefont {C.}~\bibnamefont {Bell}}, \ and\ \bibinfo
  {author} {\bibfnamefont {M.~G.}\ \bibnamefont {Blamire}},\ }\href
  {http://link.aps.org/doi/10.1103/PhysRevLett.97.177003} {\bibfield  {journal}
  {\bibinfo  {journal} {Phys. Rev. Lett.}\ }\textbf {\bibinfo {volume} {97}},\
  \bibinfo {pages} {177003} (\bibinfo {year} {2006})}\BibitemShut {NoStop}%
\bibitem [{\citenamefont {Feofanov}\ \emph {et~al.}(2010)\citenamefont
  {Feofanov}, \citenamefont {Oboznov}, \citenamefont {Bol'ginov}, \citenamefont
  {Lisenfeld}, \citenamefont {Poletto}, \citenamefont {Ryazanov}, \citenamefont
  {Rossolenko}, \citenamefont {Khabipov}, \citenamefont {Balashov},
  \citenamefont {Zorin}, \citenamefont {Dmitriev}, \citenamefont {Koshelets},\
  and\ \citenamefont {Ustinov}}]{Feofanov2010}%
  \BibitemOpen
  \bibfield  {author} {\bibinfo {author} {\bibfnamefont {A.~K.}\ \bibnamefont
  {Feofanov}}, \bibinfo {author} {\bibfnamefont {V.~A.}\ \bibnamefont
  {Oboznov}}, \bibinfo {author} {\bibfnamefont {V.~V.}\ \bibnamefont
  {Bol'ginov}}, \bibinfo {author} {\bibfnamefont {J.}~\bibnamefont
  {Lisenfeld}}, \bibinfo {author} {\bibfnamefont {S.}~\bibnamefont {Poletto}},
  \bibinfo {author} {\bibfnamefont {V.~V.}\ \bibnamefont {Ryazanov}}, \bibinfo
  {author} {\bibfnamefont {A.~N.}\ \bibnamefont {Rossolenko}}, \bibinfo
  {author} {\bibfnamefont {M.}~\bibnamefont {Khabipov}}, \bibinfo {author}
  {\bibfnamefont {D.}~\bibnamefont {Balashov}}, \bibinfo {author}
  {\bibfnamefont {A.~B.}\ \bibnamefont {Zorin}}, \bibinfo {author}
  {\bibfnamefont {P.~N.}\ \bibnamefont {Dmitriev}}, \bibinfo {author}
  {\bibfnamefont {V.~P.}\ \bibnamefont {Koshelets}}, \ and\ \bibinfo {author}
  {\bibfnamefont {A.~V.}\ \bibnamefont {Ustinov}},\ }\href
  {http://www.nature.com/doifinder/10.1038/nphys1700} {\bibfield  {journal}
  {\bibinfo  {journal} {Nat. Phys.}\ }\textbf {\bibinfo {volume} {6}},\
  \bibinfo {pages} {593} (\bibinfo {year} {2010})}\BibitemShut {NoStop}%
\bibitem [{\citenamefont {Yamashita}\ \emph {et~al.}(2005)\citenamefont
  {Yamashita}, \citenamefont {Tanikawa}, \citenamefont {Takahashi},\ and\
  \citenamefont {Maekawa}}]{Yamashita2005}%
  \BibitemOpen
  \bibfield  {author} {\bibinfo {author} {\bibfnamefont {T.}~\bibnamefont
  {Yamashita}}, \bibinfo {author} {\bibfnamefont {K.}~\bibnamefont {Tanikawa}},
  \bibinfo {author} {\bibfnamefont {S.}~\bibnamefont {Takahashi}}, \ and\
  \bibinfo {author} {\bibfnamefont {S.}~\bibnamefont {Maekawa}},\ }\href
  {http://link.aps.org/doi/10.1103/PhysRevLett.95.097001} {\bibfield  {journal}
  {\bibinfo  {journal} {Phys. Rev. Lett.}\ }\textbf {\bibinfo {volume} {95}},\
  \bibinfo {pages} {097001} (\bibinfo {year} {2005})}\BibitemShut {NoStop}%
\bibitem [{\citenamefont {{\v{Z}}uti{\'{c}}}\ and\ \citenamefont {{Das
  Sarma}}(2004)}]{Fabian2004}%
  \BibitemOpen
  \bibfield  {author} {\bibinfo {author} {\bibfnamefont {I.}~\bibnamefont
  {{\v{Z}}uti{\'{c}}}}\ and\ \bibinfo {author} {\bibfnamefont {S.}~\bibnamefont
  {{Das Sarma}}},\ }\href {http://link.aps.org/doi/10.1103/RevModPhys.76.323}
  {\bibfield  {journal} {\bibinfo  {journal} {Rev. Mod. Phys.}\ }\textbf
  {\bibinfo {volume} {76}},\ \bibinfo {pages} {323} (\bibinfo {year}
  {2004})}\BibitemShut {NoStop}%
\bibitem [{\citenamefont {Fabian}\ \emph {et~al.}(2007)\citenamefont {Fabian},
  \citenamefont {Matos-Abiague}, \citenamefont {Ertler}, \citenamefont
  {Stano},\ and\ \citenamefont {{\v{Z}}uti{\'{c}}}}]{Fabian2007}%
  \BibitemOpen
  \bibfield  {author} {\bibinfo {author} {\bibfnamefont {J.}~\bibnamefont
  {Fabian}}, \bibinfo {author} {\bibfnamefont {A.}~\bibnamefont
  {Matos-Abiague}}, \bibinfo {author} {\bibfnamefont {C.}~\bibnamefont
  {Ertler}}, \bibinfo {author} {\bibfnamefont {P.}~\bibnamefont {Stano}}, \
  and\ \bibinfo {author} {\bibfnamefont {I.}~\bibnamefont
  {{\v{Z}}uti{\'{c}}}},\ }\href
  {http://www.physics.sk/aps/pub.php?y=2007&pub=aps-07-04} {\bibfield
  {journal} {\bibinfo  {journal} {Acta Phys. Slovaca}\ }\textbf {\bibinfo
  {volume} {57}},\ \bibinfo {pages} {565} (\bibinfo {year} {2007})}\BibitemShut
  {NoStop}%
\bibitem [{\citenamefont {Ioffe}\ \emph {et~al.}(1999)\citenamefont {Ioffe},
  \citenamefont {Geshkenbein}, \citenamefont {Feigel'man}, \citenamefont
  {Fauch{\`e}re},\ and\ \citenamefont {Blatter}}]{Ioffe1999}%
  \BibitemOpen
  \bibfield  {author} {\bibinfo {author} {\bibfnamefont {L.~B.}\ \bibnamefont
  {Ioffe}}, \bibinfo {author} {\bibfnamefont {V.~B.}\ \bibnamefont
  {Geshkenbein}}, \bibinfo {author} {\bibfnamefont {M.~V.}\ \bibnamefont
  {Feigel'man}}, \bibinfo {author} {\bibfnamefont {A.~L.}\ \bibnamefont
  {Fauch{\`e}re}}, \ and\ \bibinfo {author} {\bibfnamefont {G.}~\bibnamefont
  {Blatter}},\ }\href
  {http://www.nature.com/nature/journal/v398/n6729/abs/398679a0.html}
  {\bibfield  {journal} {\bibinfo  {journal} {Nature}\ }\textbf {\bibinfo
  {volume} {398}},\ \bibinfo {pages} {679} (\bibinfo {year}
  {1999})}\BibitemShut {NoStop}%
\bibitem [{\citenamefont {Mooij}\ \emph {et~al.}(1999)\citenamefont {Mooij},
  \citenamefont {Orlando}, \citenamefont {Levitov}, \citenamefont {Tian},
  \citenamefont {van~der Wal},\ and\ \citenamefont {Lloyd}}]{Mooij1999}%
  \BibitemOpen
  \bibfield  {author} {\bibinfo {author} {\bibfnamefont {J.~E.}\ \bibnamefont
  {Mooij}}, \bibinfo {author} {\bibfnamefont {T.~P.}\ \bibnamefont {Orlando}},
  \bibinfo {author} {\bibfnamefont {L.}~\bibnamefont {Levitov}}, \bibinfo
  {author} {\bibfnamefont {L.}~\bibnamefont {Tian}}, \bibinfo {author}
  {\bibfnamefont {C.~H.}\ \bibnamefont {van~der Wal}}, \ and\ \bibinfo {author}
  {\bibfnamefont {S.}~\bibnamefont {Lloyd}},\ }\href
  {http://science.sciencemag.org/content/285/5430/1036} {\bibfield  {journal}
  {\bibinfo  {journal} {Science}\ }\textbf {\bibinfo {volume} {285}},\ \bibinfo
  {pages} {1036} (\bibinfo {year} {1999})}\BibitemShut {NoStop}%
\bibitem [{\citenamefont {Devoret}\ and\ \citenamefont
  {Schoelkopf}(2013)}]{Devoret2013}%
  \BibitemOpen
  \bibfield  {author} {\bibinfo {author} {\bibfnamefont {M.~H.}\ \bibnamefont
  {Devoret}}\ and\ \bibinfo {author} {\bibfnamefont {R.~J.}\ \bibnamefont
  {Schoelkopf}},\ }\href {http://science.sciencemag.org/content/339/6124/1169}
  {\bibfield  {journal} {\bibinfo  {journal} {Science}\ }\textbf {\bibinfo
  {volume} {339}},\ \bibinfo {pages} {1169} (\bibinfo {year}
  {2013})}\BibitemShut {NoStop}%
\bibitem [{\citenamefont {Likharev}(1979)}]{Likharev1979}%
  \BibitemOpen
  \bibfield  {author} {\bibinfo {author} {\bibfnamefont {K.~K.}\ \bibnamefont
  {Likharev}},\ }\href {http://link.aps.org/doi/10.1103/RevModPhys.51.101}
  {\bibfield  {journal} {\bibinfo  {journal} {Rev. Mod. Phys.}\ }\textbf
  {\bibinfo {volume} {51}},\ \bibinfo {pages} {101} (\bibinfo {year}
  {1979})}\BibitemShut {NoStop}%
\bibitem [{\citenamefont {Likharev}(2012)}]{Likharev2012}%
  \BibitemOpen
  \bibfield  {author} {\bibinfo {author} {\bibfnamefont {K.~K.}\ \bibnamefont
  {Likharev}},\ }\href
  {http://www.sciencedirect.com/science/article/pii/S0921453412002481}
  {\bibfield  {journal} {\bibinfo  {journal} {Phys. C}\ }\textbf {\bibinfo
  {volume} {482}},\ \bibinfo {pages} {6} (\bibinfo {year} {2012})}\BibitemShut
  {NoStop}%
\bibitem [{\citenamefont {Bychkov}\ and\ \citenamefont
  {Rashba}(1984)}]{Bychkov1984}%
  \BibitemOpen
  \bibfield  {author} {\bibinfo {author} {\bibfnamefont {Y.~A.}\ \bibnamefont
  {Bychkov}}\ and\ \bibinfo {author} {\bibfnamefont {E.~I.}\ \bibnamefont
  {Rashba}},\ }\href
  {http://stacks.iop.org/0022-3719/17/i=33/a=015?key=crossref.2f8159f54a3070499f32bac53e23f947}
  {\bibfield  {journal} {\bibinfo  {journal} {J. Phys. C}\ }\textbf {\bibinfo
  {volume} {17}},\ \bibinfo {pages} {6039} (\bibinfo {year}
  {1984})}\BibitemShut {NoStop}%
\bibitem [{\citenamefont {Gmitra}\ \emph {et~al.}(2013)\citenamefont {Gmitra},
  \citenamefont {Matos-Abiague}, \citenamefont {Draxl},\ and\ \citenamefont
  {Fabian}}]{Gmitra2013}%
  \BibitemOpen
  \bibfield  {author} {\bibinfo {author} {\bibfnamefont {M.}~\bibnamefont
  {Gmitra}}, \bibinfo {author} {\bibfnamefont {A.}~\bibnamefont
  {Matos-Abiague}}, \bibinfo {author} {\bibfnamefont {C.}~\bibnamefont
  {Draxl}}, \ and\ \bibinfo {author} {\bibfnamefont {J.}~\bibnamefont
  {Fabian}},\ }\href {http://link.aps.org/doi/10.1103/PhysRevLett.111.036603}
  {\bibfield  {journal} {\bibinfo  {journal} {Phys. Rev. Lett.}\ }\textbf
  {\bibinfo {volume} {111}},\ \bibinfo {pages} {036603} (\bibinfo {year}
  {2013})}\BibitemShut {NoStop}%
\bibitem [{\citenamefont {Dresselhaus}(1955)}]{Dresselhaus1955}%
  \BibitemOpen
  \bibfield  {author} {\bibinfo {author} {\bibfnamefont {G.}~\bibnamefont
  {Dresselhaus}},\ }\href {http://dx.doi.org/10.1103/PhysRev.100.580}
  {\bibfield  {journal} {\bibinfo  {journal} {Phys. Rev.}\ }\textbf {\bibinfo
  {volume} {100}},\ \bibinfo {pages} {580} (\bibinfo {year}
  {1955})}\BibitemShut {NoStop}%
\bibitem [{\citenamefont {Matos-Abiague}\ and\ \citenamefont
  {Fabian}(2009)}]{MatosAbiague2009}%
  \BibitemOpen
  \bibfield  {author} {\bibinfo {author} {\bibfnamefont {A.}~\bibnamefont
  {Matos-Abiague}}\ and\ \bibinfo {author} {\bibfnamefont {J.}~\bibnamefont
  {Fabian}},\ }\href {http://link.aps.org/doi/10.1103/PhysRevB.79.155303}
  {\bibfield  {journal} {\bibinfo  {journal} {Phys. Rev. B}\ }\textbf {\bibinfo
  {volume} {79}},\ \bibinfo {pages} {155303} (\bibinfo {year}
  {2009})}\BibitemShut {NoStop}%
\bibitem [{\citenamefont {Brey}\ \emph {et~al.}(2004)\citenamefont {Brey},
  \citenamefont {Tejedor},\ and\ \citenamefont
  {Fern{\'{a}}ndez-Rossier}}]{Brey2004}%
  \BibitemOpen
  \bibfield  {author} {\bibinfo {author} {\bibfnamefont {L.}~\bibnamefont
  {Brey}}, \bibinfo {author} {\bibfnamefont {C.}~\bibnamefont {Tejedor}}, \
  and\ \bibinfo {author} {\bibfnamefont {J.}~\bibnamefont
  {Fern{\'{a}}ndez-Rossier}},\ }\href
  {http://scitation.aip.org/content/aip/journal/apl/85/11/10.1063/1.1789241}
  {\bibfield  {journal} {\bibinfo  {journal} {Appl. Phys. Lett.}\ }\textbf
  {\bibinfo {volume} {85}},\ \bibinfo {pages} {1996} (\bibinfo {year}
  {2004})}\BibitemShut {NoStop}%
\bibitem [{\citenamefont {Moser}\ \emph {et~al.}(2007)\citenamefont {Moser},
  \citenamefont {Matos-Abiague}, \citenamefont {Schuh}, \citenamefont
  {Wegscheider}, \citenamefont {Fabian},\ and\ \citenamefont
  {Weiss}}]{Moser2007}%
  \BibitemOpen
  \bibfield  {author} {\bibinfo {author} {\bibfnamefont {J.}~\bibnamefont
  {Moser}}, \bibinfo {author} {\bibfnamefont {A.}~\bibnamefont
  {Matos-Abiague}}, \bibinfo {author} {\bibfnamefont {D.}~\bibnamefont
  {Schuh}}, \bibinfo {author} {\bibfnamefont {W.}~\bibnamefont {Wegscheider}},
  \bibinfo {author} {\bibfnamefont {J.}~\bibnamefont {Fabian}}, \ and\ \bibinfo
  {author} {\bibfnamefont {D.}~\bibnamefont {Weiss}},\ }\href
  {http://journals.aps.org/prl/abstract/10.1103/PhysRevLett.99.056601}
  {\bibfield  {journal} {\bibinfo  {journal} {Phys. Rev. Lett.}\ }\textbf
  {\bibinfo {volume} {99}},\ \bibinfo {pages} {056601} (\bibinfo {year}
  {2007})}\BibitemShut {NoStop}%
\bibitem [{\citenamefont {H{\"{o}}gl}\ \emph {et~al.}(2015)\citenamefont
  {H{\"{o}}gl}, \citenamefont {Matos-Abiague}, \citenamefont
  {{\v{Z}}uti{\'{c}}},\ and\ \citenamefont {Fabian}}]{Hoegl2015}%
  \BibitemOpen
  \bibfield  {author} {\bibinfo {author} {\bibfnamefont {P.}~\bibnamefont
  {H{\"{o}}gl}}, \bibinfo {author} {\bibfnamefont {A.}~\bibnamefont
  {Matos-Abiague}}, \bibinfo {author} {\bibfnamefont {I.}~\bibnamefont
  {{\v{Z}}uti{\'{c}}}}, \ and\ \bibinfo {author} {\bibfnamefont
  {J.}~\bibnamefont {Fabian}},\ }\href
  {http://journals.aps.org/prl/abstract/10.1103/PhysRevLett.115.116601}
  {\bibfield  {journal} {\bibinfo  {journal} {Phys. Rev. Lett.}\ }\textbf
  {\bibinfo {volume} {115}},\ \bibinfo {pages} {116601} (\bibinfo {year}
  {2015})}\BibitemShut {NoStop}%
\bibitem [{\citenamefont {Oreg}\ \emph {et~al.}(2010)\citenamefont {Oreg},
  \citenamefont {Refael},\ and\ \citenamefont {{von Oppen}}}]{Oreg2010}%
  \BibitemOpen
  \bibfield  {author} {\bibinfo {author} {\bibfnamefont {Y.}~\bibnamefont
  {Oreg}}, \bibinfo {author} {\bibfnamefont {G.}~\bibnamefont {Refael}}, \ and\
  \bibinfo {author} {\bibfnamefont {F.}~\bibnamefont {{von Oppen}}},\ }\href
  {http://link.aps.org/doi/10.1103/PhysRevLett.105.177002} {\bibfield
  {journal} {\bibinfo  {journal} {Phys. Rev. Lett.}\ }\textbf {\bibinfo
  {volume} {105}},\ \bibinfo {pages} {177002} (\bibinfo {year}
  {2010})}\BibitemShut {NoStop}%
\bibitem [{\citenamefont {Mourik}\ \emph {et~al.}(2012)\citenamefont {Mourik},
  \citenamefont {Zuo}, \citenamefont {Frolov}, \citenamefont {Plissard},
  \citenamefont {Bakkers},\ and\ \citenamefont {Kouwenhoven}}]{Mourik2012}%
  \BibitemOpen
  \bibfield  {author} {\bibinfo {author} {\bibfnamefont {V.}~\bibnamefont
  {Mourik}}, \bibinfo {author} {\bibfnamefont {K.}~\bibnamefont {Zuo}},
  \bibinfo {author} {\bibfnamefont {S.~M.}\ \bibnamefont {Frolov}}, \bibinfo
  {author} {\bibfnamefont {S.~R.}\ \bibnamefont {Plissard}}, \bibinfo {author}
  {\bibfnamefont {E.~P. A.~M.}\ \bibnamefont {Bakkers}}, \ and\ \bibinfo
  {author} {\bibfnamefont {L.~P.}\ \bibnamefont {Kouwenhoven}},\ }\href
  {http://science.sciencemag.org/content/336/6084/1003} {\bibfield  {journal}
  {\bibinfo  {journal} {Science}\ }\textbf {\bibinfo {volume} {336}},\ \bibinfo
  {pages} {1003} (\bibinfo {year} {2012})}\BibitemShut {NoStop}%
\bibitem [{\citenamefont {Rokhinson}\ \emph {et~al.}(2012)\citenamefont
  {Rokhinson}, \citenamefont {Liu},\ and\ \citenamefont
  {Furdyna}}]{Rokhinson2012}%
  \BibitemOpen
  \bibfield  {author} {\bibinfo {author} {\bibfnamefont {L.~P.}\ \bibnamefont
  {Rokhinson}}, \bibinfo {author} {\bibfnamefont {X.}~\bibnamefont {Liu}}, \
  and\ \bibinfo {author} {\bibfnamefont {J.~K.}\ \bibnamefont {Furdyna}},\
  }\href {http://dx.doi.org/10.1038/nphys2429} {\bibfield  {journal} {\bibinfo
  {journal} {Nature Phys.}\ }\textbf {\bibinfo {volume} {8}},\ \bibinfo {pages}
  {795} (\bibinfo {year} {2012})}\BibitemShut {NoStop}%
\bibitem [{\citenamefont {Duckheim}\ and\ \citenamefont
  {Brouwer}(2011)}]{Duckheim2011}%
  \BibitemOpen
  \bibfield  {author} {\bibinfo {author} {\bibfnamefont {M.}~\bibnamefont
  {Duckheim}}\ and\ \bibinfo {author} {\bibfnamefont {P.~W.}\ \bibnamefont
  {Brouwer}},\ }\href {http://link.aps.org/doi/10.1103/PhysRevB.83.054513}
  {\bibfield  {journal} {\bibinfo  {journal} {Phys. Rev. B}\ }\textbf {\bibinfo
  {volume} {83}},\ \bibinfo {pages} {054513} (\bibinfo {year}
  {2011})}\BibitemShut {NoStop}%
\bibitem [{\citenamefont {Nadj-Perge}\ \emph {et~al.}(2014)\citenamefont
  {Nadj-Perge}, \citenamefont {Drozdov}, \citenamefont {Li}, \citenamefont
  {Chen}, \citenamefont {Jeon}, \citenamefont {Seo}, \citenamefont {MacDonald},
  \citenamefont {Bernevig},\ and\ \citenamefont {Yazdani}}]{Nadj-Perge2014}%
  \BibitemOpen
  \bibfield  {author} {\bibinfo {author} {\bibfnamefont {S.}~\bibnamefont
  {Nadj-Perge}}, \bibinfo {author} {\bibfnamefont {I.~K.}\ \bibnamefont
  {Drozdov}}, \bibinfo {author} {\bibfnamefont {J.}~\bibnamefont {Li}},
  \bibinfo {author} {\bibfnamefont {H.}~\bibnamefont {Chen}}, \bibinfo {author}
  {\bibfnamefont {S.}~\bibnamefont {Jeon}}, \bibinfo {author} {\bibfnamefont
  {J.}~\bibnamefont {Seo}}, \bibinfo {author} {\bibfnamefont {A.~H.}\
  \bibnamefont {MacDonald}}, \bibinfo {author} {\bibfnamefont {B.~A.}\
  \bibnamefont {Bernevig}}, \ and\ \bibinfo {author} {\bibfnamefont
  {A.}~\bibnamefont {Yazdani}},\ }\href
  {http://www.sciencemag.org/content/346/6209/602.abstract} {\bibfield
  {journal} {\bibinfo  {journal} {Science}\ }\textbf {\bibinfo {volume}
  {346}},\ \bibinfo {pages} {602} (\bibinfo {year} {2014})}\BibitemShut
  {NoStop}%
\bibitem [{\citenamefont {Bergeret}\ \emph
  {et~al.}(2001{\natexlab{a}})\citenamefont {Bergeret}, \citenamefont
  {Volkov},\ and\ \citenamefont {Efetov}}]{Bergeret2001}%
  \BibitemOpen
  \bibfield  {author} {\bibinfo {author} {\bibfnamefont {F.~S.}\ \bibnamefont
  {Bergeret}}, \bibinfo {author} {\bibfnamefont {A.~F.}\ \bibnamefont
  {Volkov}}, \ and\ \bibinfo {author} {\bibfnamefont {K.~B.}\ \bibnamefont
  {Efetov}},\ }\href {http://link.aps.org/doi/10.1103/PhysRevLett.86.4096}
  {\bibfield  {journal} {\bibinfo  {journal} {Phys. Rev. Lett.}\ }\textbf
  {\bibinfo {volume} {86}},\ \bibinfo {pages} {4096} (\bibinfo {year}
  {2001}{\natexlab{a}})}\BibitemShut {NoStop}%
\bibitem [{\citenamefont {Volkov}\ \emph {et~al.}(2003)\citenamefont {Volkov},
  \citenamefont {Bergeret},\ and\ \citenamefont {Efetov}}]{Volkov2003}%
  \BibitemOpen
  \bibfield  {author} {\bibinfo {author} {\bibfnamefont {A.~F.}\ \bibnamefont
  {Volkov}}, \bibinfo {author} {\bibfnamefont {F.~S.}\ \bibnamefont
  {Bergeret}}, \ and\ \bibinfo {author} {\bibfnamefont {K.~B.}\ \bibnamefont
  {Efetov}},\ }\href {http://link.aps.org/doi/10.1103/PhysRevLett.90.117006}
  {\bibfield  {journal} {\bibinfo  {journal} {Phys. Rev. Lett.}\ }\textbf
  {\bibinfo {volume} {90}},\ \bibinfo {pages} {117006} (\bibinfo {year}
  {2003})}\BibitemShut {NoStop}%
\bibitem [{\citenamefont {Halterman}\ \emph {et~al.}(2007)\citenamefont
  {Halterman}, \citenamefont {Barsic},\ and\ \citenamefont
  {Valls}}]{Halterman2007}%
  \BibitemOpen
  \bibfield  {author} {\bibinfo {author} {\bibfnamefont {K.}~\bibnamefont
  {Halterman}}, \bibinfo {author} {\bibfnamefont {P.~H.}\ \bibnamefont
  {Barsic}}, \ and\ \bibinfo {author} {\bibfnamefont {O.~T.}\ \bibnamefont
  {Valls}},\ }\href {http://link.aps.org/doi/10.1103/PhysRevLett.99.127002}
  {\bibfield  {journal} {\bibinfo  {journal} {Phys. Rev. Lett.}\ }\textbf
  {\bibinfo {volume} {99}},\ \bibinfo {pages} {127002} (\bibinfo {year}
  {2007})}\BibitemShut {NoStop}%
\bibitem [{\citenamefont {Eschrig}\ and\ \citenamefont
  {L{\"{o}}fwander}(2008)}]{Eschrig2008}%
  \BibitemOpen
  \bibfield  {author} {\bibinfo {author} {\bibfnamefont {M.}~\bibnamefont
  {Eschrig}}\ and\ \bibinfo {author} {\bibfnamefont {T.}~\bibnamefont
  {L{\"{o}}fwander}},\ }\href {http://dx.doi.org/10.1038/nphys831} {\bibfield
  {journal} {\bibinfo  {journal} {Nature Phys.}\ }\textbf {\bibinfo {volume}
  {4}},\ \bibinfo {pages} {138} (\bibinfo {year} {2008})}\BibitemShut {NoStop}%
\bibitem [{\citenamefont {Sun}\ and\ \citenamefont {Shah}(2015)}]{Sun2015}%
  \BibitemOpen
  \bibfield  {author} {\bibinfo {author} {\bibfnamefont {K.}~\bibnamefont
  {Sun}}\ and\ \bibinfo {author} {\bibfnamefont {N.}~\bibnamefont {Shah}},\
  }\href {\doibase 10.1103/PhysRevB.91.144508} {\bibfield  {journal} {\bibinfo
  {journal} {Phys. Rev. B}\ }\textbf {\bibinfo {volume} {91}},\ \bibinfo
  {pages} {144508} (\bibinfo {year} {2015})}\BibitemShut {NoStop}%
\bibitem [{\citenamefont {Bergeret}\ and\ \citenamefont
  {Tokatly}(2013)}]{Bergeret2013}%
  \BibitemOpen
  \bibfield  {author} {\bibinfo {author} {\bibfnamefont {F.~S.}\ \bibnamefont
  {Bergeret}}\ and\ \bibinfo {author} {\bibfnamefont {I.~V.}\ \bibnamefont
  {Tokatly}},\ }\href {http://link.aps.org/doi/10.1103/PhysRevLett.110.117003}
  {\bibfield  {journal} {\bibinfo  {journal} {Phys. Rev. Lett.}\ }\textbf
  {\bibinfo {volume} {110}},\ \bibinfo {pages} {117003} (\bibinfo {year}
  {2013})}\BibitemShut {NoStop}%
\bibitem [{\citenamefont {Bergeret}\ and\ \citenamefont
  {Tokatly}(2014)}]{Bergeret2014}%
  \BibitemOpen
  \bibfield  {author} {\bibinfo {author} {\bibfnamefont {F.~S.}\ \bibnamefont
  {Bergeret}}\ and\ \bibinfo {author} {\bibfnamefont {I.~V.}\ \bibnamefont
  {Tokatly}},\ }\href {http://link.aps.org/doi/10.1103/PhysRevB.89.134517}
  {\bibfield  {journal} {\bibinfo  {journal} {Phys. Rev. B}\ }\textbf {\bibinfo
  {volume} {89}},\ \bibinfo {pages} {134517} (\bibinfo {year}
  {2014})}\BibitemShut {NoStop}%
\bibitem [{\citenamefont {Jacobsen}\ and\ \citenamefont
  {Linder}(2015)}]{Jacobsen2015}%
  \BibitemOpen
  \bibfield  {author} {\bibinfo {author} {\bibfnamefont {S.~H.}\ \bibnamefont
  {Jacobsen}}\ and\ \bibinfo {author} {\bibfnamefont {J.}~\bibnamefont
  {Linder}},\ }\href {\doibase 10.1103/PhysRevB.92.024501} {\bibfield
  {journal} {\bibinfo  {journal} {Phys. Rev. B}\ }\textbf {\bibinfo {volume}
  {92}},\ \bibinfo {pages} {024501} (\bibinfo {year} {2015})}\BibitemShut
  {NoStop}%
\bibitem [{\citenamefont {Yokoyama}\ and\ \citenamefont
  {Nazarov}(2014)}]{Yokoyama2014}%
  \BibitemOpen
  \bibfield  {author} {\bibinfo {author} {\bibfnamefont {T.}~\bibnamefont
  {Yokoyama}}\ and\ \bibinfo {author} {\bibfnamefont {Y.~V.}\ \bibnamefont
  {Nazarov}},\ }\href
  {http://iopscience.iop.org/article/10.1209/0295-5075/108/47009} {\bibfield
  {journal} {\bibinfo  {journal} {Europhys. Lett.}\ }\textbf {\bibinfo {volume}
  {108}},\ \bibinfo {pages} {47009} (\bibinfo {year} {2014})}\BibitemShut
  {NoStop}%
\bibitem [{\citenamefont {Yokoyama}\ \emph {et~al.}(2014)\citenamefont
  {Yokoyama}, \citenamefont {Eto},\ and\ \citenamefont
  {Nazarov}}]{Yokoyama2014a}%
  \BibitemOpen
  \bibfield  {author} {\bibinfo {author} {\bibfnamefont {T.}~\bibnamefont
  {Yokoyama}}, \bibinfo {author} {\bibfnamefont {M.}~\bibnamefont {Eto}}, \
  and\ \bibinfo {author} {\bibfnamefont {Y.~V.}\ \bibnamefont {Nazarov}},\
  }\href {http://iopscience.iop.org/article/10.1088/1742-6596/568/5/052035}
  {\bibfield  {journal} {\bibinfo  {journal} {J. Phys.: Conf. Ser.}\ }\textbf
  {\bibinfo {volume} {568}},\ \bibinfo {pages} {052035} (\bibinfo {year}
  {2014})}\BibitemShut {NoStop}%
\bibitem [{\citenamefont {Arjoranta}\ and\ \citenamefont
  {Heikkil\"a}(2016)}]{Arjoranta2016}%
  \BibitemOpen
  \bibfield  {author} {\bibinfo {author} {\bibfnamefont {J.}~\bibnamefont
  {Arjoranta}}\ and\ \bibinfo {author} {\bibfnamefont {T.~T.}\ \bibnamefont
  {Heikkil\"a}},\ }\href {\doibase 10.1103/PhysRevB.93.024522} {\bibfield
  {journal} {\bibinfo  {journal} {Phys. Rev. B}\ }\textbf {\bibinfo {volume}
  {93}},\ \bibinfo {pages} {024522} (\bibinfo {year} {2016})}\BibitemShut
  {NoStop}%
\bibitem [{\citenamefont {Jacobsen}\ \emph {et~al.}(2016)\citenamefont
  {Jacobsen}, \citenamefont {Kulagina},\ and\ \citenamefont
  {Linder}}]{Jacobsen2016}%
  \BibitemOpen
  \bibfield  {author} {\bibinfo {author} {\bibfnamefont {S.~H.}\ \bibnamefont
  {Jacobsen}}, \bibinfo {author} {\bibfnamefont {I.}~\bibnamefont {Kulagina}},
  \ and\ \bibinfo {author} {\bibfnamefont {J.}~\bibnamefont {Linder}},\ }\href
  {http://dx.doi.org/10.1038/srep23926} {\bibfield  {journal} {\bibinfo
  {journal} {Sci. Rep.}\ }\textbf {\bibinfo {volume} {6}},\ \bibinfo {pages}
  {23926} (\bibinfo {year} {2016})}\BibitemShut {NoStop}%
\bibitem [{\citenamefont {Josephson}(1962)}]{Josephson1962}%
  \BibitemOpen
  \bibfield  {author} {\bibinfo {author} {\bibfnamefont {B.~D.}\ \bibnamefont
  {Josephson}},\ }\href
  {http://linkinghub.elsevier.com/retrieve/pii/0031916362913690} {\bibfield
  {journal} {\bibinfo  {journal} {Phys. Lett.}\ }\textbf {\bibinfo {volume}
  {1}},\ \bibinfo {pages} {251} (\bibinfo {year} {1962})}\BibitemShut {NoStop}%
\bibitem [{\citenamefont {Josephson}(1964)}]{Josephson1964}%
  \BibitemOpen
  \bibfield  {author} {\bibinfo {author} {\bibfnamefont {B.~D.}\ \bibnamefont
  {Josephson}},\ }\href {http://link.aps.org/doi/10.1103/RevModPhys.36.216}
  {\bibfield  {journal} {\bibinfo  {journal} {Rev. Mod. Phys.}\ }\textbf
  {\bibinfo {volume} {36}},\ \bibinfo {pages} {216} (\bibinfo {year}
  {1964})}\BibitemShut {NoStop}%
\bibitem [{\citenamefont {Furusaki}\ and\ \citenamefont
  {Tsukada}(1991)}]{Furusaki1991}%
  \BibitemOpen
  \bibfield  {author} {\bibinfo {author} {\bibfnamefont {A.}~\bibnamefont
  {Furusaki}}\ and\ \bibinfo {author} {\bibfnamefont {M.}~\bibnamefont
  {Tsukada}},\ }\href
  {http://www.sciencedirect.com/science/article/pii/0038109891902016}
  {\bibfield  {journal} {\bibinfo  {journal} {Solid State Commun.}\ }\textbf
  {\bibinfo {volume} {78}},\ \bibinfo {pages} {299} (\bibinfo {year}
  {1991})}\BibitemShut {NoStop}%
\bibitem [{\citenamefont {Radovi{\'{c}}}\ \emph {et~al.}(2003)\citenamefont
  {Radovi{\'{c}}}, \citenamefont {Lazarides},\ and\ \citenamefont
  {Flytzanis}}]{Radovic2003}%
  \BibitemOpen
  \bibfield  {author} {\bibinfo {author} {\bibfnamefont {Z.}~\bibnamefont
  {Radovi{\'{c}}}}, \bibinfo {author} {\bibfnamefont {N.}~\bibnamefont
  {Lazarides}}, \ and\ \bibinfo {author} {\bibfnamefont {N.}~\bibnamefont
  {Flytzanis}},\ }\href {http://link.aps.org/pdf/10.1103/PhysRevB.68.014501}
  {\bibfield  {journal} {\bibinfo  {journal} {Phys. Rev. B}\ }\textbf {\bibinfo
  {volume} {68}},\ \bibinfo {pages} {014501} (\bibinfo {year}
  {2003})}\BibitemShut {NoStop}%
\bibitem [{\citenamefont {Keizer}\ \emph {et~al.}(2006)\citenamefont {Keizer},
  \citenamefont {Goennenwein}, \citenamefont {Klapwijk}, \citenamefont {Miao},
  \citenamefont {Xiao},\ and\ \citenamefont {Gupta}}]{Keizer2006}%
  \BibitemOpen
  \bibfield  {author} {\bibinfo {author} {\bibfnamefont {R.~S.}\ \bibnamefont
  {Keizer}}, \bibinfo {author} {\bibfnamefont {S.~T.~B.}\ \bibnamefont
  {Goennenwein}}, \bibinfo {author} {\bibfnamefont {T.~M.}\ \bibnamefont
  {Klapwijk}}, \bibinfo {author} {\bibfnamefont {G.}~\bibnamefont {Miao}},
  \bibinfo {author} {\bibfnamefont {G.}~\bibnamefont {Xiao}}, \ and\ \bibinfo
  {author} {\bibfnamefont {A.}~\bibnamefont {Gupta}},\ }\href
  {http://www.nature.com/doifinder/10.1038/nature04499} {\bibfield  {journal}
  {\bibinfo  {journal} {Nature}\ }\textbf {\bibinfo {volume} {439}},\ \bibinfo
  {pages} {825} (\bibinfo {year} {2006})}\BibitemShut {NoStop}%
\bibitem [{\citenamefont {De~Gennes}(1989)}]{DeGennes1989}%
  \BibitemOpen
  \bibfield  {author} {\bibinfo {author} {\bibfnamefont {P.~G.}\ \bibnamefont
  {De~Gennes}},\ }\href@noop {} {\emph {\bibinfo {title} {Superconductivity of
  Metals and Alloys}}}\ (\bibinfo  {publisher} {Addison Wesley},\ \bibinfo
  {year} {1989})\BibitemShut {NoStop}%
\bibitem [{\citenamefont {{\v{Z}}uti{\'{c}}}\ and\ \citenamefont
  {Valls}(1999)}]{Zutic1999}%
  \BibitemOpen
  \bibfield  {author} {\bibinfo {author} {\bibfnamefont {I.}~\bibnamefont
  {{\v{Z}}uti{\'{c}}}}\ and\ \bibinfo {author} {\bibfnamefont {O.~T.}\
  \bibnamefont {Valls}},\ }\href
  {http://link.aps.org/doi/10.1103/PhysRevB.60.6320} {\bibfield  {journal}
  {\bibinfo  {journal} {Phys. Rev. B}\ }\textbf {\bibinfo {volume} {60}},\
  \bibinfo {pages} {6320} (\bibinfo {year} {1999})}\BibitemShut {NoStop}%
\bibitem [{\citenamefont {{\v{Z}}uti{\'{c}}}\ and\ \citenamefont
  {Valls}(2000)}]{Zutic2000}%
  \BibitemOpen
  \bibfield  {author} {\bibinfo {author} {\bibfnamefont {I.}~\bibnamefont
  {{\v{Z}}uti{\'{c}}}}\ and\ \bibinfo {author} {\bibfnamefont {O.~T.}\
  \bibnamefont {Valls}},\ }\href
  {http://link.aps.org/doi/10.1103/PhysRevB.61.1555} {\bibfield  {journal}
  {\bibinfo  {journal} {Phys. Rev. B}\ }\textbf {\bibinfo {volume} {61}},\
  \bibinfo {pages} {1555} (\bibinfo {year} {2000})}\BibitemShut {NoStop}%
\bibitem [{\citenamefont {Beenakker}(1997)}]{Beenakker1997}%
  \BibitemOpen
  \bibfield  {author} {\bibinfo {author} {\bibfnamefont {C.~W.~J.}\
  \bibnamefont {Beenakker}},\ }\href
  {http://link.aps.org/doi/10.1103/RevModPhys.69.731} {\bibfield  {journal}
  {\bibinfo  {journal} {Rev. Mod. Phys.}\ }\textbf {\bibinfo {volume} {69}},\
  \bibinfo {pages} {731} (\bibinfo {year} {1997})}\BibitemShut {NoStop}%
\bibitem [{\citenamefont {Carbotte}(1990)}]{Carbotte1990}%
  \BibitemOpen
  \bibfield  {author} {\bibinfo {author} {\bibfnamefont {J.~P.}\ \bibnamefont
  {Carbotte}},\ }\href
  {http://journals.aps.org/rmp/abstract/10.1103/RevModPhys.62.1027} {\bibfield
  {journal} {\bibinfo  {journal} {Rev. Mod. Phys.}\ }\textbf {\bibinfo {volume}
  {62}},\ \bibinfo {pages} {1027} (\bibinfo {year} {1990})}\BibitemShut
  {NoStop}%
\bibitem [{\citenamefont {{De Franceschi}}\ \emph {et~al.}(1998)\citenamefont
  {{De Franceschi}}, \citenamefont {Giazotto}, \citenamefont {Beltram},
  \citenamefont {Sorba}, \citenamefont {Lazzarino},\ and\ \citenamefont
  {Franciosi}}]{DeFranceschi1998}%
  \BibitemOpen
  \bibfield  {author} {\bibinfo {author} {\bibfnamefont {S.}~\bibnamefont {{De
  Franceschi}}}, \bibinfo {author} {\bibfnamefont {F.}~\bibnamefont
  {Giazotto}}, \bibinfo {author} {\bibfnamefont {F.}~\bibnamefont {Beltram}},
  \bibinfo {author} {\bibfnamefont {L.}~\bibnamefont {Sorba}}, \bibinfo
  {author} {\bibfnamefont {M.}~\bibnamefont {Lazzarino}}, \ and\ \bibinfo
  {author} {\bibfnamefont {A.}~\bibnamefont {Franciosi}},\ }\href
  {http://scitation.aip.org/content/aip/journal/apl/73/26/10.1063/1.122926}
  {\bibfield  {journal} {\bibinfo  {journal} {Appl. Phys. Lett.}\ }\textbf
  {\bibinfo {volume} {73}},\ \bibinfo {pages} {3890} (\bibinfo {year}
  {1998})}\BibitemShut {NoStop}%
\bibitem [{\citenamefont {Vas'ko}\ \emph {et~al.}(1998)\citenamefont {Vas'ko},
  \citenamefont {Nikolaev}, \citenamefont {Larkin}, \citenamefont {Kraus},\
  and\ \citenamefont {Goldman}}]{Vasko1998}%
  \BibitemOpen
  \bibfield  {author} {\bibinfo {author} {\bibfnamefont {V.~A.}\ \bibnamefont
  {Vas'ko}}, \bibinfo {author} {\bibfnamefont {K.~R.}\ \bibnamefont
  {Nikolaev}}, \bibinfo {author} {\bibfnamefont {V.~A.}\ \bibnamefont
  {Larkin}}, \bibinfo {author} {\bibfnamefont {P.~A.}\ \bibnamefont {Kraus}}, \
  and\ \bibinfo {author} {\bibfnamefont {A.~M.}\ \bibnamefont {Goldman}},\
  }\href
  {http://scitation.aip.org/content/aip/journal/apl/73/6/10.1063/1.122020}
  {\bibfield  {journal} {\bibinfo  {journal} {Appl. Phys. Lett.}\ }\textbf
  {\bibinfo {volume} {73}},\ \bibinfo {pages} {844} (\bibinfo {year}
  {1998})}\BibitemShut {NoStop}%
\bibitem [{\citenamefont {Wan}\ \emph {et~al.}(2015)\citenamefont {Wan},
  \citenamefont {Kazakov}, \citenamefont {Manfra}, \citenamefont {Pfeiffer},
  \citenamefont {West},\ and\ \citenamefont {Rokhinson}}]{Wan2015}%
  \BibitemOpen
  \bibfield  {author} {\bibinfo {author} {\bibfnamefont {Z.}~\bibnamefont
  {Wan}}, \bibinfo {author} {\bibfnamefont {A.}~\bibnamefont {Kazakov}},
  \bibinfo {author} {\bibfnamefont {M.~J.}\ \bibnamefont {Manfra}}, \bibinfo
  {author} {\bibfnamefont {L.~N.}\ \bibnamefont {Pfeiffer}}, \bibinfo {author}
  {\bibfnamefont {K.~W.}\ \bibnamefont {West}}, \ and\ \bibinfo {author}
  {\bibfnamefont {L.~P.}\ \bibnamefont {Rokhinson}},\ }\href
  {http://dx.doi.org/10.1038/ncomms8426} {\bibfield  {journal} {\bibinfo
  {journal} {Nat. Commun.}\ }\textbf {\bibinfo {volume} {6}},\ \bibinfo {pages}
  {7426} (\bibinfo {year} {2015})}\BibitemShut {NoStop}%
\bibitem [{\citenamefont {Andreev}\ \emph {et~al.}(1991)\citenamefont
  {Andreev}, \citenamefont {Buzdin},\ and\ \citenamefont
  {Osgood}}]{Andreev1991}%
  \BibitemOpen
  \bibfield  {author} {\bibinfo {author} {\bibfnamefont {A.~V.}\ \bibnamefont
  {Andreev}}, \bibinfo {author} {\bibfnamefont {A.~I.}\ \bibnamefont {Buzdin}},
  \ and\ \bibinfo {author} {\bibfnamefont {R.~M.}\ \bibnamefont {Osgood}},\
  }\href {http://link.aps.org/doi/10.1103/PhysRevB.43.10124} {\bibfield
  {journal} {\bibinfo  {journal} {Phys. Rev. B}\ }\textbf {\bibinfo {volume}
  {43}},\ \bibinfo {pages} {10124} (\bibinfo {year} {1991})}\BibitemShut
  {NoStop}%
\bibitem [{\citenamefont {Demler}\ \emph {et~al.}(1997)\citenamefont {Demler},
  \citenamefont {Arnold},\ and\ \citenamefont {Beasley}}]{Demler1997}%
  \BibitemOpen
  \bibfield  {author} {\bibinfo {author} {\bibfnamefont {E.~A.}\ \bibnamefont
  {Demler}}, \bibinfo {author} {\bibfnamefont {G.~B.}\ \bibnamefont {Arnold}},
  \ and\ \bibinfo {author} {\bibfnamefont {M.~R.}\ \bibnamefont {Beasley}},\
  }\href {http://link.aps.org/doi/10.1103/PhysRevB.55.15174} {\bibfield
  {journal} {\bibinfo  {journal} {Phys. Rev. B}\ }\textbf {\bibinfo {volume}
  {55}},\ \bibinfo {pages} {15174} (\bibinfo {year} {1997})}\BibitemShut
  {NoStop}%
\bibitem [{\citenamefont {Brinkman}\ and\ \citenamefont
  {Golubov}(2000)}]{Brinkman2000}%
  \BibitemOpen
  \bibfield  {author} {\bibinfo {author} {\bibfnamefont {A.}~\bibnamefont
  {Brinkman}}\ and\ \bibinfo {author} {\bibfnamefont {A.~A.}\ \bibnamefont
  {Golubov}},\ }\href
  {http://journals.aps.org/prb/abstract/10.1103/PhysRevB.61.11297} {\bibfield
  {journal} {\bibinfo  {journal} {Phys. Rev. B}\ }\textbf {\bibinfo {volume}
  {61}},\ \bibinfo {pages} {11297} (\bibinfo {year} {2000})}\BibitemShut
  {NoStop}%
\bibitem [{\citenamefont {Bergeret}\ \emph
  {et~al.}(2001{\natexlab{b}})\citenamefont {Bergeret}, \citenamefont
  {Volkov},\ and\ \citenamefont {Efetov}}]{Bergeret2001a}%
  \BibitemOpen
  \bibfield  {author} {\bibinfo {author} {\bibfnamefont {F.~S.}\ \bibnamefont
  {Bergeret}}, \bibinfo {author} {\bibfnamefont {A.~F.}\ \bibnamefont
  {Volkov}}, \ and\ \bibinfo {author} {\bibfnamefont {K.~B.}\ \bibnamefont
  {Efetov}},\ }\href {\doibase 10.1103/PhysRevB.64.134506} {\bibfield
  {journal} {\bibinfo  {journal} {Phys. Rev. B}\ }\textbf {\bibinfo {volume}
  {64}},\ \bibinfo {pages} {134506} (\bibinfo {year}
  {2001}{\natexlab{b}})}\BibitemShut {NoStop}%
\bibitem [{\citenamefont {Golubov}\ \emph
  {et~al.}(2002{\natexlab{a}})\citenamefont {Golubov}, \citenamefont
  {Kupriyanov},\ and\ \citenamefont {Fominov}}]{Golubov2002}%
  \BibitemOpen
  \bibfield  {author} {\bibinfo {author} {\bibfnamefont {A.~A.}\ \bibnamefont
  {Golubov}}, \bibinfo {author} {\bibfnamefont {M.~Y.}\ \bibnamefont
  {Kupriyanov}}, \ and\ \bibinfo {author} {\bibfnamefont {Y.~V.}\ \bibnamefont
  {Fominov}},\ }\href@noop {} {\bibfield  {journal} {\bibinfo  {journal}
  {Pis'ma Zh. Eksp. Teor. Fiz.}\ }\textbf {\bibinfo {volume} {75}},\ \bibinfo
  {pages} {709} (\bibinfo {year} {2002}{\natexlab{a}})}\BibitemShut {NoStop}%
\bibitem [{\citenamefont {Golubov}\ \emph
  {et~al.}(2002{\natexlab{b}})\citenamefont {Golubov}, \citenamefont
  {Kupriyanov},\ and\ \citenamefont {Fominov}}]{Golubov2002alt}%
  \BibitemOpen
  \bibfield  {author} {\bibinfo {author} {\bibfnamefont {A.~A.}\ \bibnamefont
  {Golubov}}, \bibinfo {author} {\bibfnamefont {M.~Y.}\ \bibnamefont
  {Kupriyanov}}, \ and\ \bibinfo {author} {\bibfnamefont {Y.~V.}\ \bibnamefont
  {Fominov}},\ }\href {http://www.jetpletters.ac.ru/ps/594/article_9340.shtml}
  {\bibfield  {journal} {\bibinfo  {journal} {JETP Lett.}\ }\textbf {\bibinfo
  {volume} {75}},\ \bibinfo {pages} {588} (\bibinfo {year}
  {2002}{\natexlab{b}})}\BibitemShut {NoStop}%
\bibitem [{\citenamefont {Buzdin}\ and\ \citenamefont
  {Baladi\'e}(2003)}]{Buzdin2003}%
  \BibitemOpen
  \bibfield  {author} {\bibinfo {author} {\bibfnamefont {A.}~\bibnamefont
  {Buzdin}}\ and\ \bibinfo {author} {\bibfnamefont {I.}~\bibnamefont
  {Baladi\'e}},\ }\href {\doibase 10.1103/PhysRevB.67.184519} {\bibfield
  {journal} {\bibinfo  {journal} {Phys. Rev. B}\ }\textbf {\bibinfo {volume}
  {67}},\ \bibinfo {pages} {184519} (\bibinfo {year} {2003})}\BibitemShut
  {NoStop}%
\bibitem [{\citenamefont {Wang}\ \emph {et~al.}(2003)\citenamefont {Wang},
  \citenamefont {Xing},\ and\ \citenamefont {Sun}}]{Wang2003}%
  \BibitemOpen
  \bibfield  {author} {\bibinfo {author} {\bibfnamefont {J.}~\bibnamefont
  {Wang}}, \bibinfo {author} {\bibfnamefont {D.}~\bibnamefont {Xing}}, \ and\
  \bibinfo {author} {\bibfnamefont {H.}~\bibnamefont {Sun}},\ }\href
  {http://iopscience.iop.org/article/10.1088/0953-8984/15/27/315} {\bibfield
  {journal} {\bibinfo  {journal} {J. Phys. Condens. Matter}\ }\textbf {\bibinfo
  {volume} {15}},\ \bibinfo {pages} {4841} (\bibinfo {year}
  {2003})}\BibitemShut {NoStop}%
\bibitem [{\citenamefont {Nitta}\ \emph {et~al.}(1997)\citenamefont {Nitta},
  \citenamefont {Akazaki}, \citenamefont {Takayanagi},\ and\ \citenamefont
  {Enoki}}]{Nitta1997}%
  \BibitemOpen
  \bibfield  {author} {\bibinfo {author} {\bibfnamefont {J.}~\bibnamefont
  {Nitta}}, \bibinfo {author} {\bibfnamefont {T.}~\bibnamefont {Akazaki}},
  \bibinfo {author} {\bibfnamefont {H.}~\bibnamefont {Takayanagi}}, \ and\
  \bibinfo {author} {\bibfnamefont {T.}~\bibnamefont {Enoki}},\ }\href
  {\doibase 10.1103/PhysRevLett.78.1335} {\bibfield  {journal} {\bibinfo
  {journal} {Phys. Rev. Lett.}\ }\textbf {\bibinfo {volume} {78}},\ \bibinfo
  {pages} {1335} (\bibinfo {year} {1997})}\BibitemShut {NoStop}%
\bibitem [{\citenamefont {Koga}\ \emph {et~al.}(2002)\citenamefont {Koga},
  \citenamefont {Nitta}, \citenamefont {Akazaki},\ and\ \citenamefont
  {Takayanagi}}]{Nitta2002}%
  \BibitemOpen
  \bibfield  {author} {\bibinfo {author} {\bibfnamefont {T.}~\bibnamefont
  {Koga}}, \bibinfo {author} {\bibfnamefont {J.}~\bibnamefont {Nitta}},
  \bibinfo {author} {\bibfnamefont {T.}~\bibnamefont {Akazaki}}, \ and\
  \bibinfo {author} {\bibfnamefont {H.}~\bibnamefont {Takayanagi}},\ }\href
  {\doibase 10.1103/PhysRevLett.89.046801} {\bibfield  {journal} {\bibinfo
  {journal} {Phys. Rev. Lett.}\ }\textbf {\bibinfo {volume} {89}},\ \bibinfo
  {pages} {046801} (\bibinfo {year} {2002})}\BibitemShut {NoStop}%
\bibitem [{\citenamefont {Betthausen}\ \emph {et~al.}(2012)\citenamefont
  {Betthausen}, \citenamefont {Dollinger}, \citenamefont {Saarikoski},
  \citenamefont {Kolkovsky}, \citenamefont {Karczewski}, \citenamefont
  {Wojtowicz}, \citenamefont {Richter},\ and\ \citenamefont
  {Weiss}}]{Betthausen2012}%
  \BibitemOpen
  \bibfield  {author} {\bibinfo {author} {\bibfnamefont {C.}~\bibnamefont
  {Betthausen}}, \bibinfo {author} {\bibfnamefont {T.}~\bibnamefont
  {Dollinger}}, \bibinfo {author} {\bibfnamefont {H.}~\bibnamefont
  {Saarikoski}}, \bibinfo {author} {\bibfnamefont {V.}~\bibnamefont
  {Kolkovsky}}, \bibinfo {author} {\bibfnamefont {G.}~\bibnamefont
  {Karczewski}}, \bibinfo {author} {\bibfnamefont {T.}~\bibnamefont
  {Wojtowicz}}, \bibinfo {author} {\bibfnamefont {K.}~\bibnamefont {Richter}},
  \ and\ \bibinfo {author} {\bibfnamefont {D.}~\bibnamefont {Weiss}},\ }\href
  {http://science.sciencemag.org/content/337/6092/324} {\bibfield  {journal}
  {\bibinfo  {journal} {Science}\ }\textbf {\bibinfo {volume} {337}},\ \bibinfo
  {pages} {324} (\bibinfo {year} {2012})}\BibitemShut {NoStop}%
\bibitem [{\citenamefont {Zheng}\ and\ \citenamefont {Xing}(2009)}]{Zheng2009}%
  \BibitemOpen
  \bibfield  {author} {\bibinfo {author} {\bibfnamefont {Z.~M.}\ \bibnamefont
  {Zheng}}\ and\ \bibinfo {author} {\bibfnamefont {D.~Y.}\ \bibnamefont
  {Xing}},\ }\href
  {http://iopscience.iop.org/article/10.1088/0953-8984/21/38/385703} {\bibfield
   {journal} {\bibinfo  {journal} {J. Phys. Condens. Matter}\ }\textbf
  {\bibinfo {volume} {21}},\ \bibinfo {pages} {385703} (\bibinfo {year}
  {2009})}\BibitemShut {NoStop}%
\end{thebibliography}%

\end{document}